  \documentclass{EngC}
  \usepackage{makeidx}
  \usepackage{amsmath}% if you are using this package, it must be loaded before amsthm.sty
  \usepackage{amsthm}
  \usepackage{amssymb}
  \usepackage{mathtools}
  
  \usepackage[rightcaption,raggedright]{sidecap}% for side captions
  \usepackage{soul}           % for letterspacing in theorem-style headings

  \usepackage{rotating}
  \usepackage{floatpag}
  \rotfloatpagestyle{empty}

  \usepackage{graphicx}

  \usepackage{cite}

  \usepackage[named]{algo}
  \usepackage[noend]{algpseudocode}
  \usepackage{algorithm}
  \usepackage{color}
  
  \makeindex
  
  \theoremstyle{plain}% default
  \newtheorem{theorem}{Theorem}[section]
  
  \newtheorem{proposition}[theorem]{Proposition}

  \theoremstyle{definition}

  \theoremstyle{remark}

\begin{document}
  %Reduce spacing above and below equations
	\setlength{\abovedisplayskip}{2pt}
	\setlength{\belowdisplayskip}{2pt}

  	\title[snRadBookChapter: CUP Style]
    {LaTeX2e guide for authors using the \cambridge\ design}

  	\author{ALI WOOLLATT\\[3\baselineskip]
    This guide was compiled using \hbox{\cambridge.cls \version}\\[\baselineskip]
    The latest version can be downloaded from:
    https://authornet.cambridge.org/information/productionguide/
      LaTeX\_files/\cambridge.zip}

  %\maketitle
  % chap1.tex
% 2011/02/03, v1.10
%	\include{chap1}% introduction
% chap1.tex
% 2011/02/03, v1.10

\chapterauthor{Kumar Vijay Mishra\footnotemark\
    and Yonina C. Eldar\footnotemark
    \affil{Andrew and Erna Viterbi Faculty of Electrical Engineering, Technion - Israel Institute of Technology, Haifa, Israel}}

\chapter{Sub-Nyquist Radar: Principles and Prototypes}
\label{chap:snyq}

\footnotetext[1]{K.V.M. acknowledges partial support via Andrew and Erna Finci Viterbi Postdoctoral Fellowship and Lady Davis Postdoctoral Fellowship.}
\footnotetext[2]{This work is supported by the European Unions Horizon 2020 research and innovation program under grant agreement no. 646804-ERC-COG-BNYQ.}

\section{Introduction}
\label{sec:introduction}
Radar remote sensing advanced tremendously over the past several decades and is now applied to diverse areas such as military surveillance, meteorology, geology, collision-avoidance, and imaging \cite{skolnik2008radar}. In monostatic pulse Doppler radar systems, an antenna transmits a periodic train of known narrowband pulses within a defined coherent processing interval
(CPI). When the radiated wave from the radar interacts with moving targets, the amplitude, frequency, and polarization state of the scattered wave change. By monitoring this change, it is possible to infer the targets' size, location, and radial Doppler velocity. The reflected signal received by the radar antenna is a linear combination of echoes from multiple targets; each of these is an attenuated, time-delayed, and frequency-modulated version of the transmit signal. The delay in the received signal is linearly proportional to the target's range or its distance from the radar. The frequency modulation encodes the Doppler velocity of the target. The complex amplitude or target's reflectivity is a function of the target's size, geometry, propagation, and scattering mechanism. Radar signal processing is aimed at detecting the targets and estimating their parameters from the output of this linear, time-varying system. 

Traditional radar signal processing employs matched filtering (MF) or pulse compression \cite{peebles1998radar} in the digital domain wherein the sampled received signal is correlated with a replica of the transmit signal in the delay-Doppler plane. The MF maximizes the signal-to-noise ratio (SNR) in the presence of additive white Gaussian noise. In some specialized systems, this stage is replaced by a \textit{mismatched filter} with a different optimization metric such as minimization of peak-to-sidelobe-ratio of the output. Here, the received signal is correlated with a signal that is close but not identical to the transmit signal \cite{cilliers2007pulse,george2010implementation,mishra2012signal}. While all of these techniques reliably estimate target parameters, their resolution is inversely proportional to the support of the ambiguity function of the transmit pulse thereby restricting ability to super-resolve targets that are closely spaced.

The digital MF operation requires the signal to be sampled at or above the Nyquist sampling rate which guarantees perfect reconstruction of a bandlimited analog signal \cite{eldar2015sampling}. Many modern radar systems use wide bandwidths, typically ranging from hundreds of MHz to even GHz, in order to achieve fine radar range resolution. Since the Nyquist sampling rate is twice the baseband bandwidth, the radar receiver requires expensive, high-rate analog-to-digital converters (ADCs). The sampled signal is then also processed at high rates resulting in significant power, cost, storage, and computational overhead. Recently, in order to mitigate this rate bottleneck, new methods have been proposed that sample signals at sub-Nyquist rates\index{sub-Nyquist radar}
 and yet are able to estimate the targets' parameters \cite{eldar2012compressed,eldar2015sampling}.

Analogous trade-offs arise in other aspects of radar system design. For example, the number of transmit pulses governs the resolution in Doppler velocity. The estimation accuracy of target parameters is greatly affected by the radar's \textit{dwell time} \cite{skolnik2008radar}, i.e., the time duration a directional radar beam spends hitting a particular target. Long dwell times imply a large number of transmit pulses and, therefore, high Doppler precision. But, simultaneously, this negatively affects the ability of the radar to look at targets in other directions. Sub-Nyquist sampling approaches have, therefore, been suggested for the pulse dimension or ``slow time" domain in order to break the link between dwell time and Doppler resolution \cite{cohen2016reduced,akhtar2017compressed,mishra2014compressed}.

Finally, radars that deploy antenna arrays deal with similar sampling problems in the spatial domain. A \textit{phased array} radar antenna consists of several radiating elements that form a highly directional radiating beam pattern. Without requiring any mechanical motion, a phased array accomplishes beam-steering electronically by adjusting the relative phase of excitation in the array elements. The operational advantage is the agile scanning of the target scene, ability to track a large number of targets, and efficient search-and-track in the regions-of-interest \cite{galati1994advanced}. The beam pattern of individual array elements, array geometry, and its size define the overall antenna pattern \cite{bayliss1968design,cheng1971optimization} wherein high spatial resolution is achieved by large array apertures. As per the Nyquist Theorem, the array must not admit less than two signal samples per spatial period (i.e., radar's operating wavelength) \cite{haupt2015timed}. Otherwise, it introduces \textit{spatial aliasing} or multiple beams in the antenna pattern, thereby reducing its directivity. Often an exceedingly large number of radiating elements are required to synthesize a given array aperture in order to enhance the radar's ability to unambiguously distinguish closely spaced targets; the associated cost, weight, and area may be unacceptable. It is therefore desirable to apply sub-Nyquist techniques to thin a huge array without causing degradation in spatial resolution \cite{mishra2017high,rossi2014spatial,cohen2016summer}. 

Sub-Nyquist sampling leads to the development of low-cost, power-efficient, and small size radar systems that can scan faster and acquire larger volumes than traditional systems. Apart from design benefits, other applications of such systems have been envisioned recently including imparting hardware feasible cognitive abilities to the radar \cite{mishra2016cognitive,mishra2017performance}, a role in devising spectrally coexistent systems \cite{cohen2017spectrum}, and extension to imaging \cite{aberman2017sub}. In this chapter, we provide an overview of sub-Nyquist radars, their applications and hardware realizations.

The outline of the chapter is as follows. In the next section, we overview various reduced-rate techniques for radar system design and explain the benefits of our approach to sub-Nyquist radars. In Section~\ref{sec:snrad}, we describe the principles, algorithms, and hardware realization of temporal sub-Nyquist monostatic pulse-Doppler radar. Section~\ref{sec:rtot} presents an extension of the sub-Nyquist principle to slow-time. We then introduce the cognitive radar concept based on sub-Nyquist reception in Section~\ref{sec:specshare} and show an application to coexistence in a spectrally crowded environment. Section~\ref{sec:mimo} is devoted to spatial sub-Nyquist applications in multiple-input-multiple-output (MIMO) array radars. Finally, we consider sub-Nyquist synthetic aperture radar (SAR) imaging in Section~\ref{sec:sar}, followed by concluding remarks in Section~\ref{sec:summ}.

%\subruninhead{Notations} 
%Throughout the chapter, we reserve boldface lowercase, boldface uppercase, and calligraphic letters for vectors, matrices, and index sets, respectively. We denote the transpose and Hermitian by $(\cdot)^T$ and $(\cdot)^H$, respectively. The Kronecker and Hadamard (point-wise) products are denoted by $\otimes$ and $\odot$, respectively. The notation $\text{tr}\left\lbrace \cdot \right\rbrace $ is the trace of the matrix, $|\cdot|$ is the determinant and $\mathbb{E}\left[ \cdot \right]$ is the statistical expectation function. The notation $x \sim \mathcal{U}(u_l,u_u)$ represents a random variable drawn from the uniform distribution in the interval $[u_l,u_u]$ and $x \sim \mathcal{N}(\mu_x,\sigma_x^2)$ is the normal distribution with mean $\mu_x$ and variance $\sigma_x^2$. The functions $\text{max}$ and $\text{min}$ output the maximum and minimum value of their arguments, respectively. The function $\text{diag}(\cdot)$ outputs a diagonal matrix with the input vector along its main diagonal while $\text{vec}(\cdot)$ vectorizes a matrix by stacking its columns. The Fourier matrix $\mathbf{F}_N$ is a matrix of size $N\times N$ with $(n,k)$th entry given by $e^{\frac{-j2\pi nk}{N}}$.

\section{Prior Art and Historical Notes}
There is a large body of literature on reduced rate sampling techniques for radars. Most of these works employ compressed sensing (CS) methods which allow recovery of sparse, undersampled signals from random linear measurements \cite{eldar2012compressed}. A pre-2010 review of selected applications of CS-based radars can be found in \cite{ender2010compressive}. A qualitative, system-level commentary from the point of view of operational radar engineers is available in \cite{goodman2015pitfalls} while CS-based radar imaging studies are summarized in \cite{zhao2016race}. An excellent overview on sparsity-based SAR imaging methods is provided in \cite{cetin2014sparsity}. The review in \cite{hadi2015compressive} recaps major developments in this area from a non-mathematical perspective. In the following, we review the most significant works relevant to the sub-Nyquist formulations presented in this chapter.\\

\noindent\textbf{On-grid CS} The earliest application of CS to recover time-delays with sub-Nyquist samples in a noiseless case was formulated in \cite{baraniuk2007compressive}. CS-based parameter estimation for both delay and Doppler shifts was proposed in \cite{herman2009high} with samples acquired at the Nyquist rate. These and similar later works \cite{yoon2008compressed,tan2009range,zhang2012adaptive} discretize the delay-Doppler domain assuming that targets lie on a grid. Subsequently, these ideas were extended to colocated \cite{chen2009signal, yu2011measurement} and distributed \cite{yu2010mimo} MIMO radars where targets are located on an angle-Doppler-range grid. In practice, target parameters are typically continuous values whose discretization may introduce gridding errors \cite{chi2011sensitivity}. In particular, \cite{herman2009high} constructs a dictionary that exhaustively considers all possible delay-Doppler pairs thereby rendering the processing computationally expensive. Noise and clutter mitigation are not considered in this literature. Simulations show that such systems typically have poor performance in clutter-contaminated noisy environments.\\

\noindent\textbf{Off-grid CS} A few recent works \cite{heckel2016superrad,heckel2016super} formulate the radar parameter estimation for off-grid targets using atomic norm minimization \cite{tang2013compressed,mishra2014super}. However, these methods do not address direct analog sampling, presence of noise and clutter. Further details on this approach are available in Chapter 7 (``Super-resolution radar imaging via convex optimization'') of this book.\\

\noindent\textbf{Parametric Recovery} A different approach was suggested in \cite{bajwa2011identification} which treated radar parameter estimation as the identification of an underlying linear, time-varying system \cite{kozek2005identification}. The proposed two-stage recovery algorithm, largely based on \cite{gedalyahu2010time}, first estimates target delays and then utilizes these recovered delays to estimate Doppler velocities, and complex reflectivities. They also provide guarantees for system identification in terms of the minimum time-bandwidth product of the input signal. However, this method does not handle noise well.\\

\noindent\textbf{Matrix Completion} In some radar applications, the received signal samples are processed as data matrices which, under certain conditions, are low rank. In this context, general works have suggested retrieving the missing entries using matrix completion methods \cite{mishra2014compressed,sun2015mimo}. The target parameters are then recovered through classic radar signal processing. These techniques have not been exhaustively evaluated for different signal scenarios and their practical implementations have still not been thoroughly examined.\\

\noindent\textbf{Finite-Rate-of-Innovation (FRI) Sampling}\index{sub-Nyquist radar} The received radar signal from $L$ targets can be modeled with $3L$ degrees of freedom because three parameters - time delay, Doppler shift, and attenuation coefficient - characterize each target. The classes of signals that have finite degrees of freedom per unit of time are called finite-rate-of-innovation (FRI) signals\index{finite rate of innovation}
 \cite{vetterli2002sampling}. Low-rate sampling and signal recovery strategies for FRI signals have been studied in detail in the past \cite[Chapter 15]{eldar2015sampling}. In \cite{baransky2014prototype}, a temporal sub-Nyquist radar was proposed to recover delays relying on the FRI model. The Xampling framework \cite{eldar2015sampling} was used to obtain Fourier coefficients from low-rate samples with a practical hardware prototype. Similar techniques were later studied for delay channel estimation problems in ultra-wideband communication systems \cite{cohen2014channel,mishra2017sub} and for ultrasound imaging \cite{chernyakova2014fourier}. In \cite{barilan2014focusing}, \textit{Doppler focusing} was added to the FRI-Xampling framework to recover both delays and Dopplers. Doppler focusing is a narrowband technique which can be interpreted as low-rate beamforming in the frequency domain, and was applied earlier to ultrasound imaging \cite{tur2011innovation,wagner2012compressed}. It considers a chosen center frequency with a band of frequencies around it and coherently processes multiple echoes in this focused region to estimate the delays from low-rate samples.\\

\noindent\textbf{Extensions of Sub-Nyquist Radars} \index{sub-Nyquist radar!extensions}The system proposed in \cite{barilan2014focusing} reduces samples only in time and not in the Doppler domain. Since the set of frequencies for Doppler focusing is usually fixed \textit{a priori}, the resultant Doppler resolution is limited by the focusing; it remains inversely proportional to the number of pulses $P$ as is also the case with conventional radar. In \cite{cohen2016reduced}, sub-Nyquist processing in slow-time was introduced to recover the target range and Doppler by simultaneously transmitting few pulses in the CPI and sampling the received signals at sub-Nyquist rates. Later, \cite{eldar2015clutter} proposed a whitening procedure to mitigate the presence of clutter in a sub-Nyquist radar. Spatial-domain compressed sensing (SCS) \index{spatial compressed sensing} was examined for a MIMO array radar in \cite{rossi2014spatial} and later for phased arrays in \cite{mishra2017high}. Recently, \cite{cohen2016summer} proposed Xampling in time and space to recover delay, Doppler, and azimuth of the targets by \textit{thinning} a colocated MIMO array and collecting low-rate samples at each receive element. This sub-Nyquist MIMO radar (SUMMeR) system was also conceptually demonstrated in hardware \cite{mishra2016cognitive,cohen2017sub}. The formulation in \cite{na2018tendsur} proposes Tensor-Based 3D Sub-Nyquist Radar (TenDSuR) that performs thinning in all three domains and recovers the signal via tensor-based recovery. 
Finally, an extension to SAR was demonstrated in \cite{aberman2017sub}. Table~\ref{tbl:snyqmethods} summarizes these developments.

% For tables use
%
\begin{table}
\caption{Sub-Nyquist radars and their corresponding reduction domains}
\label{tbl:snyqmethods}       % Give a unique label
%
% Follow this input for your own table layout
%
\begin{tabular}{p{5.0cm}p{1.5cm}p{1.5cm}p{1.5cm}}
\hline\noalign{\smallskip}
Sub-Nyquist System & Temporal & Doppler & Spatial \\
\noalign{\smallskip}
\hline
\noalign{\smallskip}
Monostatic pulsed radar \cite{baransky2014prototype}  & Yes  & No & No\\
Monostatic pulse Doppler radar \cite{barilan2014focusing}  & Yes  & No & No\\
Reduced time-on-target radar \cite{cohen2016reduced} & Yes  & Yes & No\\
MIMO SCS \cite{rossi2014spatial} & No  & No & Yes\\
Phased array SCS \cite{mishra2017high} & No  & No & Yes\\
SUMMeR \cite{cohen2016summer} \cite{mishra2017high} & Yes  & No & Yes\\
TenDSuR \cite{na2018tendsur} & Yes  & Yes & Yes\\
Sub-Nyquist SAR \cite{aberman2017sub} & Yes  & No & No\\
\noalign{\smallskip}\hline\noalign{\smallskip}
\end{tabular}
\end{table}

%Classical Radar snRadar advantages

\section{Temporal Sub-Nyquist Radar}
\label{sec:snrad}
%%% Radar Tx
\index{sub-Nyquist radar!temporal sub-Nyquist}Consider a standard pulse-Doppler radar that transmits a pulse train
\begin{equation}
\label{eq:uni_model}
r_{T_X}(t)= \sum_{p=0}^{P-1} h(t-p\tau), \quad 0 \leq t \leq P \tau,
\end{equation}
consisting of $P$ uniformly spaced known pulses $h(t)$. The interpulse transmit delay $\tau$ is the pulse repetition interval (PRI) or fast time; its reciprocal is the pulse repetition frequency (PRF). The entire duration of $P$ pulses in (\ref{eq:uni_model}) is known as the CPI or slow time. The radar operates at carrier frequency $f_c$ so that its wavelength is $\lambda = c/f_c$, where $c=3\times10^8$ m/s is the speed of light.

In a conventional pulse Doppler radar, the pulse $h(t)=h_{\text{Nyq}}(t)$ is a time-limited baseband function whose continuous-time Fourier transform (CTFT) is $H_{\text{Nyq}}(f)=\int_{-\infty}^{\infty} h_{\text{Nyq}}(t) e^{-j 2\pi f t} \mathrm{d}t$. It is assumed that most of the signal's energy lies within the frequencies $\pm B_h/2$, where $B_h$ denotes the effective signal bandwidth, such that the following approximation holds:
\begin{align}
h_{\text{Nyq}}(t) \approx \int\limits_{-B_h/2}^{B_h/2} H_{\text{Nyq}}(f) e^{j 2\pi f t} \mathrm{d}f.
\end{align}
The total transmit power of the radar is defined as
\begin{equation} \label{eq:pt1}
\int_{-B_h/2}^{B_h/2} |H_{\text{Nyq}}(f)|^2\, \mathrm{d}f = P_T.
\end{equation}

\subsection{Received Signal Model}
\label{subsec:rxsig}
%%% Targets and radar Rx
Assume that the radar target scene consists of $L$ non-fluctuating point-targets, according to the Swerling-0 target model \cite{skolnik2008radar}. The transmit signal is reflected back by the $L$ targets and these echoes are received by the radar processor. The latter aims at recovering the following information about the $L$ targets from the received signal: time delay $\tau_l$, which is linearly proportional to the target's range $d_l = c\tau_l/2$; Doppler frequency $\nu_l$, proportional to the target's radial velocity $v_l = c\nu_l/4\pi f_c$; and complex amplitude $\alpha_l$. The target locations are defined with respect to the polar coordinate system of the radar and their range and Doppler are assumed to lie in the unambiguous time-frequency region, i.e., the time delays are no longer than the PRI and Doppler frequencies are up to the PRF. 

Typically, the radar assumes the following operating conditions which leads to a simplified expression
for the received signal \cite{barilan2014focusing}:
\begin{description}
\item[A1]{``Far targets" - assuming $\nu_l \ll 2\pi f_c\tau_l/P\tau$, target radar distance is large compared to the distance change
during the CPI over which attenuation $\alpha_l$ is allowed to be constant.}
\item[A2]{``Slow targets" - assuming $\nu_l \ll 2\pi f_c/P\tau B_h$, target velocity is small enough to allow for constant $\tau_l$ during the CPI. In this case, the following piecewise-constant approximation holds
$\nu_lt \approx \nu_lp\tau$, for $t \in [p\tau, (p + 1)\tau]$.}
\item[A3]{``Small acceleration" - assuming $d\nu_l/dt \ll c/2f_c(P\tau)^2$, target velocity remains approximately constant during the CPI allowing for constant $\nu_l$.}
\item[A4]{``No time or Doppler ambiguities" - The delay-Doppler pairs $(\tau_l, \nu_l)$ for all $l \in [1, L]$ lie in the radar's unambiguous region of delay-Doppler plane defined by $[0, \tau] \times
[-\pi/\tau, \pi/\tau]$.}
\item[A5]{The pairs in the set $(\tau_l, \nu_l)$ for all $l \in [1, L]$ are unique.}
\end{description}

Under the above assumptions, the received signal can be written as
\begin{equation}
\label{eq:uni_rec}
r_{R_X}(t)= \sum_{p=0}^{P-1} \sum_{l=0}^{L-1} \alpha_l h(t-\tau_l - p\tau) e^{-j \nu_l t} + w(t),
\end{equation}
for $0 \leq t \leq P\tau$, where $w(t)$ is a zero mean wide-sense stationary random signal with autocorrelation $r_w(s) = \sigma^2\delta(s)$. It is convenient to express $r_{R_X}(t)$ as a sum of single frames
\begin{equation}
\label{eq:frames_pdr}
r_{R_X}(t)= \sum_{p=0}^{P-1} r_{R_X}^p(t) + w(t),
\end{equation}
where
\begin{equation}
\label{eq:one_frame_pdr}
r_{R_X}^p(t)= \sum_{l=0}^{L-1} \alpha_l h(t-\tau_l - p\tau) e^{-j \nu_l p \tau},
\end{equation}
for $p\tau \leq t \leq (p+1) \tau$, is the return signal from the $p$th pulse. %In pulse Doppler radar, the goal is to recover the targets range and Doppler frequency, namely the time delays $\tau_l$ and Doppler shifts $\nu_l$, from the received signals $r_R^p(t), 0 \leq p \leq P-1$.

%%% Classic radar and processing
A classical radar signal processor samples each incoming frame $r_{R_X}^p(t)$ at the Nyquist rate $B_h$ to yield the digitized samples $r_{R_X}^p[n], 0 \leq n \leq N-1$, where $N=\tau B_h$. The signal enhancement process employs a MF for the sampled frames $r_{R_X}^p[n]$. This is then followed by Doppler processing where a $P$-point discrete Fourier transform (DFT) is performed on slow time samples. By stacking all the $N$ DFT vectors together, a delay-Doppler map is obtained for the target scene. Finally, the time delays $\tau_l$ and Doppler shifts $\nu_l$ of the targets are located on this map using, e.g., a constant false-alarm rate detector \cite{kay1998fundamentals}.

\subsection{Sub-Nyquist Delay-Doppler Recovery}
\label{subsec:delaydoprec}
Traditional radar systems sample the received signal at the Nyquist rate, determined by the baseband bandwidth of $h(t)$. Our goal is to recover $r_{R_X}^p(t)$ from its samples taken far below this rate. We note that over the interval $\tau$, $r_{R_X}^p(t)$ is completely specified by $\{\alpha_l, \tau_l, \nu_l\}_{l=1}^L$, and is an FRI signal with rate of innovation $3L/\tau$. Hence, in the absence of noise, one would expect to be able to accurately recover $r_{R_X}^p(t)$ from only a few samples per time $\tau$. Since radar signals tend to be sparse in the time domain, simply acquiring a few data samples at a low rate will not generally yield adequate recovery. Indeed, if the separation between samples is larger than the effective spread in time, then with high probability many of the samples will be close to zero. This implies that presampling analog processing must be performed on the frequency-domain support of the radar signal in order to smear the signal in time before low rate sampling. 

The approach we adopt follows the Xampling\index{Xampling} architecture designed for sampling and processing of analog inputs at rates far below Nyquist, whose underlying structure can be modeled as a union of subspaces (UoS). The input signal belongs to a single subspace, a priori unknown, out of multiple, possibly even infinitely many, candidate subspaces. Xampling consists of two main functions: low rate analog to digital conversion (ADC), in which the input is compressed in the analog domain prior to sampling with commercial devices, and low rate digital signal processing, in which the input subspace is detected prior to digital signal processing. The resulting sparse recovery is performed using CS techniques adapted to the analog setting. This concept has been applied to both communications \cite{mishali2010theory, mishali2011bridging, cohen2014sub, cohen2016cyclo} and radar \cite{barilan2014focusing, cohen2016towards}, among other applications.

Time-varying linear systems, which introduce both time-shifts (delays) and frequency-shifts (Doppler-shifts), such as those arising in surveillance point-target radar systems, fit nicely into the UoS model. Here, a sparse target scene is assumed, allowing to reduce the sampling rate without sacrificing delay and Doppler resolution. The Xampling-based system is composed of an ADC which filters the received signal to predetermined frequencies before taking point-wise samples. These compressed samples, or ``Xamples", contain the information needed to recover the desired signal parameters.

To demonstrate sub-Nyquist sampling, we begin by deriving an expression for the Fourier coefficients of the received signal and show that the target parameters are embodied in them. Let $\mathcal{F}_R$ and $f_{\text{Nyq}}$ be the set of all frequencies in the received signal spectrum and the corresponding Nyquist sampling rate, respectively. Consider the Fourier series representation of the aligned frames $r_{R_X}^p(t+p \tau)$, with $r_{R_X}^p(t)$ defined in (\ref{eq:one_frame_pdr}):
\begin{eqnarray}
\label{eq:fourier_coeff}
c_p[k] = \int_0^{\tau} r_{R_X}^p(t+p \tau) e^{-j 2\pi k t/\tau} \mathrm{d}t = \frac{1}{\tau} H[k] \sum_{l=0}^{L-1} \alpha_l e^{-j 2 \pi k \tau_l / \tau} e^{-j \nu_l p \tau},
\end{eqnarray}
for $k \in \kappa$, where $\kappa= \left\{k = \left. \left\lfloor \frac{f}{f_{\text{Nyq}}}N \right\rfloor \right| \, f \in \mathcal{F}_R \right\}$. From (\ref{eq:fourier_coeff}), we see that the unknown parameters $\{ \alpha_l, \tau_l, \nu_l \}_{l=0}^{L-1}$ are embodied in the Fourier coefficients $c_p[k]$. We can estimate these parameters using only a small number of Fourier coefficients which translates to a low sampling rate.

\begin{figure*}[t]
%\centering
	\includegraphics[width=1\textwidth]{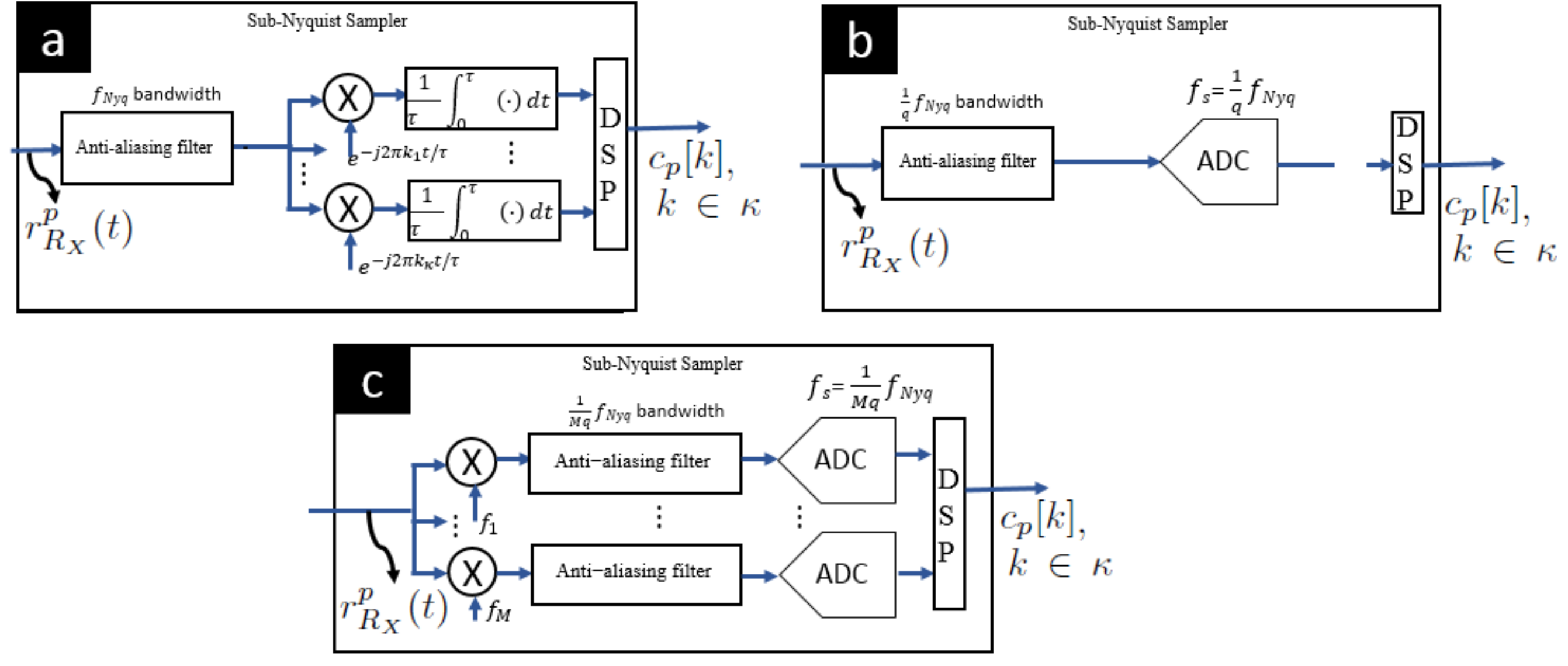}
	\caption{Sub-Nyquist sampling methods: (a) direct sampling (b) low frequencies only (c) multiband sampling.}
	\label{fig:sampling}
\end{figure*}

There are several ways to implement a sub-Nyquist sampler\index{sub-Nyquist radar!sampling methods} \cite{gedalyahu2011multichannel,mishra2017sub} in order to obtain a set of Fourier coefficients from low-rate samples of the signal. For simplicity, consider $|\kappa| = K$ such that $q=N/K$ is an integer defining the sampling reduction factor. In \textit{direct sampling} (Fig.~\ref{fig:sampling}a), the signal $r_{R_x}(t)$ obtained after the anti-aliasing filter is passed through as many analog chains as the number of sub-Nyquist coefficients $K$. Each branch is modulated by a complex exponential, followed by integration over $\tau$ and necessary digital signal processing (DSP). This technique provides the largest flexibility in choosing the Fourier coefficients, but is expensive in terms of hardware. Another approach is to limit the bandwidth of the anti-aliasing filter such that only the \textit{lowest $K$} frequencies are free of aliasing (Fig.~\ref{fig:sampling}b). We then sample these lowest $K$ frequencies. Here, the measurements are correlated and a modification in the analog hardware is also required so that the anti-aliasing filter has reduced passband. In the \textit{multiband} sampling method shown in Fig.~\ref{fig:sampling}c, $M$ disjoint randomly-chosen groups of consecutive Fourier coefficients are obtained such that the total number of sampled coefficients is $K$. This translates to splitting the signal across $M$ branches, passing the downconverted signal through reduced-passband anti-aliasing filters, and then sampling each band with a low-rate ADC. This method can be easily implemented but requires $M$ low-rate ADCs. The sub-Nyquist hardware prototypes developed in \cite{barilan2014focusing,baransky2014prototype} adopt multiband sampling using four groups of consecutive coefficients. In practice, the specific Fourier coefficients are chosen through extensive software simulations to provide low mutual coherence \cite{eldar2015sampling} for CS-based signal recovery.
%-----------------------------------------------------------------------------------
%-----------------------------------------------------------------------------------

\begin{SCfigure}[50][t]
%\centering
	\includegraphics[width=0.35\textwidth]{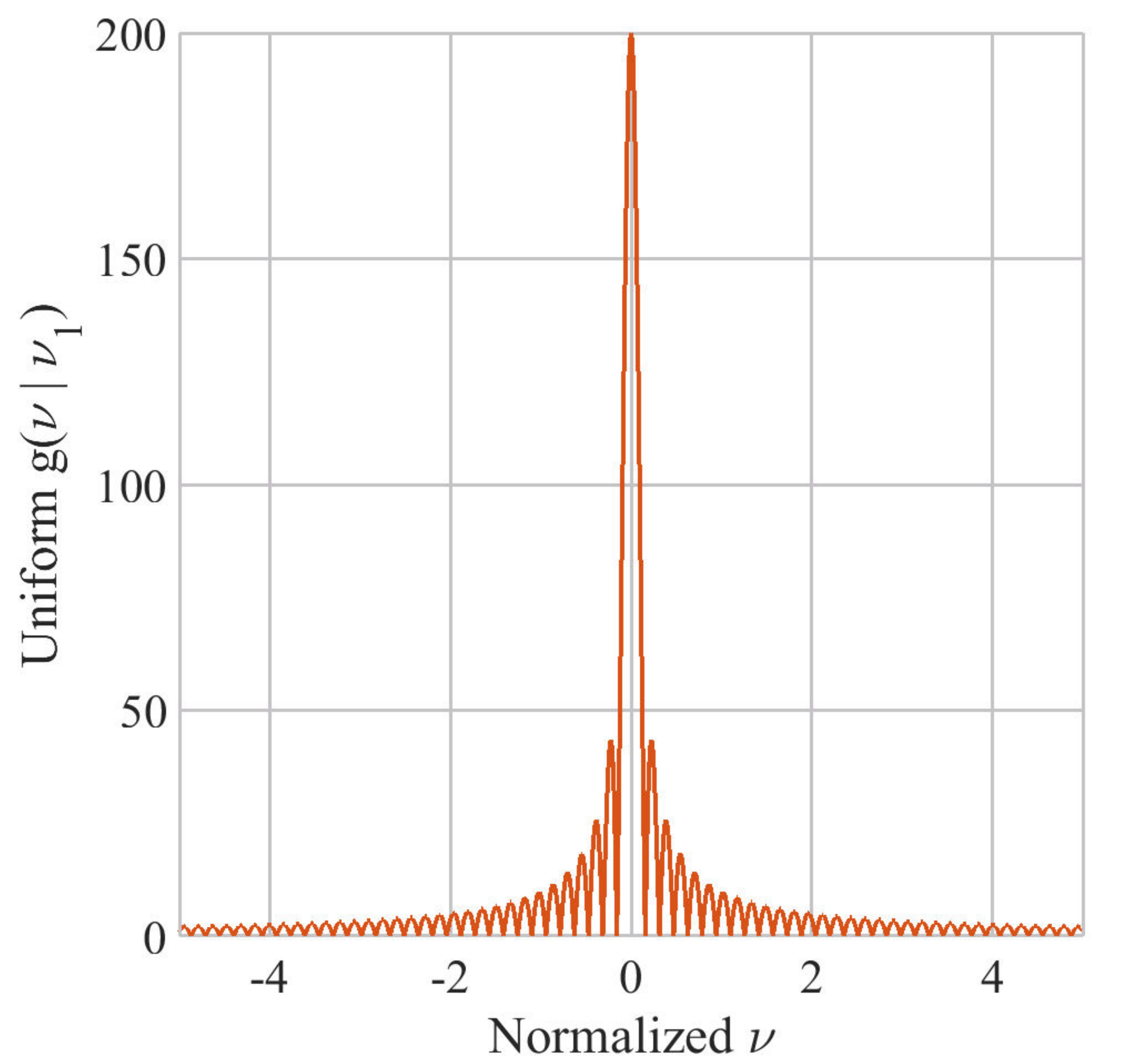}
	\caption{Sum of exponents $|g(\nu|\nu_l)|$ for $P=200$, $\tau=1$sec, and $\nu_l=0$ \cite{barilan2014focusing,cohen2017spectrum}. \textcopyright 2018 IEEE. Reprinted, with permission, from D. Cohen, K. V. Mishra, and Y. C. Eldar, ``Spectrum sharing radar: Coexistence via Xampling,'' \textit{IEEE Transactions on Aerospace and Electronic Systems}, vol. 29, no. 3, pp. 1279-1296, 2018.}
	\label{fig:sum_exp}
\end{SCfigure}

Our goal now is to recover $\{ \alpha_l, \tau_l, \nu_l \}_{l=0}^{L-1}$ from $c_p[k]$ for $k \in \kappa$ and $0 \leq p \leq P-1$. To that end,\cite{barilan2014focusing} adopts the Doppler focusing\index{Doppler focusing} approach. Consider the DFT of the coefficients $c_p[k]$ in the slow time domain:
\begin{eqnarray}
\label{eq:focused_coeff}
\tilde{\Psi}_{\nu}[k] = \sum_{p=0}^{P-1} c_p[k] e^{j \nu p \tau} = \frac{1}{\tau} H[k] \sum_{l=0}^{L-1} \alpha_l e^{-j 2 \pi k \tau_l / \tau} \sum_{p=0}^{P-1} e^{j (\nu-\nu_l) p \tau}.
\end{eqnarray}
The key to Doppler focusing follows from the approximation:
\begin{equation}
\label{eq:dop_focus_approx}
g(\nu|\nu_l) = \sum_{p=0}^{P-1} e^{j (\nu-\nu_l) p \tau} \approx \left\{ \begin{array}{ll} 
P & |\nu -\nu_l| < \pi /P \tau \\
0 & |\nu -\nu_l| \geq \pi /P \tau,
\end{array} \right.
\end{equation}
as illustrated in Fig.~\ref{fig:sum_exp}. Denote the normalized focused measurements by
\begin{equation} \label{eq:focused_coeff_norm}
\Psi_{\nu}[k] = \frac{\tau}{PH[k]} \tilde{\Psi}_{\nu}[k].
\end{equation}
%-----------------------------------------------------------------------------------
%-----------------------------------------------------------------------------------

As in traditional pulse Doppler radar, suppose we limit ourselves to the Nyquist grid so that $\tau_l/\tau=r_l/N$, where $r_l$ is an integer satisfying $0 \leq r_l \leq N-1$. Then, (\ref{eq:focused_coeff_norm}) can be approximately written in vector form as
\begin{equation}
\label{eq:doppler_foc1}
\mathbf{\Psi}_{\nu} =  \mathbf{F}_{\kappa} \mathbf{x}_\nu,
\end{equation}
where $\mathbf{\Psi}_{\nu} = \left[ \Psi_{\nu}[k_0] \dots \Psi_{\nu}[k_{K-1}] \right], k_i \in \kappa$ for $0 \leq i \leq K-1$, $\mathbf{F}_{\kappa}$ is composed of the $K$ rows of the $N \times N$ Fourier matrix indexed by $\kappa$ and $\mathbf{x}_{\nu}$ is a $L$-sparse vector that contains the values $\alpha_l$ at the indices $r_l$ for the Doppler frequencies $\nu_l$ in the ``focus zone", that is $|\nu -\nu_l| < \pi / P \tau$.
It is convenient to write (\ref{eq:doppler_foc1}) in matrix form, by vertically concatenating the vectors $\mathbf{\Psi}_{\nu}$, for $\nu$ on the Nyquist grid, namely $\nu=-\frac{1}{2\tau}+\frac{1}{P\tau}$, into the $K \times P$ matrix $\bf \Psi$, as
\begin{equation}
\label{eq:doppler_foc2}
\mathbf{\Psi} =  \mathbf{F}_{\kappa} \mathbf{X},
\end{equation}
where $\mathbf{X}$ is formed similarly by vertically concatenating the vectors $\mathbf{x}_{\nu}$. Note that the matrix $\mathbf{F}_{\kappa}$ is not square and, as a result, the system of linear equations (\ref{eq:doppler_foc2}) is under-determined. The system in (\ref{eq:doppler_foc2}) can be solved using any CS algorithm such as orthogonal matching pursuit (OMP) and $\ell_1$ minimization \cite{eldar2012compressed,eldar2015sampling}. 

A Nyquist receiver needs $B_h\tau$ samples to recover the targets. However, as stated in Theorem~\ref{th:min} below, the number of samples required by the sub-Nyquist receiver depends only on the number of targets present and not on $B_h$. This shows that a sub-Nyquist radar breaks the link between range resolution and transmit bandwidth. In general, only a few targets are present in the radar coverage region leading to a significant reduction in sampling rate.
\begin{theorem}\cite{barilan2014focusing} \label{th:min}
The minimal number of samples required for perfect recovery of $\{\alpha_l, \tau_l, \nu_l\}_{l=0}^L$ in a noiseless environment is $4L^2$. In addition, the number of samples per period is at least $2L$, and the number of periods $P \ge 2L$.
\end{theorem}

The Doppler focusing operation (\ref{eq:focused_coeff}) is a continuous operation on the variable $\nu$, and can be performed for any Doppler frequency up to the PRF. With Doppler focusing there are no inherent “blind speeds”, i.e., target velocities which are undetectable,
as occurs with classic moving target indicator \cite{skolnik2008radar}. Since strong amplitudes are indicative of true target existence as opposed to noise, Doppler focusing recovery searches
for large magnitude entries and marks them as detections. After detecting each target, its influence is removed from the set of Fourier coefficients in order to reduce masking of weaker targets and to remove spurious targets created by processing sidelobes. It is important to note that our dictionary in (\ref{eq:doppler_foc2}) is indifferent to the Doppler estimation. CS methods which estimate delay and Doppler simultaneously \cite{herman2009high}, require a dictionary which grows with the number of pulses. Here by separating delay and Doppler estimation, the CS dictionary is not a function of $P$. 

Moreover, the performance of the sub-Nyquist radar in the presence of noise improves with Doppler focusing. The following theorem states that Doppler focusing increases the per-target SNR by a factor of $P$. This linear scaling is similar to that obtained using a MF. 
\begin{theorem}\cite{barilan2014focusing} 
\label{th:snr}
Let the pre-focusing SNR of the $l$th target be $\Gamma_p^l[k] = \frac{|c_p^l[k]|^2}{\mathbb{E}[|w_p[k]|^2]}$ where $c_p^l[k]$ and $w_p[k]$ are the signal and white noise Fourier coefficients. Then, the focused SNR for the $l$th target at center frequency $\nu$ is $P\Gamma_p^l[k]$.
\end{theorem}

A continuous-value parameter recovery using Doppler focusing is described in \cite{barilan2014focusing}. For practical considerations of computational efficiency, Doppler focusing can be performed on a uniform grid of frequencies so that focused coefficients are computed efficiently using the fast Fourier Transform (FFT). Algorithm~\ref{algo:focusing} outlines this approach to solving the $P$ equations (\ref{eq:doppler_foc2}) simultaneously where, in each iteration, the maximal projection of the observation vectors onto the measurement matrix are retained. The algorithm termination criterion follows from the generalized likelihood ratio test (GLRT) framework presented in \cite{scharf1994matched}. For each iteration, the alternative and null hypotheses in the GLRT problem define the presence or absence of a candidate target, respectively. In Algorithm~\ref{algo:focusing}, $Q{\chi_2^2(\rho)}$ denotes the right-tail probability of the chi-square distribution function with $2$ degrees of freedom, $\Lambda^C$ is the complementary set of $\Lambda$ and 
\begin{equation}
\rho=\frac{P_T}{\sigma^2 |\mathcal{F}_R|}
\end{equation}
is the SNR with $\sigma^2$ the noise variance and $P_T$ the total transmit power. 

%% update algorithm + change stopping criterion (Alex' Pd)
\begin{algorithm}[!t]
\caption{Sub-Nyquist Radar Delay-Doppler Recovery\index{sub-Nyquist radar!temporal sub-Nyquist!recovery algorithm} \cite{cohen2017spectrum,barilan2014focusing}}\label{algo:focusing} 
		\begin{algorithmic}[1]
		\qinput Observations $c_p[k]$, $0 \leq p \leq P-1$ and $k \in \kappa$, probability of false alarm $P_{\text{fa}}$, noise variance $\sigma^2$, transmitted power $P_T$, total transmitted bandwidth $|\mathcal{F}_R|$
		\qoutput Estimated target parameters $\{ \hat{\alpha}_l, \hat{\tau}_l, \hat{\nu}_l \}_{l=0}^{L-1}$
		\State Create $\mathbf{\Psi}$ from $c_p[k]$ using the FFT (\ref{eq:focused_coeff}), for $k \in \kappa$ and $\nu=-1/(2\tau)+p/(P\tau)$, $0 \leq p \leq P-1$
        \State Compute detection thresholds
		$$
		\rho = \frac{P_T}{\sigma^2 |\mathcal{F}_R|}, \quad \gamma = Q^{-1}_{\chi_2^2(\rho)}(1-\sqrt[N]{1-P_{\text{fa}}})
		$$
		\State Initialization: residual $\mathbf{R}_0=\mathbf{\Psi}$, index set $\Lambda_0=\emptyset$, $t=1$
		\State Project residual onto measurement matrix:
		$$
		\mathbf{\Phi} =\mathbf{F}_{\kappa}^H \mathbf{R}_{t-1}
		$$
		\State Find the two indices $\lambda_t = [\lambda_t(1) \quad \lambda_t(2)]$ such that
		$$
		[\lambda_t(1) \quad \lambda_t(2)] = \text{ arg max}_{i,j} \left| \mathbf{\Phi}_{i,j} \right|
		$$
		\State Compute the test statistic
		$$
		\Gamma=\frac{(\mathbf{F}_{\kappa})_{\lambda_t(1)}((\mathbf{R}_{t-1})_{\lambda_t(2)})^H((\mathbf{F}_{\kappa})_{\lambda_t(1)})^H(\mathbf{R}_{t-1})_{\lambda_t(2)}}{\sigma^2}
		$$
		where $(\mathbf{M})_i$ denotes the $i$th column of $\bf M$
		\State If $\Gamma > \gamma$ continue; otherwise go to step 12
		\State Augment index set $\Lambda_t = \Lambda_t  \bigcup \{ \lambda_t \}$
		\State Find the new signal estimate
		$$
		\mathbf{\hat{X}}_{t|\Lambda_t} = (\mathbf{F}_{\kappa})_{\Lambda_t}^{\dagger} \mathbf{\Psi}, \quad 			\mathbf{\hat{X}}_{t|\Lambda_t^C} =\mathbf{0}
		$$
		\State Compute new residual
		$$
		\mathbf{R}_t= \mathbf{\Psi}-(\mathbf{F}_{\kappa})_{\Lambda_t}\mathbf{\hat{X}}
		$$
		\State Increment $t$ and return to step 4
		\State Estimated support set $\hat{\Lambda}= \Lambda_t$
		\State $\hat{\tau}_l=\frac{\tau}{N} \hat{\Lambda}(l,1)$, $\hat{\nu}_l= \frac{1}{P\tau} \hat{\Lambda}(l,2)$, $\hat{\alpha}_l=\hat{\mathbf{X}}_{\hat{\Lambda}(l,1), \hat{\Lambda}(l,2)}$
	\end{algorithmic}
\end{algorithm}

%\subsection{Performance Guarantees}
%\label{subsec:snradperf}
%\begin{theorem}\cite{barilan2014focusing} \label{th:min2}
%Suppose target Doppler frequencies are aligned to a grid $\{\tilde{\nu}_m = 2\pi m/\tau M\}_{m=-M/2}^{M/2-1}$, with no restriction on target delays. Then the minimal number of samples required for perfect recovery of $\{\alpha_l, \tau_l, \nu_l\}_{l=0}^L$ when there is no noise, is $2L\text{\emph{min}}(M,2L)$.
%\end{theorem}
%If $M\ge 2L$ then there is no gain compared with the continuous setting of Theorem~\ref{th:min}, and the minimal number of samples remains $4L^2$.
%\begin{theorem}\cite{barilan2014focusing} \label{th:min3}
%Under the conditions of Theorem~\ref{th:min2}, the minimal number of samples required for perfect recovery of $L$ targets using Doppler focusing is $2LM$, with $K \ge 2L$ and $P \ge M$.
%\end{theorem}

In Section~\ref{subsec:snrad_demo}, we introduce a sub-Nyquist prototype implementing the ideas in this section using simple hardware. Before that, we describe how to account for clutter mitigation in sub-Nyquist radar.

\subsection{Sub-Nyquist Clutter Removal}
\index{sub-Nyquist radar!clutter removal}Clutter refers to unwanted echoes from stationary objects such as buildings, trees, chaff, and ground surface as well as moving elements like weather and sea. Since strong clutter echoes hamper detection of desired targets, clutter removal has been investigated intensively. In the context of CS-based radars, \cite{tuuk2014compressed} provides a general overview of clutter rejection algorithms. In \cite{yu2013capon}, Capon beamforming is used to reject clutter and then the target is retrieved by exploiting sparse reconstruction methods. On the other hand, a few works such as \cite{sun2009novel,yang2015iterative,ma2013jointly} utilize sparsity of the clutter in the mitigation process. Along similar lines, \cite{wang2015sparsity} assumes sparse clutter and proposes a GLRT detector. However, they obtain signal samples at the Nyquist rate. 

Conventionally, clutter is modeled as a random process with Doppler frequency that follows a colored Gaussian noise distribution \cite{kay2007optimal,brennan1973theory,brennan1968optimum}. A standard operation to remove this correlated noise is to use receive filters that maximize the signal-to-clutter-plus-noise (SCNR) ratio. This method is equivalent to first whitening the received signal samples, and then performing matched-filtering with respect to a whitened pulse. Our approach \cite{eldar2015clutter} to clutter removal in sub-Nyquist radar is based on this philosophy as it fits well with our Fourier-based analysis.

In the presence of clutter and noise, the received signal $r_q(t)$ is
\begin{align}
    r(t) = r_{R_X}(t) + y(t),
\end{align}
where $r_{R_X}(t)$ is the target signal with noise as in (\ref{eq:frames_pdr}) and 
\begin{align}
    y(t) = \sum\limits_{p=0}^{P-1}\sum\limits_{c=1}^{C}\alpha_ch(t-p\tau-\tau_c)e^{j v_c p\tau},
\end{align}
is the echo from $C$ clutter targets. We assume that the mean clutter amplitude is $\mathbb{E}[|\alpha_c|^2] = \sigma^2_c$. Further, the delays $\tau_c \sim \mathcal{U}(0, \tau)$ and the clutter Doppler spectrum $v_c \sim \mathcal{N}(v_d, \sigma_d^2)$ are independent and identically distributed.

Analogous to the target signal in (\ref{eq:fourier_coeff}), the Fourier series representation of the clutter signal is given by
\begin{align}
    \tilde{c}_{p}[k] = \frac{1}{\tau} H[k]\sum\limits_{c=0}^{C-1}\alpha_ce^{-j\frac{2\pi}{\tau}k\tau_c}e^{-j v_c p\tau}.
\end{align}
Let the Fourier series coefficients of the noise be $\tilde{w}_{p}[k]$. We now form a $P\times K$ matrix $\mathbf{R}$ with $k$th column given by the Fourier coefficients $R_{p}[k] = c_{p}[k]+\tilde{c}_{p}[k]+\tilde{w}_{p}[k]$, $k\in\kappa$ such that
\begin{align}
\mathbf{R} = \mathbf{X} +\mathbf{Y} + \mathbf{N} = \mathbf{F}_P \mathbf{A} \mathbf{F}_N^K\mathbf{H} +\mathbf{B},
\end{align}
where $\mathbf{B} = \mathbf{Y} + \mathbf{N}$, $\mathbf{F}_P$ is the $P\times P$ Fourier matrix with $(l,k)$th element $e^{-j\frac{2\pi}{P}lk}$, $\mathbf{F}_N^K$ is a submatrix formed by $K$ rows of the $N\times N$ Fourier matrix with $(l,k)$th element $e^{j\frac{2\pi}{N}lk}$, $\mathbf{H} = \text{diag}(H[k])$ is a $K\times K$ diagonal matrix, $\mathbf{A}$ is a $P\times K$ sparse matrix with complex reflectivity $\alpha_l$ at the $L$ indices $(r_l,s_l)$, $\mathbf{Y}$ and $\mathbf{N}$ are $P\times K$ matrices with $(p,k)$th elements $\tilde{c}_{p}[k]$ and $\tilde{w}_{p}[k]$, respectively. As mentioned previously, noise is white over the indices $k$ (all tones). 

Our goal is to extract $\mathbf{A}$ from the measurements $\mathbf{R}$. For simplicity, we assume that $|H[k]|^2$ is unity for all $k$. The whitening transformation requires information about the statistics of clutter and noise, which are summarized in the following proposition.

\begin{proposition}\cite{eldar2015clutter}\label{prop:clutter_exp}
The mean of the clutter Fourier coefficients is $\mathbb{E}\big[\tilde{c}_{p}[k]\big] = 0$, and their correlation is given by
\begin{align}
R_{l_1}[k_1,k_2] &= \mathbb{E}\big[\tilde{c}_{p}[k_1]\overline{\tilde{c}_{p+l_1}[k_2]}\big]=C\sigma^2_c\delta_{k_1,k_2} e^{-jv_dl_1\tau-\frac{1}{2}\sigma_d^2l_1^2\tau^2 }.
\end{align}
The mean and variance of the Fourier coefficients of the noise are, respectively,
\begin{align}
\mathbb{E}\big[N_{p}[k]\big] = 0,\;\:
\mathbb{E}\big[N_{p}[k_1]\overline{N_{p+l_1}[k_2]}\big] = \frac{1}{\tau}\sigma_n^2\delta_{k_1k_2}\delta_{l_2}.
\end{align}
\end{proposition}

Our clutter mitigation technique is based on whitening all the tones of the measurements $\mathbf{R}$. It follows from Proposition~\ref{prop:clutter_exp} that the columns of $\mathbf{B}$ are uncorrelated and identically distributed. The covariance matrix $\mathbf{M}$ of the columns of $\mathbf{B}$ is a Toeplitz matrix with $m$th diagonal value
\begin{align}
\label{eq:cov_mat}
\mathbf{M}(m) = C\sigma^2_c e^{-jv_dm\tau-\frac{1}{2}\sigma_d^2m^2\tau^2 } + \frac{1}{\tau}\sigma_n^2\delta_{m}.
\end{align}
Therefore, the columns of $\mathbf{R}$ can be whitened by multiplying on the left by $\mathbf{M}^{-1/2}$:
\begin{align}
\label{eq:wht_tones}
\mathbf{M}^{-1/2}\mathbf{R} = \mathbf{M}^{-1/2}\mathbf{F}_P\mathbf{A}\mathbf{F}_N^K\mathbf{H}+\mathbf{M}^{-1/2}\mathbf{B},
\end{align}
where $\mathbf{M}^{-1/2}\mathbf{B}$ corresponds to white noise. From here, we proceed with Doppler focusing on $\mathbf{M}^{-1/2}\mathbf{R}$ by taking a Hermitian transpose of (\ref{eq:wht_tones}) and multiplying on the right by $\mathbf{M}^{-1/2}\mathbf{F}_P$:
\begin{align}
\label{eq:mat_sketch}
\mathbf{\Psi} = \mathbf{H}(\mathbf{F}_N^K)^H\mathbf{A}^H\mathbf{F}_P^H\mathbf{M}^{-1}\mathbf{F}_P +\mathbf{B}^H\mathbf{M}^{-1}\mathbf{F}_P = \widetilde{\mathbf{\Psi}} + \widetilde{\mathbf{W}}
\end{align}
where $\widetilde{\mathbf{W}}$ is white noise for each focused frequency. This equation represents a sparse matrix recovery problem. For known matrices $\mathbf{D}_1=\mathbf{H}(\mathbf{F}_N^K)^H$ and $\mathbf{D}_2=\mathbf{F}_P^H\mathbf{M}^{-1}\mathbf{F}_P$, we are given measurements $\mathbf{\Psi} =\mathbf{D}_1\mathbf{X}\mathbf{D}_2$, and the goal is to retrieve the sparse matrix $\mathbf{X}=\mathbf{A}^H$. These problems are solved by matrix sketching algorithms as described in \cite{wimalajeewa2013recovery}. It has been shown in \cite{eldar2015clutter} that whitened Doppler focusing generally increases the SCNR \cite{eldar2015clutter}. %If $\mathbf{M}$ is a circulant matrix, then SCNR has been shown to linearly increase by $P$ after whitening and focusing \cite{eldar2015clutter}.
Compared to other CS-based radars \cite{herman2009high,baraniuk2007compressive}, this technique is robust to the presence of clutter despite sampling at low-rates.
\subsection{Sub-Nyquist Hardware Prototype}
\label{subsec:snrad_demo}
\index{sub-Nyquist radar!hardware!temporal}\index{sub-Nyquist radar|temporal sub-Nyquist|hardware}The first sub-Nyquist radar hardware implementation was presented in \cite{baransky2014prototype}. It was then developed further to incorporate Doppler focusing and clutter removal in \cite{barilan2014focusing} and \cite{eldar2015clutter}, respectively. Since sub-Nyquist techniques manifest themselves mostly in the radar receiver, this prototype emulates receive processing.

%-----------------------------------------------------------------------------------
%\begin{figure}
%\sidecaption
\begin{SCfigure}[50][t]
  \includegraphics[scale=0.45]{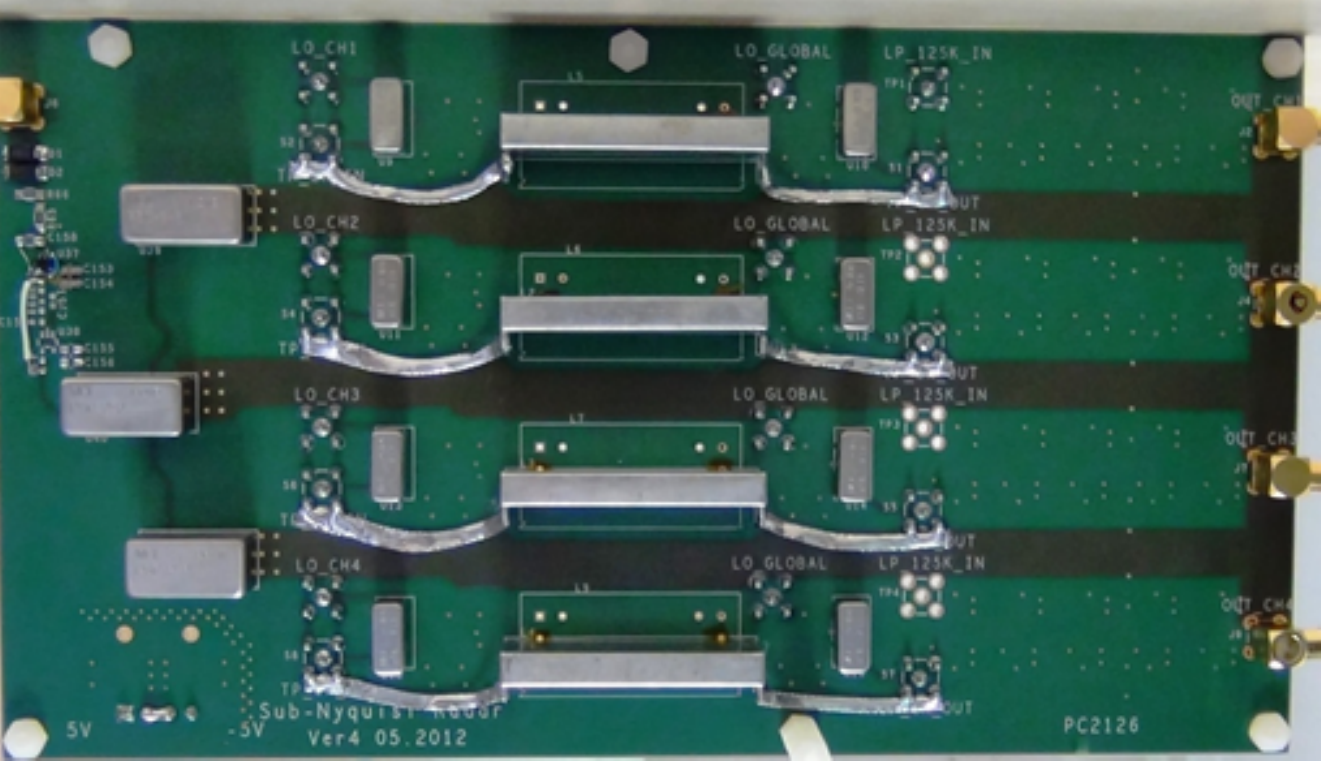}
  \caption{The four-channel Xampler board \cite{baransky2014prototype}. \textcopyright 2014 IEEE. Reprinted, with permission, from E. Baransky, G. Itzhak, I. Shmuel, N. Wagner, E. Shoshan, and Y. C. Eldar, ``A sub-Nyquist radar prototype: Hardware and algorithms,'' \textit{IEEE Transactions on Aerospace and Electronic Systems}, vol. 50, pp. 809-822, 2014.}
	\label{fig:xampler}
\end{SCfigure}
%\end{figure}
%-----------------------------------------------------------------------------------
%\begin{figure}
%\sidecaption
%  \includegraphics[scale=0.65]{ni_xampler_box.pdf}
%  \caption{Sub-Nyquist hardware prototype showing connections between the Xampler board and NI chassis.}
%	\label{fig:ni_xampler_box}
%\end{figure}
\begin{SCfigure}[50][t]
\includegraphics[scale=0.65]{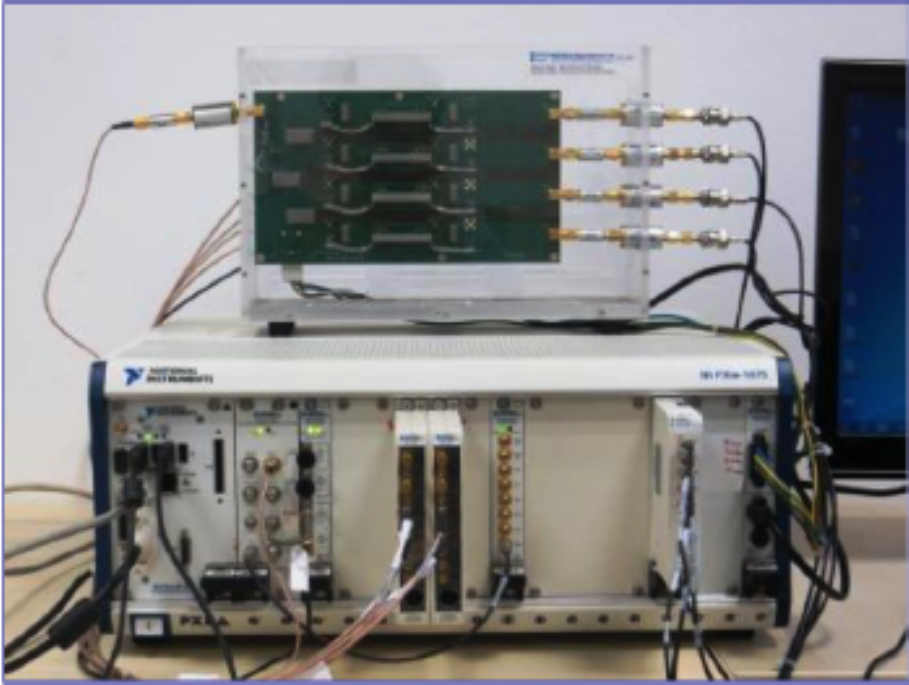}
  \caption{Sub-Nyquist hardware prototype showing connections between the Xampler board and NI chassis \cite{baransky2014prototype,barilan2014focusing,cohen2016reduced}. \textcopyright 2016 IEEE. Reprinted, with permission, from D. Cohen and Y. C. Eldar, ``Reduced time-on-target in pulse Doppler radar: Slow time domain compressed sensing,'' in \textit{IEEE Radar Conference}, 2016, pp. 1-4.}
	\label{fig:ni_xampler_box}
\end{SCfigure}
%-----------------------------------------------------------------------------------
\begin{SCfigure}[50][t]
%\centering
  \includegraphics[scale=0.35]{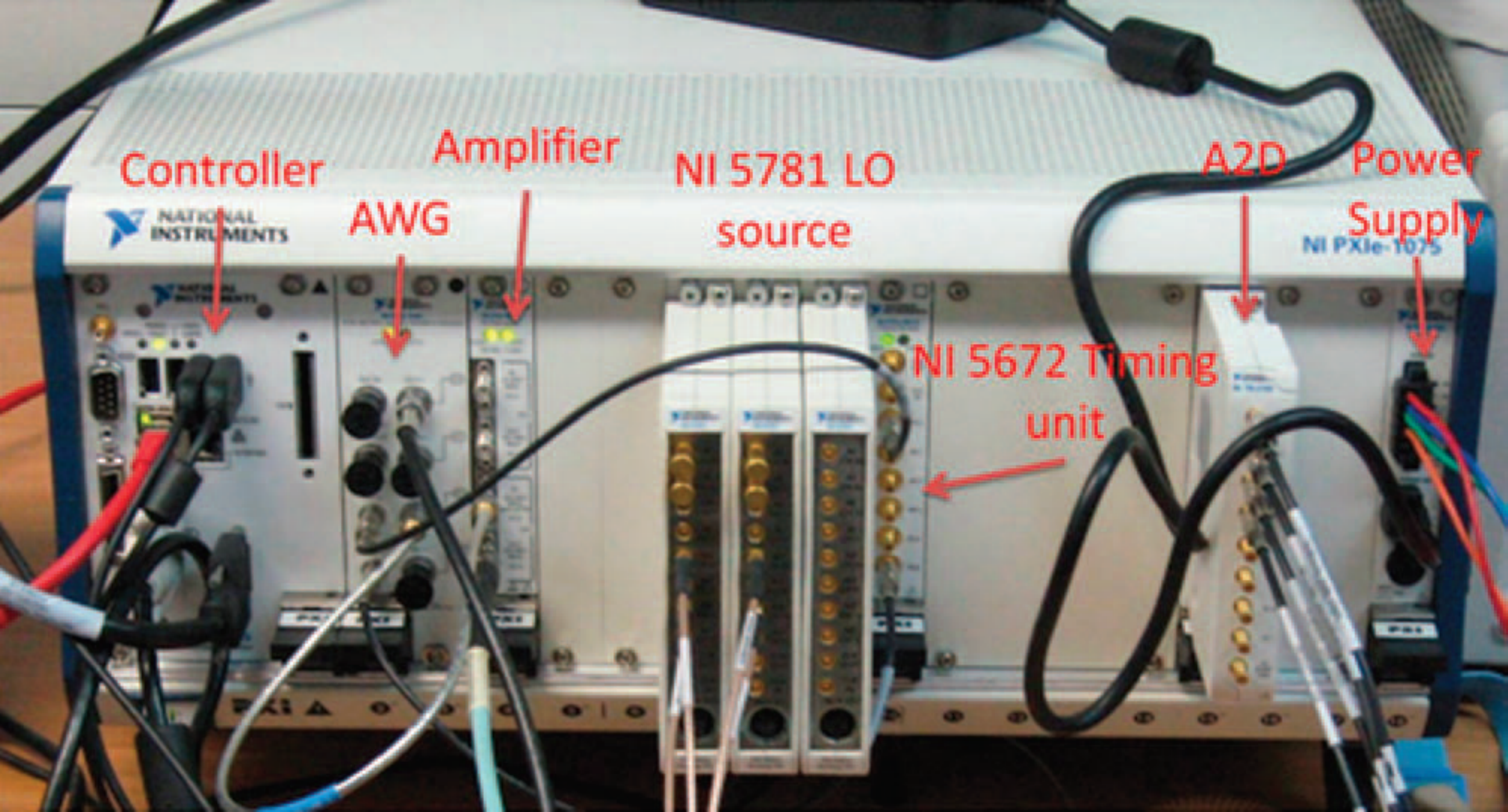}
  \caption{NI chassis showing various signal generation and synchronization components.}
	\label{fig:ni_box}
\end{SCfigure}
%-----------------------------------------------------------------------------------
\begin{SCfigure}[50][t]
%\centering
  \includegraphics[scale=0.20]{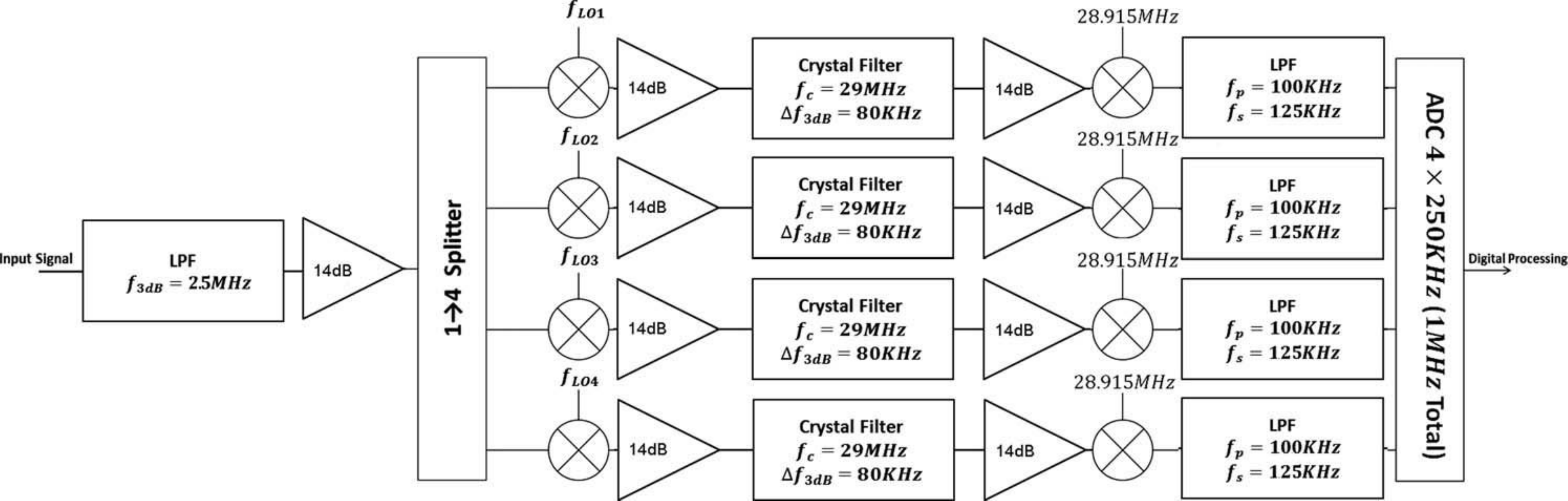}
  \caption{Block diagram of 4-channel solid-state receiver with four up-modulating local oscillators with respective center frequencies of 28.375 MHz, 28.275 MHz, 27.65 MHz, and 27.391 MHz \cite{baransky2014prototype}. \textcopyright 2014 IEEE. Reprinted, with permission, from E. Baransky, G. Itzhak, I. Shmuel, N. Wagner, E. Shoshan, and Y. C. Eldar, ``A sub-Nyquist radar prototype: Hardware and algorithms,'' \textit{IEEE Transactions on Aerospace and Electronic Systems}, vol. 50, pp. 809-822, 2014.}
	\label{fig:snrad_hw_block_diagram}
\end{SCfigure}
%-----------------------------------------------------------------------------------
\begin{figure*}[t]
\centering
  \includegraphics[scale=0.35]{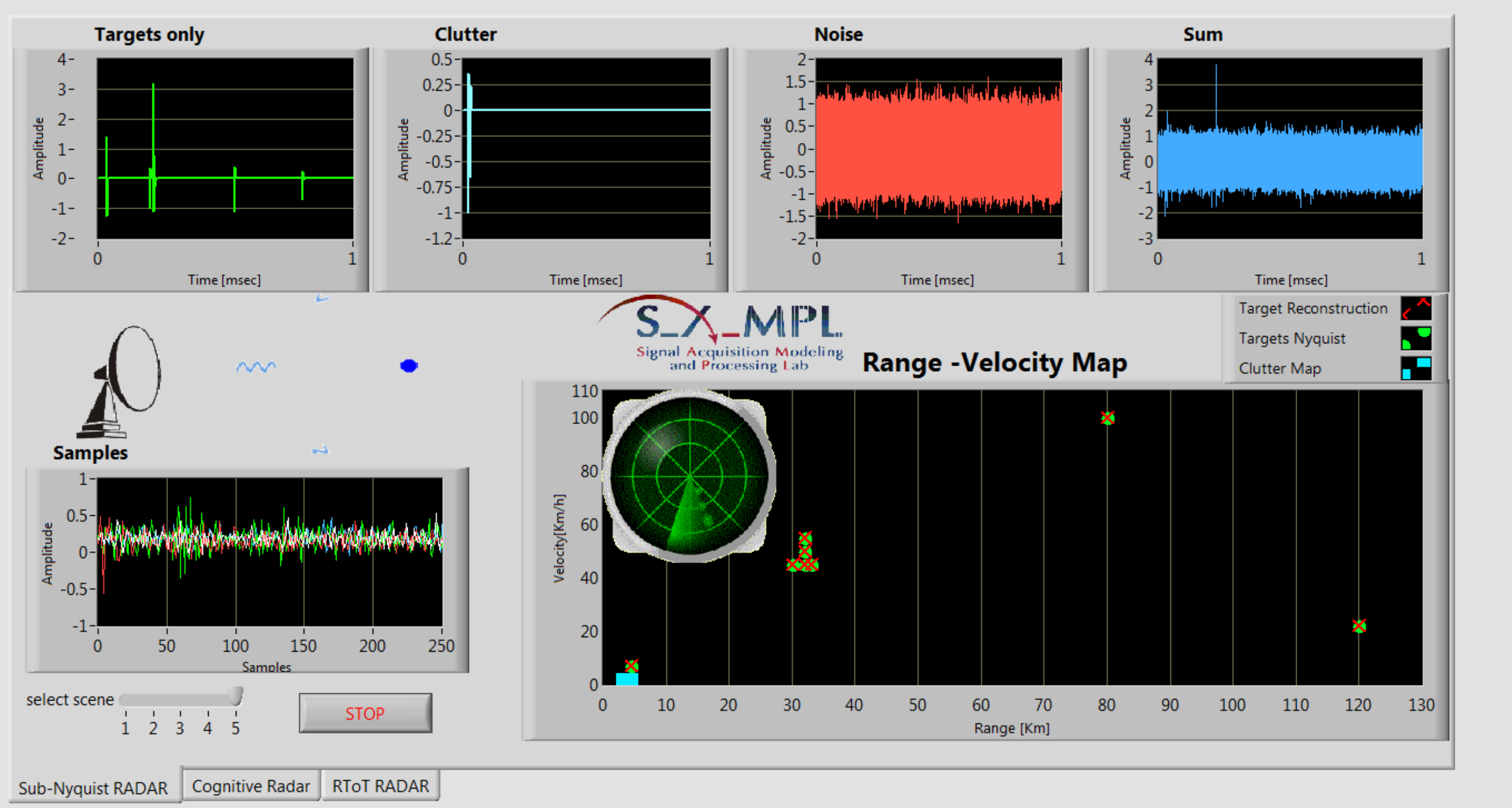}
  \caption{Sub-Nyquist prototype experiment \cite{eldar2015clutter}. Top left to right: Signal corresponding to targets, clutter, noise, and all three combined. Bottom left to right: Low rate samples at receiver and delay-Doppler map with true and recovered targets.}
	\label{fig:snrad_demo_res}
\end{figure*}
%-----------------------------------------------------------------------------------
The basic prototype (Fig.~\ref{fig:ni_xampler_box}) consists of an analog front-end (Fig.~\ref{fig:xampler}), fed by a synthetized RF signal using National Instruments (NI) hardware and followed by digital delay-Doppler map recovery. To evaluate the Xampler board, we make use of NI equipment for both system synchronization and RF signal sources. Figure~\ref{fig:ni_box} shows the entire assembly wrapped in the NI chassis. We transmit 50 pulses with bandwidth 20 MHz. At the receiver, a multiple bandpass sampling approach was chosen, where four groups of consecutive Fourier coefficient subsets are selected. Each channel is fed by a local oscillator (LO), which modulates the desired frequency band of the channel to the central frequency of a narrow 80 KHz bandwidth band pass filter (BPF). A fifth LO, common to all 4 channels, modulates the BPF output to a low frequency band, and sampled with a standard low rate ADC operating at $250$ kHz frequency. The digital samples are acquired by the chassis controller and a MATLAB function is launched that runs Doppler focusing. The digital reconstruction algorithm, performed at a low rate of $250$ kSps, allows recovery of the unknown delays and Doppler frequencies of the targets. A block diagram of the system is shown in Fig.~\ref{fig:snrad_hw_block_diagram}.

Real-time analog experiments show that the system is able to maintain good detection capabilities, while sampling radar signals that require Nyquist rate of about $30$ MHz at a total rate of $1$ MHz, i.e., $1/30$th of the Nyquist rate. We conducted several experiments in order to test the accuracy of our system under various conditions. For example, Fig.~\ref{fig:snrad_demo_res} shows results for a hardware experiment where the target scene has seven scatterers with different delays and Doppler frequencies. A few cases of closely spaced targets in the delay-Doppler plane are also included. Clutter is also added to the scene and identified by the system. Our low-rate processing rejects the clutter and successfully detects only targets in the delay-Doppler plane despite sampling at $1/30$th of the Nyquist rate. The digital recovery algorithm is efficient as it involves only solving 1D delay recovery problems post FFT-based Doppler focusing and without increasing the size of the dictionary.

\section{Doppler Sub-Nyquist Radar}
\label{sec:rtot}
\index{sub-Nyquist radar!Doppler sub-Nyquist}\index{reduced time-on-target}
The temporal sub-Nyquist processing in the fast time domain described in the previous section breaks the link between signal bandwidth, sampling rate, and range resolution. The Xampling framework can also be extended in the slow time or Doppler frequency domain. The Doppler resolution in classical radar processing is given by $2\pi/P_1\tau$ where $P_1$ is the number of pulses transmitted during the CPI. In Doppler domain sub-Nyquist processing, we non-uniformly transmit $P_2 < P_1$ pulses and reduce the power consumption and dwell time in a particular direction without loss of Doppler resolution. The advantage is gaining the ability to look at other directions within the same CPI by interleaving transmissions in different directions.

A few other CS-based works \cite{mishra2014compressed,akhtar2017compressed} have considered reduced time-on-target (RToT) scenarios without addressing analog sampling. The Doppler sub-Nyquist processing that we review here was introduced in \cite{cohen2016reduced}, and is based on the prototype and principles presented in the previous section.

%\subsection{RToT Signal Model}
%\label{subsec:rtot_sig}
We consider a non-uniformly transmitting pulse-Doppler radar such that the $p$th pulse is sent at time $m_p\tau$, where $\{m_p\}_{p=0}^{P_2-1}$ is an ordered set of integers such that $m_p \ge p$. Then, (\ref{eq:uni_model}) is written as
\begin{equation}
\label{eq:nonuni_model}
r_{T_X}(t)= \sum_{p=0}^{P_2-1} h(t-m_p\tau), \quad 0 \leq t \leq P_1 \tau.
\end{equation}
The received signal $r_{R_X}(t)$ is accordingly expressed as a sum of single frames
\begin{equation}
\label{eq:frames_pdr_nonuni}
r_{R_X}(t)= \sum_{p=0}^{P_2-1} r_{R_X}^p(t),
\end{equation}
where
\begin{equation}
\label{eq:one_frame_pdr_nonuni}
r_{R_X}^p(t)= \sum_{l=0}^{L-1} \alpha_l h(t-\tau_l - m_p\tau) e^{-j \nu_l m_p \tau},
\end{equation}
for $0 \leq t \leq P_1\tau$, is the return signal from the $p$th pulse. Our goal is to recover the targets range and Doppler frequency from the received signals $r_{R_X}^p(t)$, with reduced number of transmit pulses $P_2 < P_1$ as well as low-rate samples per pulse.

\subsection{Xampling in CPI and Delay-Doppler Recovery}
\label{subsec:xamp_cpi}
As before, we consider the Fourier series representation of the aligned frames $r_{R_X}^p(t+m_p\tau)$:
\begin{eqnarray}
\label{eq:fourier_coeff_rtot}
X_p[k] = \frac{1}{\tau} H[k] \sum_{l=0}^{L-1} \alpha_l e^{-j 2 \pi k \tau_l / \tau} e^{-j \nu_l m_p \tau},\phantom{1}0\le k \le N-1,
\end{eqnarray}
where $N = B_h\tau$. From (\ref{eq:fourier_coeff_rtot}), the Fourier coefficients embody all the information about the unknown parameters $\{\alpha_l, \tau_l, \nu_l\}_{l-0}^{L-1}$. The goal is then to recover these parameters from $X_p[k]$, $0 \le p \le P_2 -1$. The low rate sampling technique is as described earlier in Section~\ref{subsec:delaydoprec} but the processing steps to recover the target parameters is different to account for sub-Nyquist sampling in Doppler. %The Xampling framework allows obtaining an arbitrary set $\kappa$ of $K$ Fourier coefficients from $K$ point-wise samples of the received signal after appropriate analog preprocessing.

Let $\mathbf{X}$ be the $K \times P$ matrix with $p$th column given by the Fourier coefficients $X_p[k]$, $k \in \kappa$. Then $\mathbf{X}$ can be expressed as
\begin{align}
\label{eq:rtot_prob}
\mathbf{X} = \mathbf{H}\mathbf{F}_N^K\mathbf{A}(\mathbf{F}_{P_1}^{P_2})^T,
\end{align}
where $\mathbf{H}=\frac{1}{\tau}\text{diag}(H[k])$, $\mathbf{F}_N^K$ is a $K \times N$ partial Fourier matrix, $\mathbf{F}_{P_1}^{P_2}$ is a $P_2 \times P_1$ partial Fourier matrix indexed by the values of $m_p$, $1\le p \le P_2$, and $\mathbf{A}$ is an $N \times P_1$ sparse matrix with $\alpha_l$ values at the $L$ indices $\{s_l,\tau_l\}$. We would like to recover $\mathbf{A}$ from the measurements $\mathbf{X}$. 

The system of linear equations (\ref{eq:rtot_prob}) can be solved by CS techniques. However, this problem is different than the temporal sub-Nyquist formulation of Section~\ref{subsec:delaydoprec} where only the range sensing matrix $\mathbf{F}_N^K$ is a partial DFT. In Doppler sub-Nyquist radar, both range and Doppler sensing matrices (i.e., $\mathbf{F}_N^K$ and $\mathbf{F}_{P_1}^{P_2}$, respectively) are partial DFTs. Analogous to Theorem~\ref{th:min}, we have the following result for the Doppler sub-Nyquist radar. 
\begin{theorem}\cite{cohen2016reduced}
\label{th:min_rtot}
The minimal number of samples required for perfect recovery of $\mathbf{A}$ for L targets in noiseless settings is $4L^2$. In addition, the number of samples per period is at least $2L$, and the number of periods $P_2 \ge 2L$.
\end{theorem}
Note that the number of periods $P_2$ here is for non-uniform transmission while the minimum number of periods in Theorem~\ref{th:min} pertain to uniformly spaced pulses in the CPI. Theorem~\ref{th:min_rtot} indicates the lower limit of rate reduction in temporal and Doppler domains.

To solve for the sparse matrix $\mathbf{A}$ in (\ref{eq:rtot_prob}) one can use 
%can be obtained by solving the optimization problem
%\begin{align}
%\label{eq:rtot_opt1}
% & \underset{\mathbf{A}}{\text{minimize}}\;\; ||\mathbf{A}||_1 \nonumber\\
% & \text{subject to} \;\; \mathbf{Y} = \mathbf{F}_N^K \mathbf{A} \left(\mathbf{F}_{P_1}^{P_2}\right)^T,
%\end{align}
%where $||\mathbf{A}||_1$ is the $\ell_1$-norm of $\text{vec}(\mathbf{A})$. 
the matrix version of OMP or $\ell_1$ minimization \cite{eldar2015sampling}.
% When the signal is embedded in noise, fast iterative shrinkage-thresholding algorithm (FISTA) can be employed to instead solve the following problem:
%\begin{align}
%\label{eq:rtot_opt2}
% & \underset{\mathbf{A}}{\text{minimize}}\;\; \frac{1}{2}||\mathbf{Y} - \mathbf{F}_N^K \mathbf{A} \left(\mathbf{F}_{P_1}^{P_2}\right)^T||_{\mathcal{F}} + \mu||\mathbf{A}||_1,
%\end{align}
%where $\mu$ is a regularization parameter and $||\cdot||_\mathcal{F}$ denotes the Frobenius norm. 
Alternatively, Doppler focusing is still approximately applicable. The non-uniform discrete Fourier transform of the coefficients $X_p[k]$ is
\begin{eqnarray}
\label{eq:focused_coeff_nonuni}
\Psi_{\nu}[k] = \sum_{p=0}^{P_2-1} X_p[k] e^{j \nu m_p \tau} = \frac{1}{\tau} H[k] \sum_{l=0}^{L-1} \alpha_l e^{-j 2 \pi k \tau_l / \tau} \sum_{p=0}^{P_2-1} e^{j (\nu-\nu_l) m_p \tau}.
\end{eqnarray}
This can be approximated similar to (\ref{eq:dop_focus_approx}) % to yield the vector form
%\begin{align}
%\label{eq:dop_foc_vec_nonuni}
%\mathbf{\Psi}_{\nu} = P_2\mathbf{H}\mathbf{F}_N^K\mathbf{x}_{\nu},
%\end{align}
%where $\mathbf{\Psi}_{\nu}= \begin{bmatrix} \Psi_{\nu}[k_0] & \cdots & \Psi_{\nu}[k_{K-1}]\end{bmatrix}$, $k_i \in \kappa$ $\forall$ $0 \le i \le K-1$ and $\mathbf{x}_{\nu}$ is an $L$-sparse vector with coefficients $\alpha_l$ at the indices $r_l$ for the focused Doppler frequencies. These $P_1$ equations can now be 
and solved by Algorithm~\ref{algo:focusing} described earlier. However, this is a poor approximation because the $P_2$ points in the sum of exponents $\sum_{p=0}^{P_2-1} e^{j (\nu-\nu_l) m_p \tau}$ are not equally spaced over the unit circle.

\subsection{RToT Hardware Prototype}
\label{subsec:rtot_demo}
\begin{SCfigure}[50][t]
%\centering
  \includegraphics[scale=0.4]{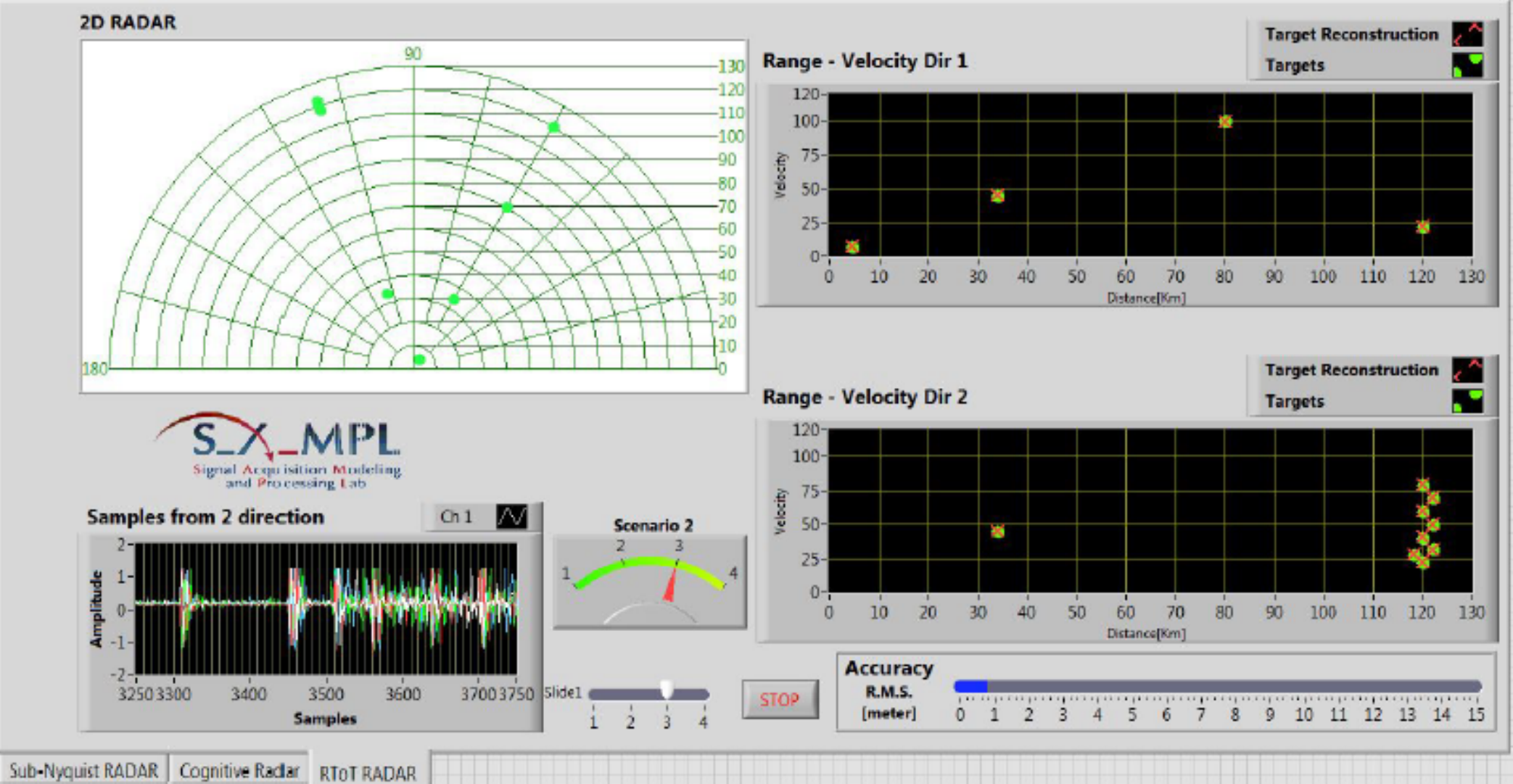}
  \caption{RToT sub-Nyquist radar prototype \cite{cohen2016reduced}. Top left: Targets at two different azimuths. Bottom left: Echoes from both directions are acquired via non-uniform pulses. Top and bottom right: Delay-Doppler maps showing reconstruction for both directions. \textcopyright 2016 IEEE. Reprinted, with permission, from D. Cohen and Y. C. Eldar, ``Reduced time-on-target in pulse Doppler radar: Slow time domain compressed sensing,'' in \textit{IEEE Radar Conference}, 2016, pp. 1-4.}
	\label{fig:rtot_hw_res}
\end{SCfigure}

\index{sub-Nyquist radar!hardware!Doppler}\index{sub-Nyquist radar!Doppler sub-Nyquist!hardware}A hardware implementation of the RToT concept is described in \cite{cohen2016reduced}. It uses the sub-Nyquist hardware prototype presented in Section~\ref{subsec:snrad_demo}. %The analog front-end or Xampler card was fed with a synthesized RF signal from National Instruments (NI) AWR tool. The system samples four distinct $80$ kHz wide subbands of the radar signal spectral content resulting in the acquisition of four sets of consecutive Fourier coefficients. The overall sampling rate is 1 MHz which is only 5\% of the Nyquist rate. 
We evaluated the prototype for a scenario wherein targets are located at two distinct azimuths. Here, $P_1=50$ pulses were chosen such that a quarter of them were sent in one direction and the rest in another. The target scenario for both is then simultaneously recovered within the same original CPI (Fig.~\ref{fig:rtot_hw_res}). In this experiment, the reduction in temporal domain is the same as in Section~\ref{subsec:snrad_demo}, i.e., $1/30$ of the Nyquist rate. In the Doppler domain, pulses in the two directions are reduced by $75\%$ and $25\%$, respectively.
%-----------------------------------------------------------------------------------
%-----------------------------------------------------------------------------------

\section{Cognitive Sub-Nyquist Radar and Spectral Coexistence}
\label{sec:specshare}
In the previous two sections, we focused on processing the received signal. The receiver design in the sub-Nyquist framework can be exploited to also alter the behavior of the radar transmitter. In this section, we discuss the opportunistic control of the transmitter to impart cognition to the radar and leverage it for spectrum sharing applications. For alternative, non-sub-Nyquist approaches to cognition in radars, we refer the reader to Chapters 9 (``Spectrum sensing for cognitive radar via model sparsity exploitation") and 10 (``Cooperative spectrum sharing between sparse sensing based radar") of this book.

The unhindered operation of a radar that shares its spectrum with communication systems has captured a great deal of attention within the operational radar community in recent years \cite{griffiths2015radar}. The interest in such spectrum sharing radars is largely due to electromagnetic spectrum being a scarce resource and almost all services having a need for a greater access to it. %With the allocation of available spectrum to newer communications technologies, the radio-frequency (RF) interference in radar bands is on the rise. Spectrum sharing radars aim to use the information from coexisting wireless and navigation services to manage this interference.

Recent research in spectrum sharing radars has focused on S and C-bands, where the spectrum has seen increasing cohabitation by Long Term-Evolution (LTE) cellular/wireless commercial communication systems. Many synergistic efforts by major agencies are underway for efficient radio spectrum utilization. %The Enhancing Access to the Radio Spectrum (EARS) project by the National Science Foundation (NSF) \cite{bernhard2010final} brings together many different users for a flexible access to the electromagnetic spectrum. 
A significant recent development is the announcement of the Shared Spectrum Access for Radar and Communications (SSPARC)\index{SSPARC}
program \cite{jacyna2016ssparc} by the Defense Advanced Research Projects Agency (DARPA). This program is focused on S-band military radars and views spectrum sharing as a cooperative arrangement where the radar and communication services actively exchange information. It defines spectral \textit{coexistence} as equipping existing radar systems with spectrum sharing capabilities and \textit{spectral co-design} as developing new systems that utilize opportunistic spectrum access \cite{guerci2015joint}. For a review of spectral interference from different services at IEEE radar bands, see \cite{cohen2017spectrum}.

A variety of system architectures have been proposed for spectrum sharing radars. Most put emphasis on optimizing the performance of either radar or communications while ignoring the performance of the other. The radar-centric architectures \cite{cohen2017spectrum,mishra2017auto} usually assume fixed interference levels from communication systems and design the system for high probability of detection ($P_d$). Similarly, the communications-centric systems attempt to improve performance metrics like the error vector magnitude and bit/symbol error rate for interference from radar. With the introduction of the SSPARC program, joint radar-communication performance is being investigated \cite{chiriyath2016inner}. In nearly all cases, real-time exchange of information between radar and communications hardware has not yet been integrated into the system architectures. %Exceptions to this are automotive solutions where the same waveform is used for both target detection and communications \cite{mishra2017auto}. 
In a similar vein, our proposed method, described later in this section, incorporates handshaking of spectral information between the two systems.

\begin{SCfigure}[50][t]
%\centering
  \includegraphics[scale=0.20]{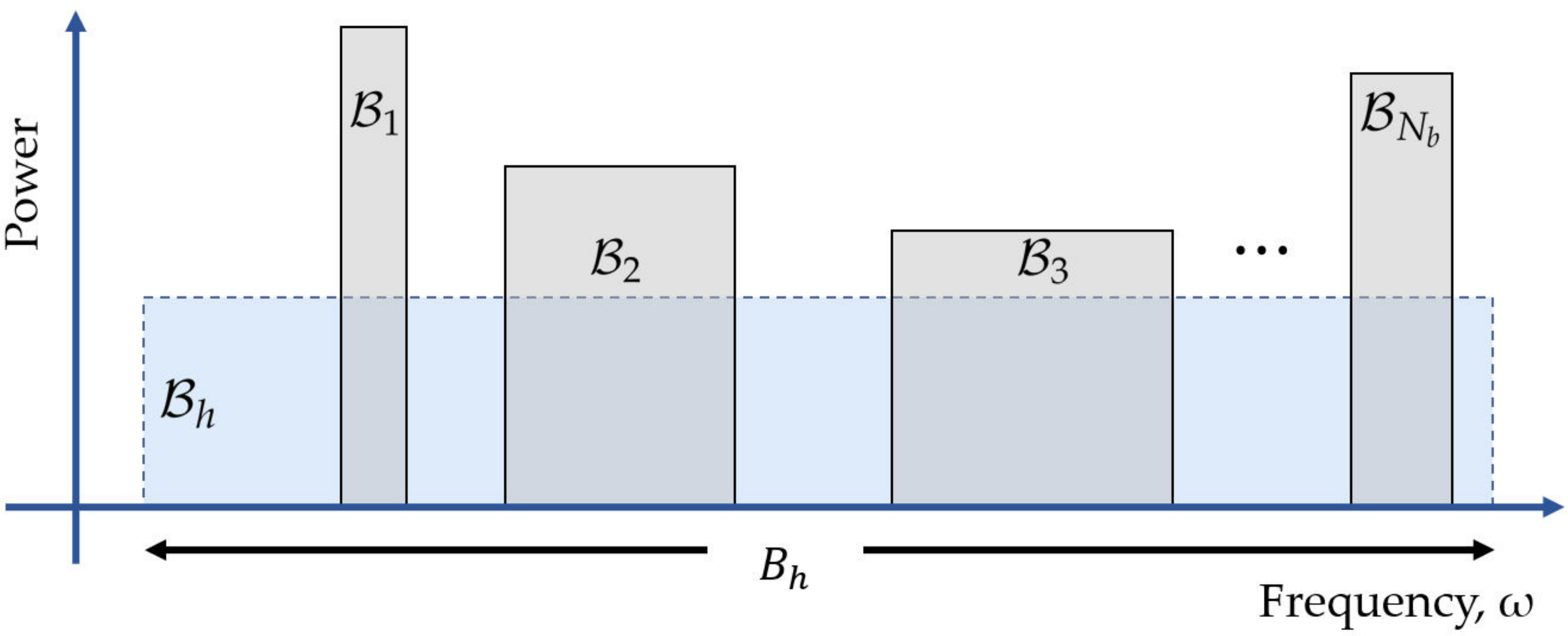}
  \caption{A conventional radar with bandwidth $B_h$ transmits in the band $\mathcal{B}_h$. A cognitive radar transmits only in subbands $\{\mathcal{B}_i\}_{i=1}^{N_b}$, but with increased in-band power. The sub-Nyquist receiver samples and processes only these subbands \cite{mishra2017performance}. \textcopyright 2017 IEEE. Reprinted, with permission, from K. V. Mishra and Y. C. Eldar, ``Performance of time delay estimation in a cognitive radar,'' in \textit{IEEE International Conference on Acoustics, Speech and Signal Processing}, 2017, pp. 3141-3145.}
	\label{fig:cogspec}
\end{SCfigure}

Conventional receiver processing techniques to remove RF interference in radar employ notch filters at hostile frequencies. Typically, spectrum sharing is achieved by notching out the radar waveform's bandwidth causing a decrease in range resolution. Our spectrum sharing solution departs from this baseline. The approach we adopt follows the Xampling architecture on which the sub-Nyquist radar prototype described earlier in Section~\ref{sec:snrad} is based. We recall that the sub-Nyquist receiver samples and processes only small narrow subbands of the received signal. Hence, we capitalize on the simple observation that if only narrow spectral bands are sampled and processed, then one can restrict the transmit signal to these bands. The concept of transmitting only a few subbands that the receiver processes is one way to formulate a cognitive radar (CRr)\index{sub-Nyquist radar!cognitive}\index{cognitive radar!sub-Nyquist} \cite{cohen2016towards}. The delay-Doppler recovery is then performed as presented earlier in Section~\ref{sec:snrad}. The range resolution obtained through this multiband signal spectrum fragmentation can be the same as that of a wideband traditional radar. Furthermore, by concentrating all the available power in the transmitted narrow bands rather than over a wide bandwidth, the CRr increases SNR as illustrated in Fig.~\ref{fig:cogspec}. 

In the CRr system \cite{cohen2016towards}, the support of subbands varies with time to allow for dynamic and flexible adaptation to the environment. Such a system also enables the radar to disguise the transmitted signal as an electronic counter measure or cope with crowded spectrum by using a smaller interference-free portion.
%-----------------------------------------------------------------------------------
%-----------------------------------------------------------------------------------
The CRr configuration is key to spectrum sharing since the radar transceiver adapts its transmission to available bands, achieving coexistence with communication signals. To detect vacant bands, a communication receiver is needed, that performs spectrum sensing over a large bandwidth. Such systems have recently received tremendous interest in communications research, which faces a bottleneck in terms of spectrum availability. To increase the efficiency of spectrum managing, dynamic opportunistic exploitation of temporarily vacant spectral bands by secondary users has been considered, under the name of Cognitive Radio (CRo)\index{cognitive radio}\cite{cohen2017magazine}. Here, we use a CRo receiver to detect the occupied communication bands, so that our radar transmitter can exploit the spectral holes. One of the main challenges of spectrum sensing in the context of CRo is the sampling rate bottleneck due to the wide signal bandwidth. In this context, we use the Xampling framework to subsample and process the signal \cite{mishali2010theory,cohen2017spectrum}.

Denote the set of all frequencies of the available common spectrum by $\mathcal{F}$. The communication and radar systems occupy subsets $\mathcal{F}_C$ and $\mathcal{F}_R$ of $\mathcal{F}$, respectively, such that $\mathcal{F}_C \cap \mathcal{F}_R = \emptyset$. Once the CRo receiver has identified $\mathcal{F}_C$, it provides the radar with spectral occupancy information. Equipped with this spectral map as well as a known radio environment map (REM) detailing typical interference, the CRr transmitter chooses narrow frequency subbands that minimize interference for its transmission. The radar conveys the frequencies $\mathcal{F}_R$ to the communication receiver as well, so that it can ignore the radar bands while sensing the spectrum. The combined CRo-CRr system results in spectral coexistence via the Xampling (SpeCX)\index{SpeCX}\index{spectral coexistence!Xampling}\index{sub-Nyquist radar!spectral coexistence}\index{cognitive radar!spectral coexistence} framework, which optimizes the radar's performance without interfering with existing communication transmissions. Our hardware prototype for SpeCX presented in Section~\ref{subsec:hwdemo} performs real-time recovery of CRo and CRr signals sharing a common spectrum at SNRs as low as $-5$ dB.

\subsection{Cognitive Radio}
\label{subsec:cogradio}
%-----------------------------------------------------------------------------------
\begin{SCfigure}[50][t]
  %\centering
    \includegraphics[width=0.65\columnwidth]{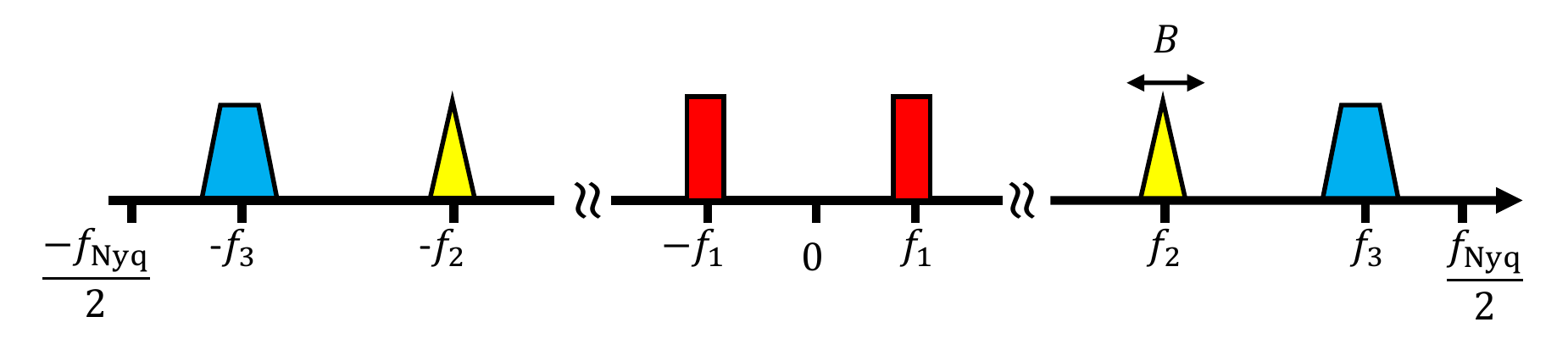}
    \caption{Multiband model with $K=6$ bands \cite{cohen2017spectrum}. \textcopyright 2018 IEEE. Reprinted, with permission, from D. Cohen, K. V. Mishra, and Y. C. Eldar, ``Spectrum sharing radar: Coexistence via Xampling,'' \textit{IEEE Transactions on Aerospace and Electronic Systems}, vol. 29, no. 3, pp. 1279-1296, 2018.}
    \label{fig:multiband}
\end{SCfigure}
%-----------------------------------------------------------------------------------
\index{cognitive radio}We first introduce the signal model, processing, and prototype of CRo in the context of SpeCX. Let $x_{C}(t)$ be a real-valued continuous-time communication signal, supported on $\mathcal{F} = [-1/2T_{\text{Nyq}}, +1/2T_{\text{Nyq}}]$ and composed of up to $N_{\text{sig}}$ transmit waveforms such that 
\begin{equation}
\label{eq:xmodel}
x_C(t)=\sum_{i=1}^{N_{\text{sig}}} s_i(t),
\end{equation}
where $s(t)$ has unknown carrier frequency $f_i$, and $X_c(f)$ is the Fourier transform of $x_C(t)$. %, defined by
%\begin{equation}
%X_C(f)=\lim_{T \rightarrow \infty} \frac{1}{\sqrt{T}}\int_{-T/2}^{T/2}x_C(t)e^{-j2\pi f t} \mathrm{d} t,
%\end{equation}
%is zero for every $f \notin \mathcal{F}$. 
We denote by $f_{\text{Nyq}} = 1/T_{\text{Nyq}}$ the Nyquist rate of $x_C(t)$. The waveforms, respective carrier frequencies and bandwidths are unknown. We only assume that the single-sided bandwidth $B_c^i$ for the $i$th transmission does not exceed an upper limit $B$. Such sparse wideband signals belong to the so-called \textit{multiband signal model} \cite{mishali2009multicoset,mishali2010theory}. Figure~\ref{fig:multiband} illustrates the two-sided spectrum of a multiband signal with $K=2N_{\text{sig}}$ bands centered around unknown carrier frequencies $|f_i| \leq f_{\text{Nyq}}/2$. 

\begin{SCfigure}[50][t]
\includegraphics[width=0.6\linewidth]{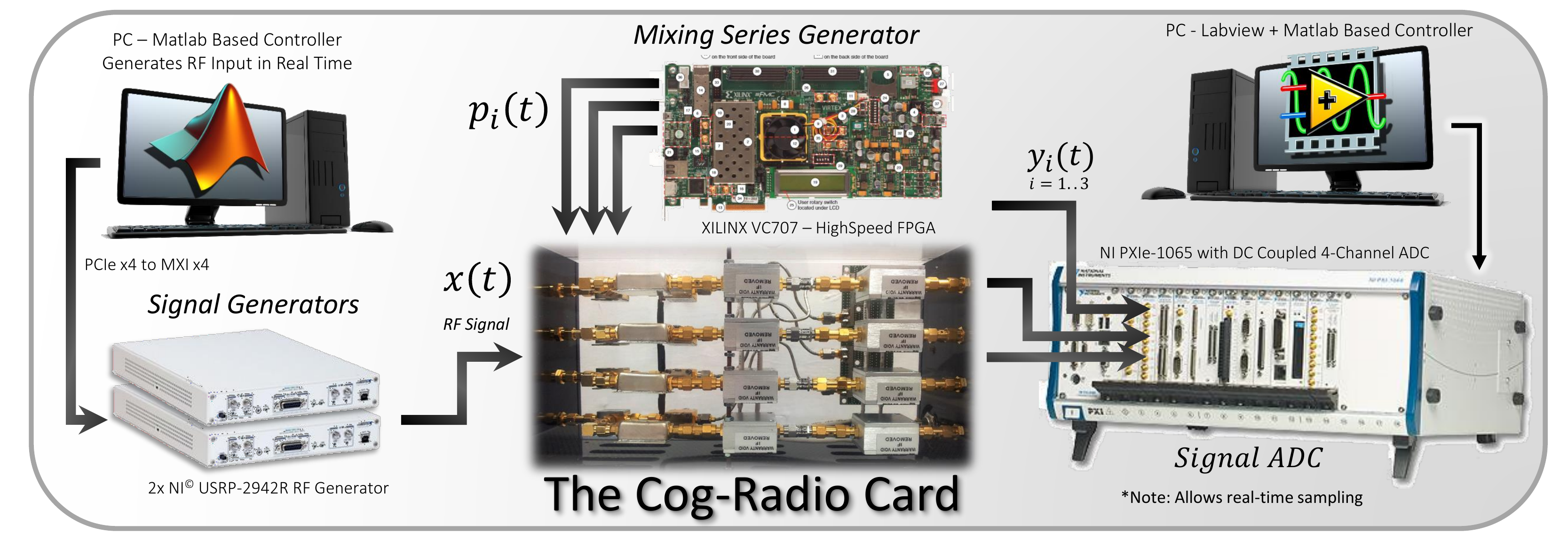}
\caption{CRo system \cite{cohen2017analog}. Reprinted/adapted by permission from Springer Nature: ``Analog to digital cognitive radio,'' in \textit{Handbook of Cognitive Radio}, W. Zhang, Ed. by  D. Cohen, S. Tsiper, and Y. C. Eldar \textcopyright (2017).}%\textcopyright 2017 Springer. Reprinted, with permission, from D. Cohen, S. Tsiper, and Y. C. Eldar, ``Analog to digital cognitive radio,'' in \textit{Handbook of Cognitive Radio}, W. Zhang, Ed. Springer Singapore, 2017.}
\label{fig:crohardware}
\end{SCfigure}

Let $\mathcal{F}_C \subset \mathcal{F}$ be the unknown support of $x_C(t)$. % where
%\begin{equation}
%\mathcal{F}_C=\{f | |f-f_i|< B_c^i/2, \text{ for all } 1 \leq i \leq N_{\text{sig}}\}.
%\end{equation}
The goal of the CRo communication receiver is to retrieve $\mathcal{F}_C$, while sampling and processing $x_C(t)$ at low rates in order to reduce system cost and resources. A CRo system was developed earlier \cite{mishali2010theory,cohen2017analog} for blind sensing (see Fig.~\ref{fig:crohardware}). Next, we explain the details on combining this system with the sub-Nyquist radar to implement SpeCX.
%-----------------------------------------------------------------------------------
%-----------------------------------------------------------------------------------

The input signal at the communication receiver of the SpeCX system is 
\begin{equation}
x(t)=x_C(t)+x_R(t),
\end{equation}
where $x_R(t)=r_{T_X}(t)+r_{R_X}(t)$ is the radar signal sensed by the communication receiver, composed of the transmitted and received radar signals defined in (\ref{eq:uni_model}) and (\ref{eq:uni_rec}), respectively.
Since the frequency support of $x_C(t)$ is unknown, a classic processor would sample such a signal at its Nyquist rate, which can be prohibitively high. In this work, we instead use the modulated wideband converter (MWC)\index{modulated wideband converter} \cite{mishali2010theory}, a sub-Nyquist sampling technique that achieves the lower sampling rate bound for perfect blind recovery of multiband signals, namely twice the Landau rate, and is also practically feasible. The MWC is composed of $M$ parallel channels. In each channel, an analog mixing front-end, where $x_C(t)$ is multiplied by a mixing function $p_i(t)$, aliases the spectrum, such that each band appears in baseband. The mixing functions $p_i(t)$ are periodic with period $T_p$ such that $f_p=1/T_p \ge B$ and have thus the following Fourier expansion:
\begin{equation}
p_i(t) =\sum_{l=-\infty}^{\infty} c_{il} e^{j\frac{2\pi}{T_p} lt}.
\end{equation}

In each channel, the signal next goes through a lowpass filter (LPF) with cut-off frequency $f_s/2$ and is sampled at rate $f_s \ge f_p $, resulting in samples $z_i[n]$. Define $N=2\left\lceil \frac{f_{\text{Nyq}}+f_s}{2f_p} \right\rceil$ and $\mathcal{F}_s=[-f_s/2,f_s/2]$. Following the calculations in \cite{mishali2010theory}, the relation between the known discrete time Fourier transform of the samples $z_i[n]$ and the unknown $X_C(f)$ is given by
\begin{equation} \label{eq:mwc}
\mathbf{z}(f)=\mathbf{A}(\mathbf{x}_C(f) + \mathbf{x}_R(f)), \qquad f \in \mathcal{F}_s,
\end{equation}
where $\mathbf{z}(f)$ is a vector of length $M$ with $i$th element $z_i(f)=Z_i(e^{j2\pi fT_s})$ and the unknown vector $\mathbf{x}_C(f)$ is given by
\begin{equation}
{\mathbf{x}_C}_i(f)=X_C(f+(i-\lceil N/2 \rceil)f_p), \quad f \in \mathcal{F}_s,
\end{equation}
for $1 \leq i \leq N$. The vector ${\mathbf{x}_R}_i(f)$ is defined similarly. The $M \times N$ matrix $\mathbf{A}$ contains the known coefficients $c_{il}$ such that
$
\mathbf{A}_{il} = c_{i,-l}=c^*_{il}
$.

\begin{figure}
	\centering
		\includegraphics[width=0.8\textwidth]{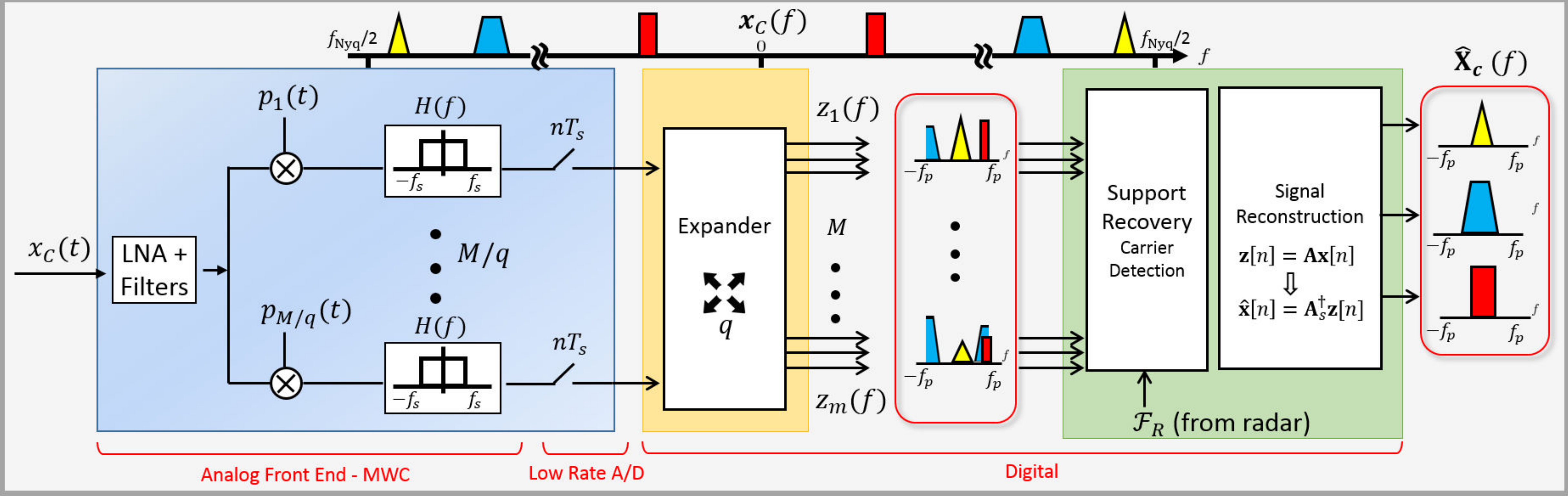}
		\caption{Schematic implementation of the MWC analog sampling front-end and digital signal recovery from low rate samples \cite{cohen2017spectrum}. The CRo inputs are the communication signal $x_C(t)$ and radar support $\mathcal{F}_R$. The communication support output $\mathcal{F}_C$ is shared with the radar transmitter. \textcopyright 2018 IEEE. Reprinted, with permission, from D. Cohen, K. V. Mishra, and Y. C. Eldar, ``Spectrum sharing radar: Coexistence via Xampling,'' \textit{IEEE Transactions on Aerospace and Electronic Systems}, vol. 29, no. 3, pp. 1279-1296, 2018.}
		\label{fig:HighLevel}
\end{figure}

The MWC analog mixing front-end, shown in Fig.~\ref{fig:HighLevel}, results in folding the spectrum to baseband with different weights for each frequency interval. The CRo's goal is now to recover the support of $\mathbf{x}_C(f)$ from the low rate samples $\mathbf{z}(f)$. The recovery of $\mathbf{x}_C(f)$ for each $f$ independently is inefficient and not robust to noise. Instead, the support recovery paradigm from \cite{mishali2010theory} exploits the fact that the bands occupy continuous spectral intervals so that $\mathbf{x}_C(f)$ are jointly sparse for $f \in \mathcal{F}_p$. The continuous to finite block \cite{mishali2010theory} then produces a finite system of equations, called multiple measurement vectors (MMV) from the infinite number of linear systems (\ref{eq:mwc}).

From (\ref{eq:mwc}), we have
\begin{equation}
\mathbf{Q = \Phi Z} \mathbf{\Phi}^H,
\end{equation}
where
\begin{equation} \label{eq:q_ctf}
\mathbf{Q}= \int_{f \in \mathcal{F}_p} \mathbf{z}(f) \mathbf{z}^H(f) \mathrm{d}f,\phantom{1}\mathbf{Z}= \int_{f \in \mathcal{F}_p} \mathbf{x}(f) \mathbf{x}^H(f) \mathrm{d}f,
\end{equation}
%is a $M \times M$ matrix and \par\noindent\small
%\begin{equation}
%\mathbf{Z}= \int_{f \in \mathcal{F}_p} \mathbf{x}_C(f) \mathbf{x}_C^H(f) \mathrm{d}f
%\end{equation}\normalsize
%is a $N \times N$ matrix. 
are $M \times M$ and $N \times N$ matrices, respectively. Here, $\mathbf{x}(f)=\mathbf{x}_C(f) + \mathbf{x}_R(f)$. The matrix $\bf Q$ is then decomposed to a frame $\bf V$ such that $\mathbf{Q=VV}^H$. Clearly, there are many ways to select $\bf V$. One possibility is to construct it by performing an eigendecomposition of $\bf Q$ and choosing $\bf V$ as the matrix of eigenvectors corresponding to the non zero eigenvalues. The finite dimensional MMV system is then given by 
\begin{equation} \label{eq:CTF}
\mathbf{V}=\mathbf{A}(\mathbf{U}_C + \mathbf{U}_R).
\end{equation}
The support of the unique sparsest solution of (\ref{eq:CTF}) is the same as the support of our original set of equations (\ref{eq:mwc}) \cite{mishali2010theory}. Therefore, the support of $\mathbf{U}_C$ and $\mathbf{U}_R$ are disjoint. 

The frequency support $\mathcal{F}_R$ of $x_R(t)$ is known at the communication receiver. From $\mathcal{F}_R$, we derive the support $S_R$ of the radar slices $\mathbf{x}_R(f)$, which is identical to the support of $\mathbf{U}_R$, such that
\begin{equation} \label{eq:sr}
S_R= \left\{ n \left|  \left|n - \frac{f_R^i}{f_p}-\left\lceil N/2 \right\rceil \right| \right. <\frac{f_s+B_R^i}{2f_p} \right\},
\end{equation}
for $1 \leq i \leq N_b$.
Our goal can then be stated as recovering the support of $\mathbf{U}_C$ from $\bf V$, given the known support $S_R$ of $\mathbf{U}_R$. This can be formulated as a sparse recovery with partial support knowledge, studied under the framework of modified CS \cite{vaswani2010modified}. Modified-CS has been used to adapt CS recovery algorithms to exploit partial known support. In particular, greedy algorithms, such as OMP, have been modified to OMP with partial known support \cite{stankovi2009compressive}. Instead of starting with an initial empty support set, one starts with $S_R$ as the initial support. In the first iteration, we compute the estimate
\begin{equation}\mathbf{\hat{U}}_1^{S_R} = \mathbf{A}_{S_R}^{\dagger}\mathbf{V}, \quad
\mathbf{\hat{U}}_{1_i} = \mathbf{0}, \quad \forall i \notin {S_R},
\end{equation}
and residual
\begin{equation}
\mathbf{V}_1=\mathbf{V}-\mathbf{A}_{S_R}\mathbf{\hat{U}}_1.
\end{equation}
The remainder of the algorithm is then identical to OMP. 

Once the overall support $S_C \bigcup S_R$ is known, we have%the slices of $x(t)$ are recovered by reducing the system of equations (\ref{eq:mwc}) to $S_C \bigcup S_R$, so that
\begin{eqnarray} \label{eq:recs}
\mathbf{\hat{x}}^{S_C \bigcup S_R}[n] &=& \mathbf{A}_{S_C \bigcup S_R}^{\dagger} \mathbf{z}[n],  \\
\mathbf{\hat{x}}_i[n] &=& 0, \quad \forall i \notin {S_C \bigcup S_R}. \nonumber
\end{eqnarray}
Here, $\mathbf{x}^{S_C \bigcup S_R}(f)$ denotes the vector $\mathbf{x}(f)$ reduced to its support, $\mathbf{A}_{S_C \bigcup S_R}$ is composed of the columns of $\bf A$ indexed by $S_C \bigcup S_R$ and $\dagger$ is the Moore-Penrose pseudo-inverse. The occupied communication support is then
\begin{equation}  \label{eq:Fc}
\mathcal{F}_C=\{ f | |f-(i+\left\lceil N/2 \right\rceil)f_p| \leq \frac{f_p}{2}, \text{ for all } i \in S_C \}.
\end{equation} 
%-----------------------------------------------------------------------------------
%-----------------------------------------------------------------------------------

\subsection{Cognitive Radar}
\label{subsec:cograd}
\index{sub-Nyquist radar!spectral coexistence}\index{cognitive radar!spectral coexistence}After CRo detects the communication signal support, the CRr transmits 
%We refer the reader to Section~\ref{sec:snrad} for the signal model of pulse Doppler radar. %As mentioned earlier, the bandwidth $B_h$ of the transmitted pulses governs the range resolution of the radar. Large bandwidth is necessary to obtain high resolution, but such a spectral requirement is at odds with the coexisting communication. We, therefore, propose an alternative efficient spectral utilization method wherein the radar transmits several narrow frequency bands instead of a full-band radar signal. In particular, we propose exploiting only a fraction of the bandwidth $B_h$ for both transmission and reception of the radar signal, without degrading its range resolution. 
a pulse $h(t)$ in the unused parts of the spectrum. The transmit signal is supported over $N_b$ disjoint frequency bands, with bandwidths $\{B_r^i\}_{i=1}^{N_b}$ centered around the respective frequencies $\{f_r^i\}_{i=1}^{N_b}$, such that $\sum_{i=1}^{N_b} B_r^i < B_h$. The number of bands $N_b$ is known to the receiver and does not change during operation. The location and extent of the bands $B_r^i$ and $f_r^i$ are determined by the radar transmitter through an optimization procedure to identify the least contaminated bands (see Section~\ref{subsubsec:wf_opt}). The resulting transmitted radar signal CTFT is
\begin{equation}
H_R(f)
 = \left\{ \begin{array}{ll} 
    \beta_i H_{\text{Nyq}}(f), & f \in \mathcal{F}_R^i, \text{ for } 1 \leq i \leq N_b\\
    0, & \text{otherwise},
   \end{array} \right.
\end{equation}
where $\mathcal{F}_R^i = [f_r^i -B_r^i/2, f_r^i +B_r^i/2]$ is the set of frequencies in the $i$th band such that $\mathcal{F}_R= \bigcup_{i=1}^{N_b}\mathcal{F}_R^i$. The parameters $\beta_i >1$ are chosen such that the total transmit power $P_T$ of the spectrum sharing radar remains the same as that of the conventional radar: 
\begin{equation} \label{eq:pt}
\int_{-B_h/2}^{B_h/2} |H_{\text{Nyq}}(f)|^2\, \mathrm{d}f = \sum_{i=1}^{N_b}\int\limits_{\mathcal{F}_r^i} |H_R(f)|^2\, \mathrm{d}f = P_T.
\end{equation}
%In particular, if we choose $\beta_i=\beta$ for all $1 \leq i \leq N_b$ \cite{mishra2017performance}, then
%\begin{equation} \label{eq:beta_same}
%\beta=\sqrt{\frac{\int_{-B_h/2}^{B_h/2} |H_{\text{Nyq}}(f)|^2 \mathrm{d}f}{\int\limits_{\mathcal{F}_R} |H_R(f)|^2\, \mathrm{d}f}}, 
%\end{equation}
%where
%\begin{equation} \label{eq:omega}
%\mathcal{F}_R= \bigcup_{i=1}^{N_b}\mathcal{F}_R^i.
%\end{equation}

The radar identifies an appropriate transmit frequency set $\mathcal{F}_R \subset \mathcal{F}\setminus\mathcal{F}_C$ such that the radar's probability of detection $P_d$ is maximized. For a fixed probability of false alarm $P_{\text{fa}}$ the $P_d$ increases with higher signal to interference and noise ratio \cite{kay1998fundamentals}. At the spectrum sharing radar receiver, we employ the sub-Nyquist approach described in Section~\ref{subsec:delaydoprec}, where the delay-Doppler map is recovered from the subset of Fourier coefficients defined by $\mathcal{F}_R$.

\subsubsection{Optimal Radar Transmit Bands}
\label{subsubsec:wf_opt}
We now explain the procedure through which a CRr selects transmit subbands that have minimal spectral interference. The REM is assumed to be known to the radar transmitter in the form of typical interfering energy levels with respect to frequency bands, represented by a vector $\mathbf{y} \in \mathbb{R}^q$, where $q$ is the number of frequency bands with bandwidth $b_y \triangleq |\mathcal{F}|/q$. %The radar measures the REM vector $\mathbf{y}$ (in dBm) in passive mode by sweeping over $\mathcal{F}$.
In addition, the information from the CRo indicates that the radar waveform must avoid all frequencies in the set $\mathcal{F}_C$. Therefore, we set $\bf y$ to be equal to $\infty$ in these bands. Our goal is to select subbands from the set $\mathcal{F}\setminus\mathcal{F}_C$ with minimal interference. We do that by seeking a block-sparse frequency vector $\mathbf{w} \in \mathbb{R}^p$ with unknown block lengths, where $p$ is the number of discretized frequencies, whose support indicates frequency bands with low interference for the radar. Each entry of $\bf w$ represents a subband of bandwidth $b_w \triangleq |\mathcal{F}|/p$.

To this end, we use the structured sparsity framework of \cite{huang2011learning} based on the one-dimensional graph sparsity structure whose nodes denote the $p$ frequency points of $\mathbf{w}$. %The $p$ nodes are the ordered entries of $\mathbf{w}$, so that neighbor nodes are indexed by adjacent frequency bands. Block sparsity is enforced by encouraging the graph to contain connected regions, which, in the context of our problem, correspond to low inference frequency subbands for radar transmission. In contrast to traditional block sparsity approaches \cite{eldar2012compressed}, this formulation does not require \textit{a priori} knowledge of the location of the non-zero blocks. This is achieved by replacing the traditional sparse recovery $\ell_0$ constraint by a more general term $c(\mathbf{w})$, referred to as the coding complexity, such that
%\begin{equation} \label{eq:c_def} 
%c(\mathbf{w})= \min_F \{c(F) | \text{supp}(\mathbf{w}) \subset F \},
%\end{equation} 
In order to find the desired block-sparse $\mathbf{w}$, the formulation in \cite{huang2011learning} replaces the traditional sparse recovery $\ell_0$ constraint by a more general term $c(\mathbf{w})$, referred to as the coding complexity such that $c(F)=g\log p +|F|$, where $F \subset \{1, \dots, p\}$ is a sparse subset of the index set of the coefficients of $\bf w$ and $g$ is the number of connected regions or blocks of $F$. This coding complexity, which accounts for both the number of discretized frequencies $|F|$ and the number of connected regions $g$, favors blocks within the graph. In our setting, this reduces to solving the following optimization problem for finding the block-sparse frequency vector $\mathbf{w}$ with $(\mathbf{y}_{\text{inv}})_i = 1/\mathbf{y}_i$:
\begin{equation}
	\label{eq:band_select1}
	 \text{minimize}_{\mathbf{w}} \; ||\mathbf{y}_{\text{inv}}-\mathbf{D}\mathbf{w}||_2^2 + \lambda c(\mathbf{w}),
\end{equation}
where $\lambda$ is a regularization parameter and $c(\mathbf{w})$ is defined by $c(\mathbf{w})= \min_F \{c(F) |$ $\text{supp}(\mathbf{w}) \subset F \}$. The matrix $\mathbf{D}$ is $q \times p$ matrix and maps each discrete frequency in $\bf w$ to the corresponding band in $\mathbf{y}_{\text{inv}}$. That is, the $(i,j)$th entry of $\mathbf{D}$ is equal to $1$ if the $j$th frequency in $\bf w$ belongs to the $i$th band in $\bf y$; otherwise, it is equal to $0$. Problem (\ref{eq:band_select1}) can be solved using structured OMP \cite{huang2011learning}.
\begin{algorithm}
\caption{Cognitive Radar Band Selection \cite{cohen2017spectrum}}\label{algo:sel} 
	\begin{algorithmic}[1]
		\qinput REM vector $\mathbf{y}$ and subbands bandwidth $b_y=|\mathcal{F}|/q$, shared support $\mathcal{F}$, communication support $\mathcal{F}_C$, mapping matrix $\bf D$, number of discretized frequencies $p$, number of bands $N_b$
		\qoutput Block sparse vector $\bf w$, radar support $\mathcal{F}_R$
		\State Set $\mathbf{y}_i = \infty$, for each $i$th subband not in $\mathcal{F}_C$ and compute $(\mathbf{y}_{\text{inv}})_i=1/\mathbf{y}_i$
		\State Initialization $F_0 = \emptyset$, $\bf w=0$, $t=1$
        \State Find the index $\lambda_t$ so that $\lambda_t = \arg \max \phi(i)$, where
        $$ \phi(i)=\frac{||\mathbf{P}_i(\mathbf{D}\mathbf{\hat{w}}_{t-1}-\mathbf{y}_{\text{inv}})||_2^2}{c(i \bigcup F_{t-1})-c(F_{t-1})}$$ with $\mathbf{P}_i=\mathbf{D}_i(\mathbf{D}_i^T\mathbf{D}_i)^{\dagger}\mathbf{D}_i^T$
        \State Augment index set $F_t = \lambda_t \bigcup F_{t-1}$
         \State Find the new estimate $\mathbf{\hat{w}}_{t|F_t}=\mathbf{D}_{F_t}^{\dagger} \mathbf{y}_{\text{inv}}, \quad \hat{\mathbf{w}}_{t|F_t^C}= \mathbf{0}$
         \State If the number of blocks, or connected regions, $g(\mathbf{w}) >N_b$, go to step 7. Otherwise, return to step 3
         \State Remove the last index $\lambda_t$ so that $F_t = F_{t-1}$ and $\mathbf{\hat{w}}_t= \mathbf{\hat{w}}_{t-1}$ 
         \State Compute the radar support $\mathcal{F}_R= \bigcup_{j \in F_t} [jb_w-|\mathcal{F}|/2, (j+1)b_w-|\mathcal{F}|/2]$ with $b_w=|\mathcal{F}|/p$
	\end{algorithmic}
\end{algorithm}

\subsubsection{Delay-Doppler Recovery}
\label{subsubsec:ddr_crr}
In order to recover the delay-Doppler map from only $N_b$ transmitted narrow bands, CRr employs a sub-Nyquist receiver that we explained earlier in Section~\ref{subsec:delaydoprec}. The radar receiver first filters the CRr subbands supported on $\mathcal{F}_R$ %given by (\ref{eq:omega}) 
and computes the Fourier coefficients of the received signal. Our resulting spectrum sharing SpeCX framework is summarized in Algorithm \ref{algo:sharing}\index{SpeCX!algorithm}\index{sub-Nyquist radar!spectral coexistence!algorithm}.
\begin{algorithm}
\caption{Spectral Coexistence via Xampling (SpeCX) \cite{cohen2017spectrum}}\label{algo:sharing} 
	\begin{algorithmic}[1]
	\qinput Communication signal $x_C(t)$
	\qoutput Estimated target parameters $\{ \hat{\alpha}_l, \hat{\tau}_l, \hat{\nu}_l \}_{l=0}^{L-1}$
	\State Initialization: perform spectrum sensing at the CRo receiver on $x_C(t)$ following the procedure in Section~\ref{subsec:cogradio}
	\State Choose the least noisy subbands for the radar transmit spectrum with respect to detected $\mathcal{F}_C$ using Algorithm \ref{algo:sel}
	\State Send $\mathcal{F}_R$ to communication and radar receivers
    \State Perform target delay and Doppler estimation using Algorithm \ref{algo:focusing}
	\State Perform spectrum sensing at the communication receiver on $x(t)=x_C(t)+x_R(t)$ following the procedure in Section~\ref{subsec:cogradio}
	\State If $\mathcal{F}_C$ changes, then the radar transmitter goes back to step 2
	\end{algorithmic}
\end{algorithm}

%Theorem \ref{th:min} translates into requirements on the total bandwidth of the transmitted bands, such that \cite{cohen2017spectrum}
%\begin{equation} \label{eq:min_bandwidth}
%B_{\text{tot}}=N \sum_{i=1}^{N_b} \left\lceil \frac{B_r^i}{B_h} \right\rceil
%\geq 2L.
%\end{equation}
For time delay estimation, \cite{mishra2017performance} compares the performance of conventional and cognitive radars using the extended Ziv-Zakai lower bound (EZB)\index{Ziv-Zakai bound}. In a conventional radar, the EZB for a single target delay estimate $\hat{\tau_0}$ is 
\begin{align}
\label{eq:ezzlb_con}
EZB_R(\hat{\tau_0}) = \sigma^2_{\tau_0}\cdot 2Q\left(\sqrt{\dfrac{SNR}{2}}\right) + \dfrac{\Gamma_{3/2}\left(\dfrac{SNR}{4}\right)}{SNR\cdot\overline{F}^2},
\end{align}
where $Q(\cdot)$ denotes the right tail Gaussian probability function, $\Gamma_{a}(b)$ is the incomplete gamma function with parameter $a$ and upper limit $b$, and $\overline{F}$ is the root-mean-square (rms) bandwidth of the full-band signal. The bound for CRr is given in the following theorem.

\begin{theorem}\cite{mishra2017performance}
\label{thm:ezb_crr}
The extended Ziv-Zakai lower bound (EZB) for delay estimation in a cognitive radar is 
\begin{flalign}
\label{eq:ezzlb_cog}
EZB_{CRr}(\hat{\tau}_0) &= \sigma^2_{\tau_0}\cdot 2Q\left(\sqrt{\dfrac{\widetilde{SNR}} {2}}\right) + \dfrac{\Gamma_{3/2}\left(\dfrac{\widetilde{SNR}}{4}\right)}{\sum\limits_{i=1}^{N_b}SNR_i\cdot\overline{F_i^2}},
\end{flalign}
where $SNR_i$ and $\overline{F_i}$ are the in-band SNR and rms bandwidth of the $i$th subband and $\widetilde{SNR}$ is the total SNR. 
\end{theorem}

As noted in \cite{mishra2017performance}, since $\sum\limits_{i=1}^{N_b}B_r^i \subset B_h$, we have $\widetilde{SNR} > SNR$ for given $P_T$. Therefore, the SNR threshold for asymptotic performance of $EZB_{CRr}$ is lower than $EZB_{R}$. As the noise increases and power remains constant for both radars, the asymptotic performance of $EZB_{CRr}$ is more tolerant to noise than $EZB_{R}$. 

The multiband design strategy, besides allowing a dynamic form of the transmitted signal spectrum over only a small portion of the whole bandwidth to enable spectrum sharing, has two additional advantages. First, as we show in hardware experiments (Section VI.B), our CS reconstruction achieves the same resolution as traditional Nyquist processing over a significantly smaller bandwidth. Second, the entire transmit power is concentrated in small narrow bands. Therefore, the SNR in the sampled bands is improved which leads to better parameter estimation as indicated by Theorem~\ref{thm:ezb_crr}.

\subsection{SpeCX Prototype}
\label{subsec:hwdemo}
\index{SpeCX!hardware}\index{sub-Nyquist radar!spectral coexistence!hardware}\index{cognitive radar!spectral coexistence!hardware}
%-----------------------------------------------------------------------------------
\begin{SCfigure}[50][t]
	%\centering
		\includegraphics[width=0.7\textwidth]{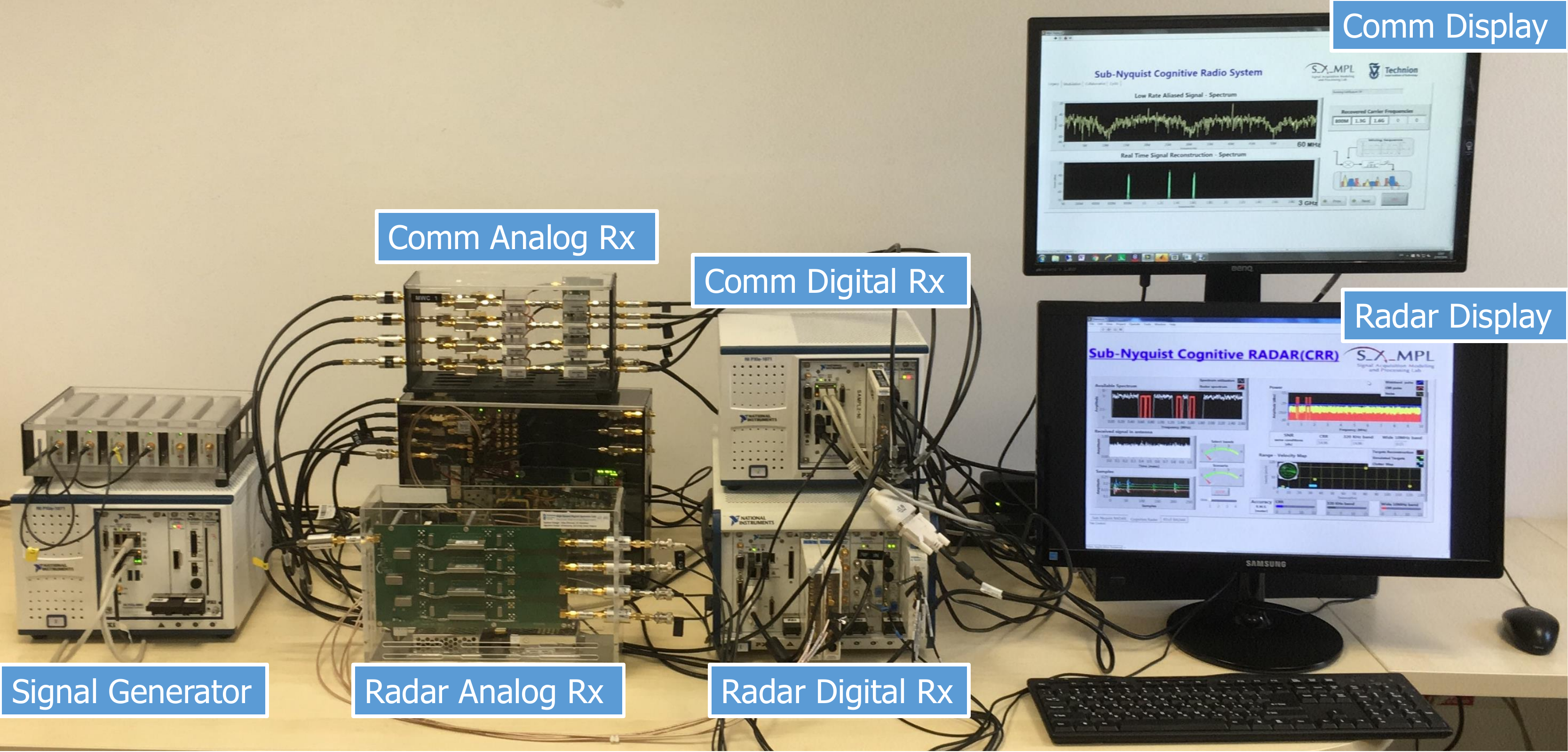}
		\caption{Shared spectrum prototype \cite{cohen2017spectrum}. The system is composed of a signal generator, a CRo receiver based on the MWC, a communication digital receiver, and a CRr analog and digital receiver. \textcopyright 2018 IEEE. Reprinted, with permission, from D. Cohen, K. V. Mishra, and Y. C. Eldar, ``Spectrum sharing radar: Coexistence via Xampling,'' \textit{IEEE Transactions on Aerospace and Electronic Systems}, vol. 29, no. 3, pp. 1279-1296, 2018.} 
		\label{fig:proto}
\end{SCfigure}
%-----------------------------------------------------------------------------------
Figure~\ref{fig:proto} shows our SpeCX prototype, composed of a CRo receiver and a CRr transceiver. The CRo hardware realizes the system shown in Fig.~\ref{fig:HighLevel}. At the heart of the system lies our proprietary MWC board \cite{mishali2011hardware} that implements the sub-Nyquist analog front-end receiver. The card first splits the wideband signal into $M=4$ hardware channels, with an expansion factor of $q=5$, yielding $Mq=20$ virtual channels after digital expansion. In each channel, the signal is then mixed with a periodic sequence $p_i(t)$, generated on a dedicated FPGA, with $f_p=20\, \text{MHz}$. The sequences are chosen as truncated versions of Gold Codes. These were heuristically found to give good detection results \cite{mishali2009expected}, primarily due to small bounded cross-correlations within a set.

Next, the modulated signal passes through a Chebyshev LPF of 7th order with a cut-off frequency ($-3\,\text{dB}$) of $50$ MHz. Finally, the low rate analog signal is sampled by a National Instruments ADC operating at $f_s=(q+1)f_p=120\,\text{MHz}$, leading to a total sampling rate of $480\,\text{MHz}$. The digital receiver is implemented on a National Instruments PXIe-1065 computer with DC coupled ADC. Since the digital processing is performed at the low rate $120 \, \text{MHz}$, very low computational load is required in order to achieve real time recovery. MATLAB and LabVIEW platforms are used for digital recovery operations.
%-----------------------------------------------------------------------------------
\begin{SCfigure}[50][t]
	%\centering
		\includegraphics[width=0.6\textwidth]{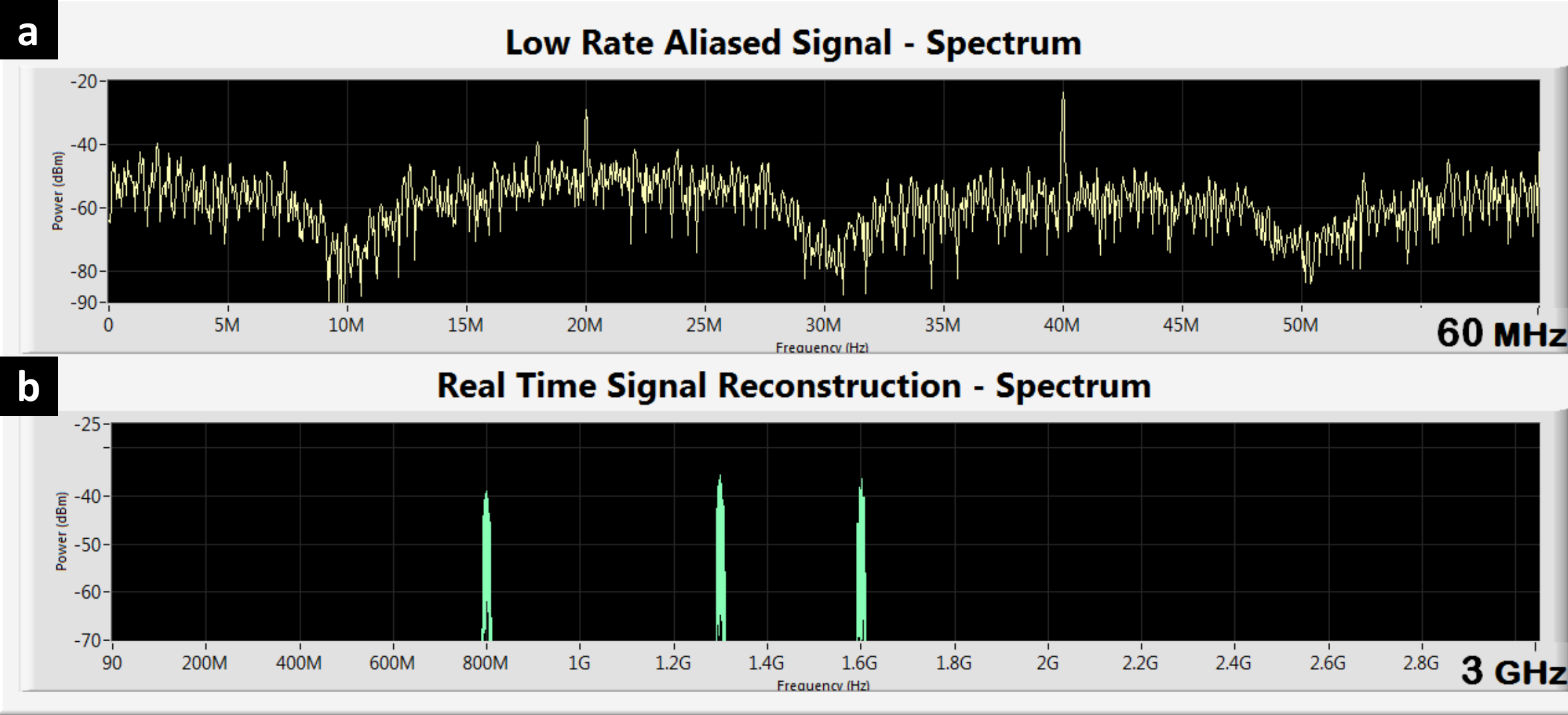}
		\caption{SpeCX communication system display \cite{cohen2017spectrum} showing (a) low rate samples acquired from one MWC channel at rate $ 120\,\text{MHz}$, and (b) digital reconstruction of the entire spectrum from sub-Nyquist samples. \textcopyright 2018 IEEE. Reprinted, with permission, from D. Cohen, K. V. Mishra, and Y. C. Eldar, ``Spectrum sharing radar: Coexistence via Xampling,'' \textit{IEEE Transactions on Aerospace and Electronic Systems}, vol. 29, no. 3, pp. 1279-1296, 2018.}
		\label{fig:LegacySim}
\end{SCfigure}
%-----------------------------------------------------------------------------------

The prototype is fed with RF signals composed of up to $N_{\text{sig}}=5$ real communication transmissions, namely $K=10$ spectral bands with total bandwidth occupancy of up to $200\,\text{MHz}$ and varying support, with Nyquist rate of $6\,\text{GHz}$. To test the system's support recovery capabilities, an RF input is generated using vector signal generators, each producing a modulated data channel with individual bandwidth of up to $20\,\text{MHz}$, and carrier frequencies ranging from $250\,\text{MHz}$ up to $3.1\,\text{GHz}$. The input transmissions then go through an RF combiner, resulting in a dynamic multiband input signal, that enables fast carrier switching for each of the bands. This input is specially designed to allow testing the system's ability to rapidly sense the input spectrum and adapt to changes, as required by modern CRo and shared spectrum standards, e.g. in the SSPARC program. The system's effective sampling rate, equal to $480\,\text{MHz}$, is only $8\%$ of the Nyquist rate and 2.4 times the Landau rate. The main advantage of the Xampling framework, demonstrated here, is that sensing is performed in real-time from sub-Nyquist samples for the entire spectral range. 

Support recovery is digitally performed on the low rate samples. The prototype successfully recovers the support of the CRo transmitted bands, as demonstrated in Fig.~\ref{fig:LegacySim}. The signal is then reconstructed in real-time. Reconstruction does not require interpolation to the Nyquist rate and the active transmissions are recovered at the low rate of $20\,\text{MHz}$, corresponding to the bandwidth of the slices $\mathbf{z}(f)$ defined in (\ref{eq:mwc}). By combining spectrum sensing and signal reconstruction, the MWC serves as two separate communication devices. The first is a state-of-the-art CRo that performs real time spectrum sensing at sub-Nyquist rates, and the second is a receiver able to decode multiple data transmissions simultaneously, regardless of their carrier frequencies, while adapting to real time spectral changes.
%-----------------------------------------------------------------------------------
\begin{SCfigure}[50][t]
%\centering
	\includegraphics[width=0.5\textwidth]{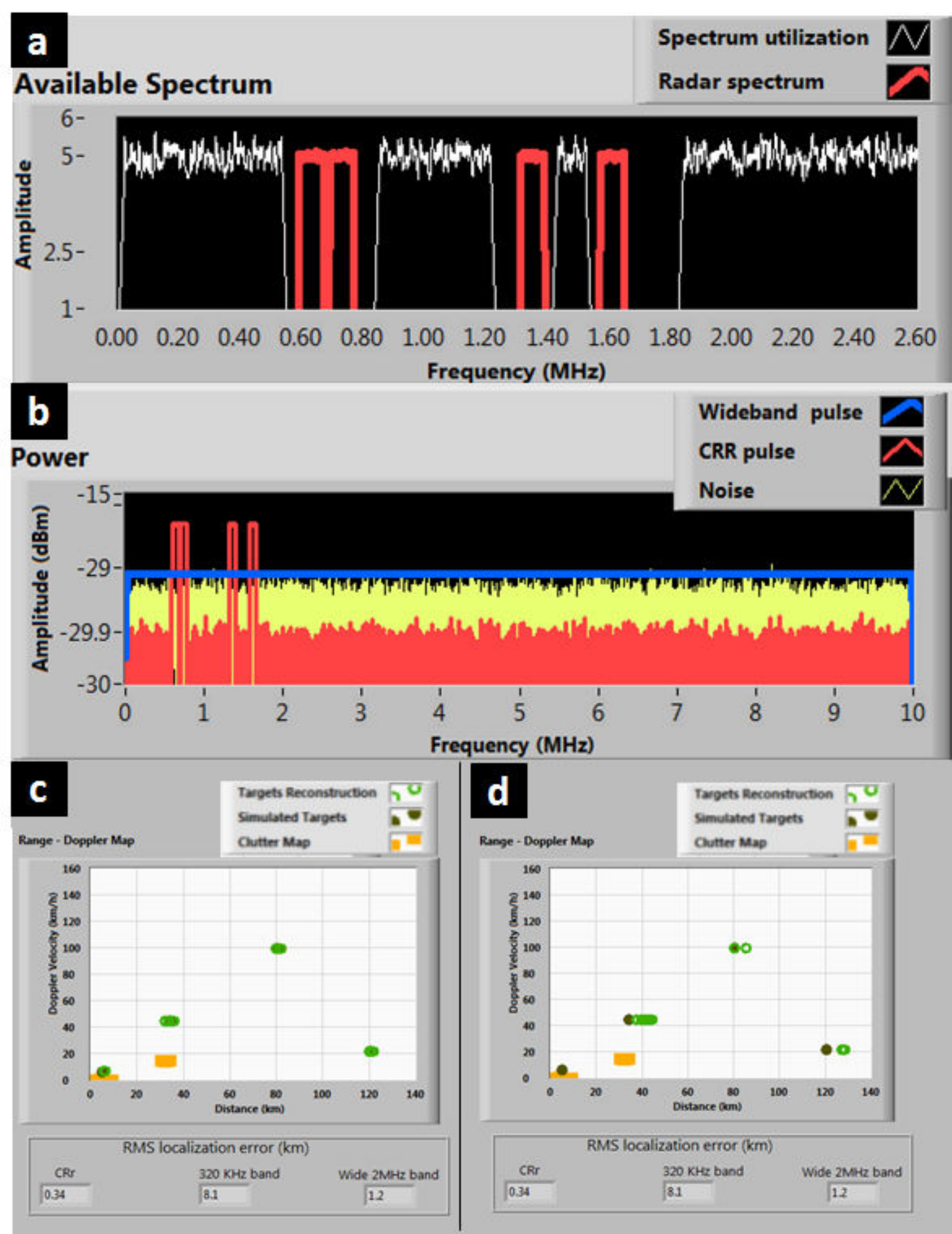}
	\caption{SpeCX radar display \cite{cohen2017spectrum} showing (a) coexisting CRo and CRr (b) CRr spectrum compared with the full-band radar spectrum. The range-Doppler display of detected and true locations of the targets for the case of (a) CRr (four disjoint bands) and (d) all four transmit subbands together forming a contiguous 320 kHz band. \textcopyright 2018 IEEE. Reprinted, with permission, from D. Cohen, K. V. Mishra, and Y. C. Eldar, ``Spectrum sharing radar: Coexistence via Xampling,'' \textit{IEEE Transactions on Aerospace and Electronic Systems}, vol. 29, no. 3, pp. 1279-1296, 2018.}
	\label{fig:hardware}
\end{SCfigure}
%-----------------------------------------------------------------------------------

The CRr system is based on the sub-Nyquist radar receiver board described in Section~\ref{subsec:snrad_demo}. The prototype simulates transmission of $P=50$ pulses towards $L=9$ targets. The CRr transmits over $N_b=4$ bands, selected according to the procedure presented in Section \ref{subsubsec:wf_opt}, after the spectrum sensing process has been completed by the communication receiver. We compare the target detection performance of our CRr with a traditional wideband radar with bandwidth $B_h=20\, \text{MHz}$. The CRr transmitted bandwidth is thus equal to $3.2\%$ of the wideband.

Figure~\ref{fig:hardware} shows windows from the graphical user interface (GUI) of our CRr system. Figure~\ref{fig:hardware}(a) illustrates the coexistence between the radar transmitted bands (thick curve) and the existing communication bands (thin curve). The gain in power is demonstrated in Fig.~\ref{fig:hardware}(b) which plots the wideband radar spectrum, CRr, and noise. The true and recovered range-Doppler maps for the CRr (whose transmit signal consists of four disjoint subbands) are shown in Fig.~\ref{fig:hardware}(c). All 9 targets are perfectly recovered and clutter is discarded. Fig.~\ref{fig:hardware}(d) shows the performance when the four subbands are joined together to result in a 320 kHz contiguous band for the radar transmitter. There are many missed detections and false alarms in this case. Let the true and estimated ranges of the $i$th target be $d_{i}$ and $\hat{d}_i$, respectively. Then the root-mean-square localization error (RMSLE) of $L$ targets is given by 
\begin{align}
\label{eq:rmsle}
\text{RMSLE} = \sqrt{\frac{1}{L}\sum\limits_{i=1}^L(d_i-\hat{d}_i)^2}.
\end{align}
In Fig.~\ref{fig:hardware}(c)-(d), the RMSLE is shown as follows: CRr ($0.34$km), 320 kHz band or 4 adjacent bands with same bandwidth ($8.1$km), and wideband ($1.2$km). The poor resolution of the 4 adjacent bands scenario is due to its small aperture. The native range resolution in case of $2$ MHz wideband scenario is 75 m. In Fig.~\ref{fig:hardware}(c), the CRr is able to detect 9 targets at locations $6.097$, $31.764$, $35.046$, $35.451$, $35.479$, $81.049$, $81.570$, $121.442$, and $120.922$ km. Here, the distance between two closely spaced targets is less than $75$ m.

\section{Spatial Sub-Nyquist: Application to MIMO Radar}
\label{sec:mimo}
\index{sub-Nyquist radar!spatial sub-Nyquist}\index{sub-Nyquist radar!MIMO}\index{MIMO!sub-Nyquist}\index{sub-Nyquist radar!SUMMeR}
We now consider extending sub-Nyquist processing to the spatial domain for the particular case of MIMO radar \cite{fishler2004mimo}. MIMO radars use an array of several transmit and receive antenna elements, with each transmitter radiating a different, mutually orthogonal waveform. Waveform orthogonality can be in time, frequency or code. Our system is based on the collocated MIMO configuration \cite{li2007mimo}, in which the elements are close to each other so that the radar cross-section of a target appears identical in all elements. The MIMO receiver separates and coherently processes the target echoes corresponding to each transmitter. The angular resolution of MIMO using the classic virtual ULA is the same as a phased array with equivalent virtual aperture but many more antenna elements.

Conventional MIMO radar's spatial (angular) and range resolutions are limited by the number of elements and the receiver sampling rate, respectively. Here, we extend the Xampling framework for temporal sub-Nyquist radar in Section~\ref{sec:snrad} to both space and time by simultaneously \textit{thinning} an antenna array and sampling received signals at sub-Nyquist rates. This sub-Nyquist collocated MIMO radar (SUMMeR) recovers the target range, azimuth, and Doppler velocity without loss of any of the aforementioned radar resolutions. In SUMMeR, the radar antenna elements are randomly placed within the aperture, and signal orthogonality is achieved by frequency division multiplexing (FDM). The FDM-based sub-Nyquist MIMO mitigates the range-azimuth coupling by randomizing the element locations in the aperture \cite{cohen2017high}.

%In this section, we describe the SUMMeR prototype with Doppler processing and simultaneously recover all three parameters, range, azimuth and Doppler. We also present a hardware prototype that demonstrates these concepts in real-time and uses cognitive transmission techniques similar to those in Section~\ref{subsec:cograd} for each transmit element. Cognitive transmission imparts two advantages to the SUMMeR hardware. First, the spatial sub-Nyquist processing of large arrays can be easily designed without replicating the pre-filtering operation for each subband in the hardware. Second, since the total transmit power remains the same, a cognitive signal has more in-band power resulting in an increase in signal-to-noise ratio (SNR) as discussed in Section~\ref{subsubsec:ddr_crr}.

\subsection{Sub-Nyquist Collocated MIMO Radar Model}
\label{subsec:mimo}
\begin{SCfigure}[50][t]
\includegraphics[width=0.55\textwidth]{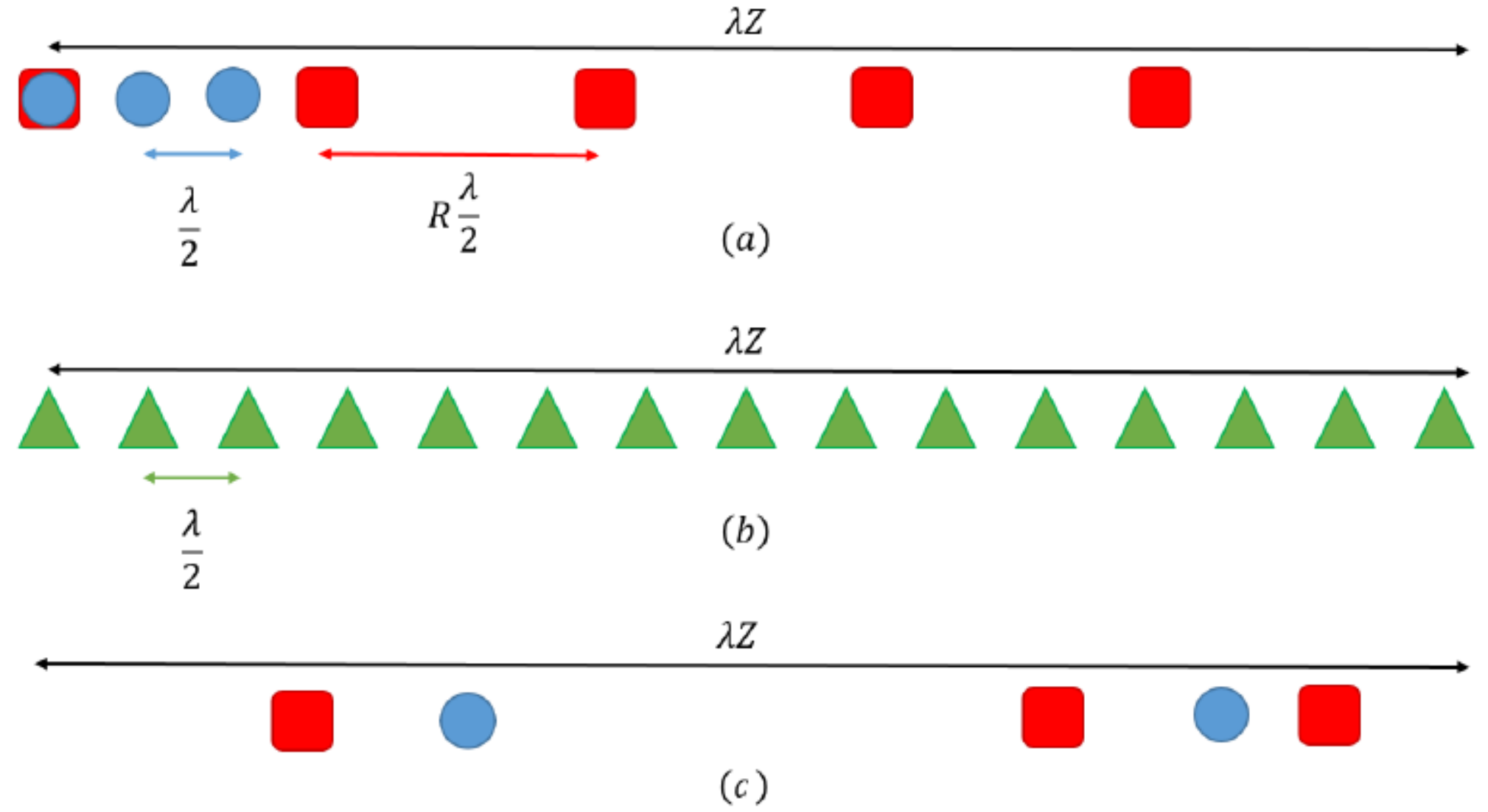}
\caption{Location of transmit (diamonds) and receive (triangles) antenna elements within the same physical aperture for (a) conventional MIMO array with $T=5$ transmitters and $R=4$ receivers, (b) virtual ULA with $TR=20$ antenna elements, and (c) randomly thinned MIMO array with $M=4$ transmitters and $Q=3$ receivers.}
\label{fig:all_arrays}
\end{SCfigure}

Let the operating wavelength of the radar be $\lambda$ and the total number of transmit and receive elements be $T$ and $R$ respectively. The classic approach to collocated MIMO adopts a virtual ULA structure, where the receive antennas spaced by $\frac{\lambda}{2}$ and transmit antennas spaced by $R\frac{\lambda }{2}$ form two ULAs (or vice versa). Here, the coherent processing of a total of $TR$ channels in the receiver creates a virtual equivalent of a phased array with $TR$ $\frac{\lambda }{2}$-spaced receivers and normalized aperture $Z=\frac{TR}{2}$. This standard array structure and the corresponding receiver virtual array are illustrated in Fig.~\ref{fig:all_arrays}(a)-(b) for $T=5$ and $R=4$.
%---------------------------------------------------------------------------------
%---------------------------------------------------------------------------------

Consider a collocated MIMO radar system that has $M<T$ transmit and $Q<R$ receive antennas. The locations of these antennas are chosen uniformly at random within the aperture of the virtual array mentioned above, as in Fig.~\ref{fig:all_arrays}(c). The $m$th transmitting antenna sends $P$ pulses
\begin{equation}
\label{eq:trMth}
    \begin{array}{lll}
     {{s}_{m}}\left( t \right)=\sum\limits_{p=0}^{P-1}{{{h}_{m}}\left( t - p\tau\right)}{{e}^{j2\pi {{f}_{c}}t}},\quad0\le t\le P\tau,
    \end{array}
\end{equation} 
where $\{{{h}_{m}}\left( t \right)\}_{m=0}^{M-1}$ is a set of narrowband, orthogonal FDM pulses each with CTFT 
\begin{equation}
\label{eq:ctft}
H_m(\omega) = \int\limits_{-\infty}^{\infty}h_m(t)e^{-j\omega t}dt.
\end{equation} 
For simplicity, we assume that $f_c \tau$ is an integer. The pulse time support is denoted by $T_p$.

Consider a target scene with $L$ non-fluctuating point targets following the Swerling-0 model \cite{skolnik2008radar} whose locations are given by their ranges $R_l$, Doppler velocity $v_l$, and azimuth angles $\theta_l$, $1 \le l \le L$. The pulses transmitted by the radar are reflected back by the targets and collected at the receive antennas. When the received waveform is downconverted from RF to baseband, we obtain the following signal at the $q$th antenna, 
\begin{equation}
x_q \left( t \right) = \sum\limits_{p=0}^{P-1}\sum\limits_{m=0}^{M-1} \sum\limits_{l=1}^{L} \alpha_l h_m \left( t-p\tau-\tau _{l} \right) e^{j2 \pi \beta_{mq} \vartheta _l}e^{j 2\pi f^D_l p \tau},
\end{equation} 
where $\alpha_l$ denotes the complex-valued reflectivity of the $l$th target, $\tau_l = 2R_l/c$ is the range-time delay the $l$th target, $f_l^D = \frac{2v_l}{c}f_c$ is the frequency in the Doppler spectrum, $\vartheta_l = \sin\theta_l$ is the azimuth parameter, and $\beta_{mq}$ is governed by the array structure. We express $x_q(t)$ as a sum of single frames 
\begin{equation}
\label{eq:frames}
x_q(t)= \sum_{p=0}^{P-1} x_q^p(t),
\end{equation} 
where 
\begin{equation}
\label{eq:one_frame}
x_q^p(t)= \sum_{m=0}^{M-1} \sum_{l=1}^{L} \alpha_l h(t-\tau_l - p\tau) e^{j2 \pi \beta_{mq} \vartheta _l} e^{j 2\pi f^D_l p \tau}.
\end{equation} 
Our goal is to estimate the time delay ${{\tau}_{l}}$, azimuth ${{\theta }_{l}}$, and Doppler shifts $f_l^D$ of each target from low rate samples of $x_q(t)$, for $0 \leq q \leq Q-1$, and a small number of $M$ channels and $Q$ antennas.

\subsection{Xampling in Time and Space}
\label{subsec:xamp_mimo}
The application of Xampling in both space and time enables recovery of range, direction, and velocity at sub-Nyquist rates. The sampling technique is the same as in Section~\ref{subsec:delaydoprec} but now the low-rate samples are obtained in both range and azimuth domains. The received signal $x_q(t)$ is separated into $M$ channels, aligned, and normalized. The Fourier coefficients of the received signal corresponding to the channel that processes the $m$th transmitter echo at the $q$th receiver are given by 
\begin{equation}
\label{coffAlligned}
   y_{m,q}^p[k]=   \sum_{l=1}^{L} \alpha_l e^{j2\pi \beta_{mq} \vartheta_l}e^{-j\frac{2\pi }{\tau}k\tau_l} e^{-j2\pi f_m \tau_l} e^{j 2 \pi f^D_l p \tau},
\end{equation} 
where $-\frac{N}{2} \leq k \leq -\frac{N}{2}-1$, $f_m$ is the (baseband) carrier frequency of the $m$th transmitter and $N$ is the number of Fourier coefficients per channel. %Xampling obtains a set $\kappa$ of arbitrarily chosen Fourier coefficients from low rate samples of the received channel signal such that $|\kappa| = K < N$. More details can be found in \cite{baransky2014prototype}.

As in traditional MIMO, assume that the time delays, azimuths, and Doppler frequencies are aligned to a grid. In particular, $\tau_l = \frac{\tau}{TN}s_l$, $\vartheta_l = -1+\frac{2}{TR} r_l$, and $f^D_l = -\frac{1}{2\tau}+\frac{1}{P\tau}u_l$, where $s_l$, $r_l$ ,and $u_l$ are integers satisfying $0 \leq s_l \leq TN-1$, $0 \leq r_l \leq TR -1$, and $0 \leq u_l \leq P-1$, respectively.
Let $\mathbf{Z}^m$ be the $KQ \times P$ matrix with $q$th column given by the vertical concatenation of $y_{m,q}^p[k], k \in \kappa$, for $0 \leq q \leq Q-1$. We can then write $\mathbf{Z}^m$ as 
\begin{equation}
\label{eq:model2}
\mathbf{Z}^m = \left( \mathbf{{B}}^m \otimes  \mathbf{A}^m \right) \mathbf{X}_D \mathbf{F}_P^H.
\end{equation} 
Here, $\mathbf{A}^m$ denotes the $K \times TN$ matrix whose $(k,n)$th element is $e^{- j \frac{2 \pi}{TN} \kappa_kn} e^{-j2\pi \frac{f_m}{B_h} \frac{n}{T}}$ with $\kappa_k$ the $k$th element in $\kappa$, $\mathbf{B}^m$ is the $Q \times TR$ matrix with $(q,p)$th element $e^{-j2 \pi \beta_{mq} (-1 +\frac{2}{TR}p)}$, and $\mathbf{F}_P$ denotes the $ P \times P$ Fourier matrix. The matrix $\mathbf{X}_D$ is a $T^2NR \times P$ sparse matrix that contains the values $\alpha_l$ at the $L$ indices $\left( r_l TN +s_l, u_l \right)$. 

The range and azimuth dictionaries $\mathbf{A}^m$ and $\mathbf{B}^m$ are not square matrices due to low-rate sampling of Fourier coefficients at each receiver and reduction in antenna elements, respectively. Therefore, the system of equations in (\ref{eq:model2}) is undetermined in azimuth and range. Our goal is to recover $\mathbf{X}_D$ from the measurement matrices $\mathbf{Z}^m, 0 \leq m \leq M-1$. The temporal, spatial, and frequency resolution stipulated by $\mathbf{X}_D$ are $\frac{1}{TB_h}$, $\frac{2}{TR}$, and $\frac{1}{P \tau}$ respectively.
%Theorem \ref{th:cond2} presents necessary conditions on the minimal number of channels $MQ$, samples per receiver $MK$ and pulses per transmitter $P$ for perfect recovery of $\mathbf{X}_D$ from (\ref{eq:model2}) under the grid assumption.

\begin{theorem}\cite{cohen2016summer}
\label{th:cond2}
The minimal number of transmit and receive array elements, i.e., M and Q, respectively, required for perfect recovery of $\mathbf{X}_D$ with $L$ targets in a noiseless setting are determined by $MQ \geq 2L$. In addition, the number of samples per receiver is at least $MK \geq 2L$ where $K$ is the number of Fourier coefficients sampled per receiver and the number of pulses per transmitter is $P \geq 2L$.
\end{theorem}
Theorem~\ref{th:cond2} shows that the number of SUMMeR transmit and receive elements as well as samples $K$ depend only on the number of targets present. These design parameters, therefore, can be substantially lesser than the requirements of a Nyquist MIMO array. Similar results for temporal and Doppler sub-Nyquist radars were obtained in Theorems~\ref{th:min} and \ref{th:cond2}. 

\subsection{Range-Azimuth-Doppler Recovery}
\label{subsec:rad_rec}
To jointly recover the range, azimuth, and Doppler frequency of the targets, we apply the concept of Doppler focusing from Section~\ref{subsec:delaydoprec} to our MIMO setting. Doppler focusing for a specific frequency $\nu$ yields
\begin{eqnarray}
\label{eq:dop_reduced}
   \Phi^{\nu}_{m,q}[k] &=&\sum_{p=0}^{P-1} y_{m,q}^p[k] e^{-j2\pi \nu p \tau} \\
 &=&   \sum_{l=1}^{L} \alpha_l e^{j2\pi \beta_{mq} \vartheta_l}e^{-j\frac{2\pi }{\tau}(k+f_m \tau)\tau_l} \sum_{p=0}^{P-1} e^{j 2 \pi (f^D_l -\nu) p\tau}, \nonumber
\end{eqnarray} 
for $-\frac{N}{2} \leq k \leq -\frac{N}{2}-1$. Following Section~\ref{subsec:delaydoprec}, it holds that 
\begin{equation}
\label{eq:dop_foc}
\sum_{p=0}^{P-1} e^{j 2 \pi (f^D_l -\nu) p\tau} \cong \left\{ \begin{array}{ll} P & |f^D_l -\nu| < \frac{1}{2P\tau}, \\
0 & \text{otherwise}. \end{array} \right.
\end{equation} 
Then, for each focused frequency $\nu$,  (\ref{eq:dop_reduced}) reduces to a 2D problem, which can be solved using CS recovery techniques, as summarized in Algorithm \ref{algo:omp2}. Note that step 1 can be performed using the FFT. In the algorithm description, $\text{vec}(\mathbf{Z})$ concatenates the columns of $\mathbf{Z}^m$, for $0 \leq m \leq M-1$, $\mathbf{e}_t(l)= \left[ (\mathbf{e}_t^0(l))^T \, \cdots \, (\mathbf{e}_t^{M-1}(l))^T \right]^T$
where 
\begin{equation}
\mathbf{e}_t^m(l) = \text{vec}\left( (\mathbf{\bar{B}}^m \otimes \mathbf{A}^m)_{\Lambda_t(l,2)TN+\Lambda_t(l,1)} \left((\mathbf{\bar{F}}^m)^T_{\Lambda_t(l,3)} \right)^T\right),
\end{equation}
with $\Lambda_t(l,i)$ the $(l,i)$th element in the index set $\Lambda_t$ at the $t$th iteration, and $\mathbf{E}_t= [\mathbf{e}_t(1) \, \dots \, \mathbf{e}_t(t)]$.
Once $\mathbf{X}_D$ is recovered, the delays, azimuths, and Dopplers are estimated as
\begin{equation}
\hat{\tau}_l = \frac{\tau \Lambda_L(l,1)}{TN}, \,  \hat{\vartheta}_l = -1+\frac{2 \Lambda_L(l,2)}{TR}, \, \hat{f}_l^D = -\frac{1}{2\tau}+\frac{\Lambda_L(l,3)}{P\tau}.
\end{equation}
Since in real scenarios, targets delays, Dopplers, and azimuths are not necessarily aligned to a grid, a finer grid can be used around detection points on the coarse grid to reduce quantization error. This technique adds a step after support detection in each iteration (step 4 in Algorithm \ref{algo:omp2})\index{sub-Nyquist radar!spatial sub-Nyquist!recovery algorithm}. 

\begin{algorithm}
\caption{Simultaneous sparse 3D recovery based OMP with focusing\index{sub-Nyquist radar!spatial sub-Nyquist!recovery algorithm}\index{sub-Nyquist radar!SUMMeR!recovery algorithm} \cite{cohen2016summer}}\label{algo:omp2} 
\begin{algorithmic}[1]
\qinput Observation matrices $\mathbf{Z}^{m}$, measurement matrices $\mathbf{A}^{m}$, $\mathbf{B}^{m}$, for all $0 \leq m \leq M-1$
\qoutput Index set $\Lambda$ containing the locations of the non zero indices of $\mathbf{X}$, estimate for sparse matrix $\mathbf{\hat{X}}$
\State Perform Doppler focusing for $0 \leq i \leq K-1$ and $0 \leq j \leq Q-1$:
$$
\mathbf{\Phi}^{(m,\nu)}_{i,j} = \sum_{p=0}^{P-1} \mathbf{\mathbf{Z}}^{m}_{i+jK,p} e^{j 2 \pi \nu p \tau}.
$$
\State Initialization: residual $\mathbf{R}_0^{(m,\nu)}=\mathbf{\Phi}^{(m,\nu)}$, index set $\Lambda_0=\emptyset$, $t=1$
\State Project residual onto measurement matrices for $0 \leq p \leq P-1$:
$$
\mathbf{\Psi}^{\nu} =\mathbf{A}^H \mathbf{R}^{\nu} \mathbf{B},
$$
where
$
\mathbf{A} = [ \mathbf{A}^{0^T} \, \mathbf{A}^{1^T} \, \cdots \, \mathbf{A}^{(M-1)^T}]^T,
$
$
\mathbf{B} = [ \mathbf{B}^{0^T} \, \mathbf{B}^{1^T} \, \cdots \, \mathbf{B}^{(M-1)^T}]^T,
$
and $\mathbf{R}^{\nu} = \text{diag} \left( [ \mathbf{R}^{(0,\nu)}_{t-1} \, \cdots \, \mathbf{R}^{(M-1, \nu)}_{t-1}] \right)$ is block diagonal
\State Find the three indices $\lambda_t = [\lambda_t(1) \, \lambda_t(2) \, \lambda_t(3)]$ such that
$$
[\lambda_t(1) \quad \lambda_t(2) \quad \lambda_t(3)] = \text{ arg max}_{i,j,\nu} \left| \mathbf{\Psi}^{\nu}_{i,j} \right|
$$
\State Augment index set $\Lambda_t = \Lambda_t  \bigcup \{ \lambda_t \}$
\State Find the new signal estimate
$$
%\arg \min\limits_{\alpha_1, \dots, \alpha_l} || \text{vec}(\mathbf{Y}) - \sum_{l=1}^t \alpha_l \mathbf{d}_t(l)  ||_2
\mathbf{\hat{\alpha}}=[\hat{\alpha}_{1} \, \dots \, \hat{\alpha}_{t}]^T = ( \mathbf{E}_t^T \mathbf{E}_t )^{-1} \mathbf{E}_t^T \text{vec}(\mathbf{Z}) 
$$
\State Compute new residual
$$
\mathbf{R}_t^{(m,\nu)}= \mathbf{Z}^m- \sum_{l=1}^t \alpha_l e^{j2 \pi \left(-\frac{1}{2}+\frac{ \Lambda_t(l,3)}{P} \right) p} \mathbf{a}^m_{\Lambda_t(l,1)} \left( \mathbf{\bar{b}}^m_{\Lambda_t(l,2)} \right)^T$$
\State If  $t < L$, increment $t$ and return to step 2; otherwise stop
\State Estimated support set $\hat{\Lambda}= \Lambda_L$ \\
Estimated matrix $\mathbf{\hat{X}}_D$: $\left( \Lambda_L(l,2) TN+ \Lambda_L(l,1) ,  \Lambda_L(l,3)\right)$-th component is given by $\hat{\alpha}_l$ while rest of the elements are zero
\end{algorithmic}
\end{algorithm}

\subsection{Multi-Carrier and Cognitive Transmission}
\label{subsec:mc_summer}
\index{sub-Nyquist radar!spatial sub-Nyquist!cognition}\index{MIMO!cognitive}\index{cognitive radar!MIMO}\index{sub-Nyquist radar!SUMMeR!cognition}The frequency bands left vacant can be exploited to increase the system's performance without expanding the total bandwidth of $B_{\text{tot}}=TB_h$. Denote by $\gamma=T/M$ the compression ratio of the number of transmitters. In multi-carrier SUMMeR, every transmit antenna sends $\gamma$ pulses, each belonging to a different frequency band, in one PRI. The total number of user bands is $M\gamma B_h=TB_h$. The $i$th pulse of the $p$th PRI is transmitted at time $i\frac{\tau}{\gamma}+p\tau$, for $0 \leq i < \gamma$ and $0 \leq p \leq P-1$. The samples are then acquired and processed as described above. Besides increasing the detection performance, this method multiplies the Doppler dynamic range by a factor of $\gamma$ with the same Doppler resolution since the CPI, equal to $P\tau$, is unchanged. Preserving the CPI allows to maintain the targets' stationarity.

Cognitive transmission described in Section~\ref{subsec:cograd} can also be extended to a SUMMeR system wherein the spectrum of each of the transmitted waveforms is limited to a few non-overlapping frequency bands while keeping the transmit power per transmitter the same. Cognitive transmission imparts two advantages to the SUMMeR hardware. First, the spatial sub-Nyquist processing of large arrays can be easily designed without replicating the pre-filtering operation for each subband in the hardware. Second, since the total transmit power remains the same, a cognitive signal has more in-band power resulting in an increase in SNR as discussed in Section~\ref{subsubsec:ddr_crr}.

\subsection{Cognitive SUMMeR Hardware Prototype}
\label{subsec:hw_mimo}
\index{sub-Nyquist radar!spatial sub-Nyquist!hardware}\index{cognitive radar!MIMO!hardware}\index{sub-Nyquist radar!SUMMeR!hardware}
 %-----------------------------------------------------------------------------------
 \begin{table}[t]
 \centering
 \caption{Technical characteristics of the cognitive SUMMeR prototype}
 \label{tbl:techmodes}
 \vspace{13pt}
 	\begin{tabular}{ l | c | c | c | c}
 		\hline
          \noalign{\vskip 1pt}    
          	Parameters & Mode 1 & Mode 2 & Mode 3 & Mode 4\\[1pt]
 		\hline
 		\hline
         \noalign{\vskip 1pt}    
 		   	\#Tx, \#Rx & 8,10 & 8,10 & 4,5 & 8,10\\[1pt]
 		   	Element placement & Uniform & Random & Random & Random\\[1pt]
 		   	Equivalent aperture & 8x10 & 8x10 & 8x10 & 20x20\\[1pt]            
             Angular resolution (sine of DoA) & 0.025 & 0.025 & 0.025 & 0.005\\[1pt]
 		\hline
         \noalign{\vskip 1pt}    		            
 	        Range resolution & \multicolumn{4}{c}{1.25 m}\\[1pt]
             Signal bandwidth per Tx & \multicolumn{4}{c}{12 MHz (15 MHz including guard-bands)}\\[1pt]
 		   	Pulse width & \multicolumn{4}{c}{4.2 $\mu$s}\\[1pt]
             Carrier frequency & \multicolumn{4}{c}{10 GHz}\\[1pt]
             Unambiguous range & \multicolumn{4}{c}{15 km}\\[1pt]
             Unambiguous DoA & \multicolumn{4}{c}{180$^{\circ}$ (from -90$^{\circ}$ to 90$^{\circ}$)} \\[1pt]   
         \hline
         \noalign{\vskip 1pt}    		            
             PRI & \multicolumn{4}{c}{100 $\mu$s}\\[1pt]
             Pulses per CPI & \multicolumn{4}{c}{10}\\[1pt]
            Unambiguous Doppler & \multicolumn{4}{c}{from $-75$ m/s to $75$ m/s} \\[1pt]
 		\hline
 		%\hline
 	\end{tabular}
 \end{table}
%-----------------------------------------------------------------------------------
A cognitive SUMMeR prototype was first presented in \cite{mishra2018cognitive}. The system realizes a receiver with a maximum of 8 transmit (Tx) and 10 receive (Rx) antenna elements. A scenario includes modeling of pulse transmission, accurate power loss due to wave propagation in a realistic medium, and interaction of a transmit signal with the target. A large variety of scenarios, consisting of different target parameters, i.e., delays, Doppler frequencies, and amplitudes, and array configurations, i.e., number of transmitters and receivers and antenna locations, can be examined using the prototype. The waveform generator board produces an analog signal corresponding to the synthesized radar environment, which is amplified and routed to the MIMO radar receiver board. The prototype then samples and processes the signal in real-time. The physical array aperture and simulated target response correspond to an X-band ($f_c = 10$ GHz) radar. 

A conventional 8x10 MIMO radar receiver would require simultaneous hardware processing of 80 (or 160 I/Q) data streams. Since a separate sub-Nyquist receiver for each of these 80 channels is expensive, we implement the eight channel analog processing chain for only one receive antenna element and serialize the received signals of all 10 elements through this chain. This approach allows the prototype to implement a number of receivers greater than 10 as the eight-channel hardware only limits the number of transmitters.

If we use the same pre-filtering approach as in Section~\ref{subsec:snrad_demo} for each of the eight channels of our sub-Nyquist MIMO prototype, then the hardware design would need a total of $4\times8=32$ BPFs and ADCs excluding the analog filters to separate transmit channels. We sidestep this requirement by adopting cognitive transmission wherein the analog signal of each channel exists only in certain pre-determined subbands and consequently, a BPF stage is not required. More importantly, for each channel, a single low-rate ADC subsamples this narrow-band signal as long as the subbands are \textit{coset} bands so that they do not alias after sampling \cite{cohen2014channel}. This implementation needs only eight low-rate ADCs, one per channel. Another advantage of this approach is flexibility of the prototype in selecting the Xampling slices. Unlike Section~\ref{subsec:snrad_demo}, the number and spectral locations of slices are not permanently fixed, and they can be changed.

\begin{SCfigure}[50]
%\centering
%\includegraphics[width=1\textwidth]{blockandphoto.pdf}
\includegraphics[width=0.6\textwidth]{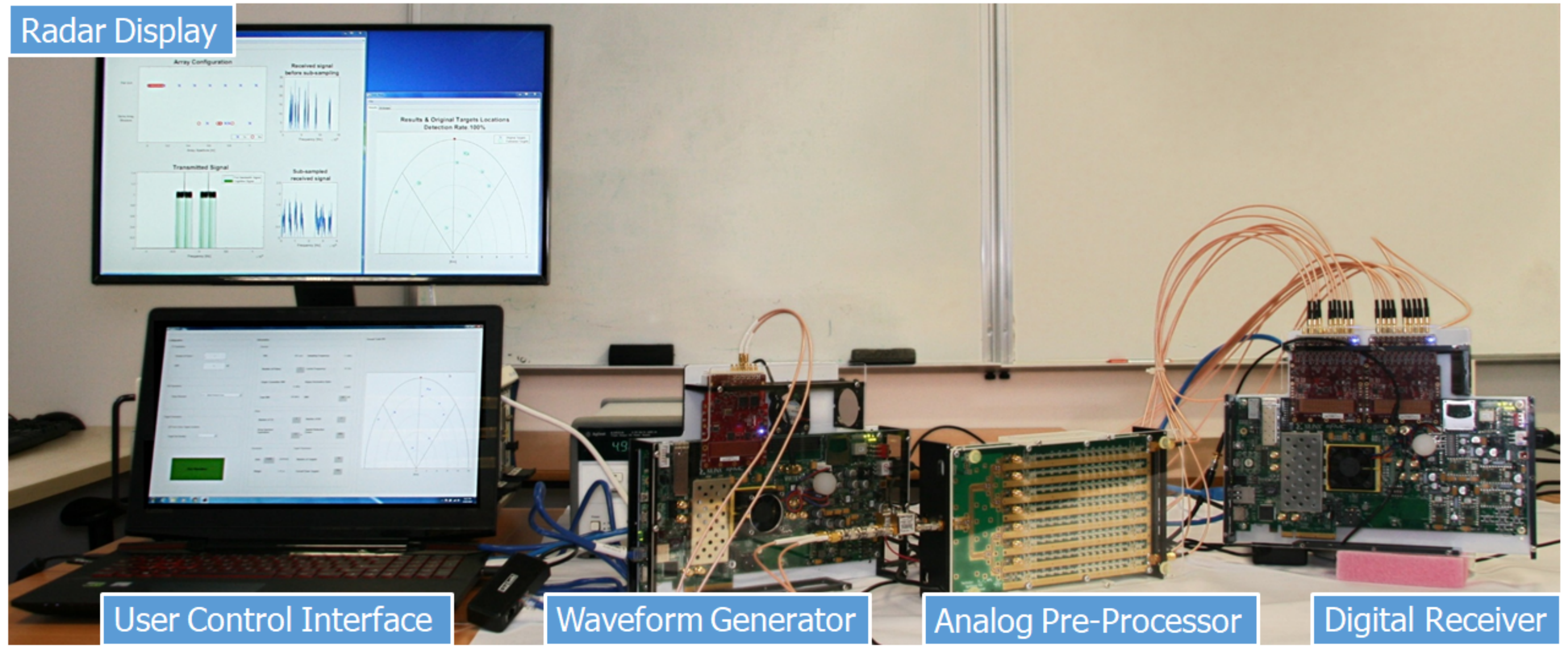}
\caption{Sub-Nyquist MIMO prototype and user interface. The analog pre-processor module consists of two cards mounted on opposite sides of a common chassis \cite{mishra2016cognitive}. \textcopyright 2016 IEEE. Reprinted, with permission, from K. V. Mishra, E. Shoshan, M. Namer, M. Meltsin, D. Cohen, R. Madmoni, S. Dror, R. Ifraimov, and Y. C. Eldar, ``Cognitive sub-Nyquist hardware prototype of a collocated MIMO radar,'' in \textit{International Workshop on Compressed Sensing Theory and its Applications to Radar, Sonar and Remote Sensing}, 2016, pp. 56-60.}
\label{fig:mimoblockphoto}
\end{SCfigure}

\begin{SCfigure}[50]
\includegraphics[width=0.5\textwidth]{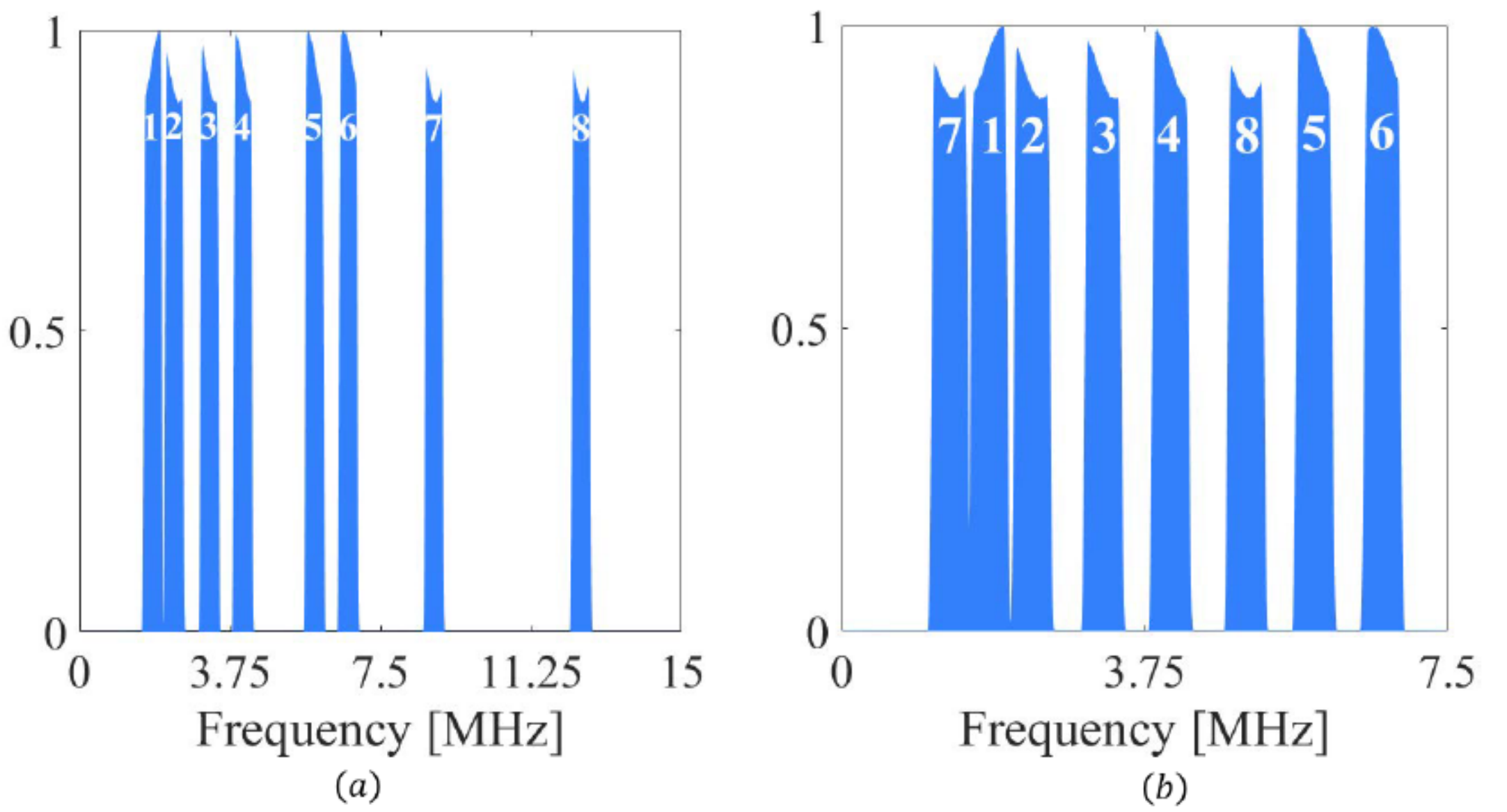}
\caption{The normalized one-sided spectrum of one channel of a given receiver (a) before and (b) after subsampling with a 7.5 MHz ADC. Each of the subbands spans 375 kHz and is marked with a numeric label. In a non-cognitive processing, the signal occupies the entire 15 MHz spectrum before sampling \cite{mishra2016cognitive}. \textcopyright 2016 IEEE. Reprinted, with permission, from K. V. Mishra, E. Shoshan, M. Namer, M. Meltsin, D. Cohen, R. Madmoni, S. Dror, R. Ifraimov, and Y. C. Eldar, ``Cognitive sub-Nyquist hardware prototype of a collocated MIMO radar,'' in \textit{International Workshop on Compressed Sensing Theory and its Applications to Radar, Sonar and Remote Sensing}, 2016, pp. 56-60.}
\label{fig:slicespec}
\end{SCfigure}

Table~\ref{tbl:techmodes} lists detailed technical characteristics of the prototype. The system can be configured to operate in various array configurations or \textit{modes}. Mode 3 and 4 are sub-Nyquist MIMO modes; the hardware switches off the inactive channels and does not sample any data over the corresponding ADCs. Figure~\ref{fig:mimoblockphoto} shows the sub-Nyquist MIMO prototype, user interface and radar display.
%-----------------------------------------------------------------------------------
%\begin{figure}
%\sidecaption
\begin{SCfigure}[50]
\includegraphics[scale=0.35]{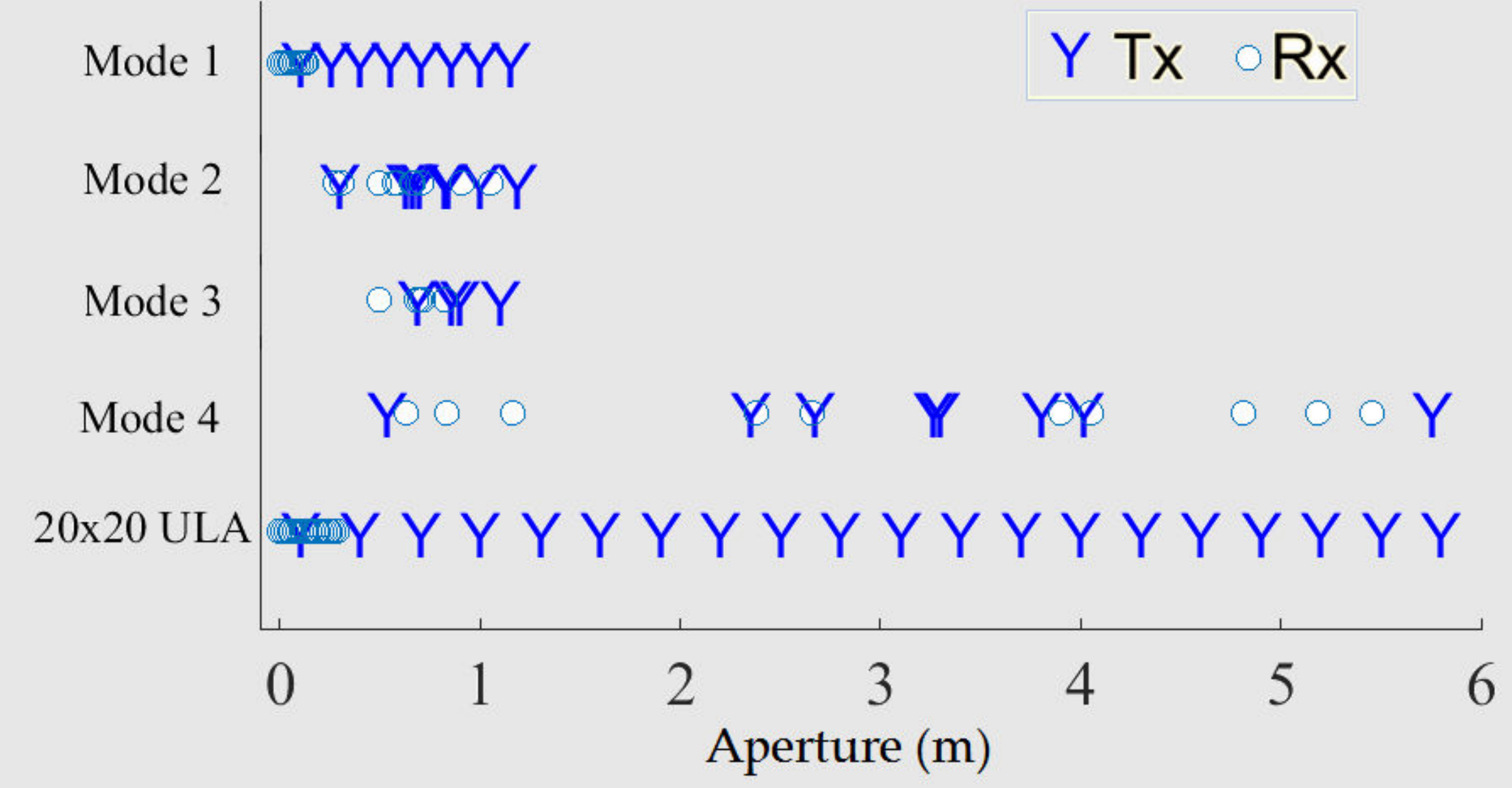}
  \caption{Tx and Rx element locations for the hardware prototype modes over a 6 m antenna aperture. Mode 4's virtual array equivalent is the $20\times20$ ULA \cite{mishra2016cognitive}. \textcopyright 2016 IEEE. Reprinted, with permission, from K. V. Mishra, E. Shoshan, M. Namer, M. Meltsin, D. Cohen, R. Madmoni, S. Dror, R. Ifraimov, and Y. C. Eldar, ``Cognitive sub-Nyquist hardware prototype of a collocated MIMO radar,'' in \textit{International Workshop on Compressed Sensing Theory and its Applications to Radar, Sonar and Remote Sensing}, 2016, pp. 56-60.}
\label{fig:arrays}
\end{SCfigure}
%\end{figure}
%-----------------------------------------------------------------------------------
%-----------------------------------------------------------------------------------
As shown in Fig.~\ref{fig:slicespec}a, the cognitive radar signal occupies only certain subbands in a 15 MHz band. Here, the sliced transmit signal has eight subbands each of width 375 kHz with the frequency range of 1.63-2, 2.16-2.53, 3.05-3.42, 3.88-4.25, 5.66-6.03, 6.51-6.88, 8.64-9.01, and 12.32-12.69 MHz before subsampling.
The total signal bandwidth is $0.375 \times 8 = 3$ MHz. This signal is subsampled at 7.5 MHz and the subbands locations were chosen so that there is no aliasing between different subbands (Fig.~\ref{fig:slicespec}b). A non-cognitive signal would have occupied the entire 15 MHz spectrum requiring a Nyquist sampling rate of 30 MHz. Therefore, the use of cognitive transmission enables spectral sampling reduction by a factor of $4$ ($=30$ MHz$/7.5$ MHz) for each channel. Depending on whether the guard-bands of the non-cognitive transmission are included in the computation or not, the effective signal bandwidth is reduced by a factor of $5$ ($=15$ MHz$/3$ MHz) or $4$ ($=12$ MHz$/3$ MHz) respectively for each channel. Mode 3 has 50\% spatial sampling reduction when compared with Mode 1 or 2. Table~\ref{tbl:cogsum_red} summarizes the reduction of various resources in Mode 3 when compared with Mode 1.
%-----------------------------------------------------------------------------------
%-----------------------------------------------------------------------------------
\begin{table}
\caption{Cognitive SUMMeR Prototype: Comparison of Resource Reduction}
\label{tbl:cogsum_red}       % Give a unique label
%
% Follow this input for your own table layout
%
\begin{tabular}{p{4.0cm}p{1.8cm}p{2.6cm}p{1.0cm}}
\hline\noalign{\smallskip}
Resource & Nyquist Mode 1 & Sub-Nyquist Mode 3 & Reduction\\
\noalign{\smallskip}
\hline
\noalign{\smallskip}
Bandwidth usage per Tx (including guard-bands) & $15$ MHz  & $3$ MHz & $80$\%\\
Bandwidth usage per Tx (excluding guard-bands) & $12$ MHz  & $3$ MHz & $75$\%\\
Temporal sampling rate per channel & $30$ MHz & $7.5$ MHz & $75$\%\\
Spatial sampling rate & $8\times 10$  & $4\times 5$ & $50$\%\\
Tx/Rx hardware channels & $80$  & $20$ & $75$\%\\
\noalign{\smallskip}\hline\noalign{\smallskip}
\end{tabular}
\end{table}
%-----------------------------------------------------------------------------------
\begin{SCfigure}[50]
\centering
\includegraphics[width=0.5\columnwidth]{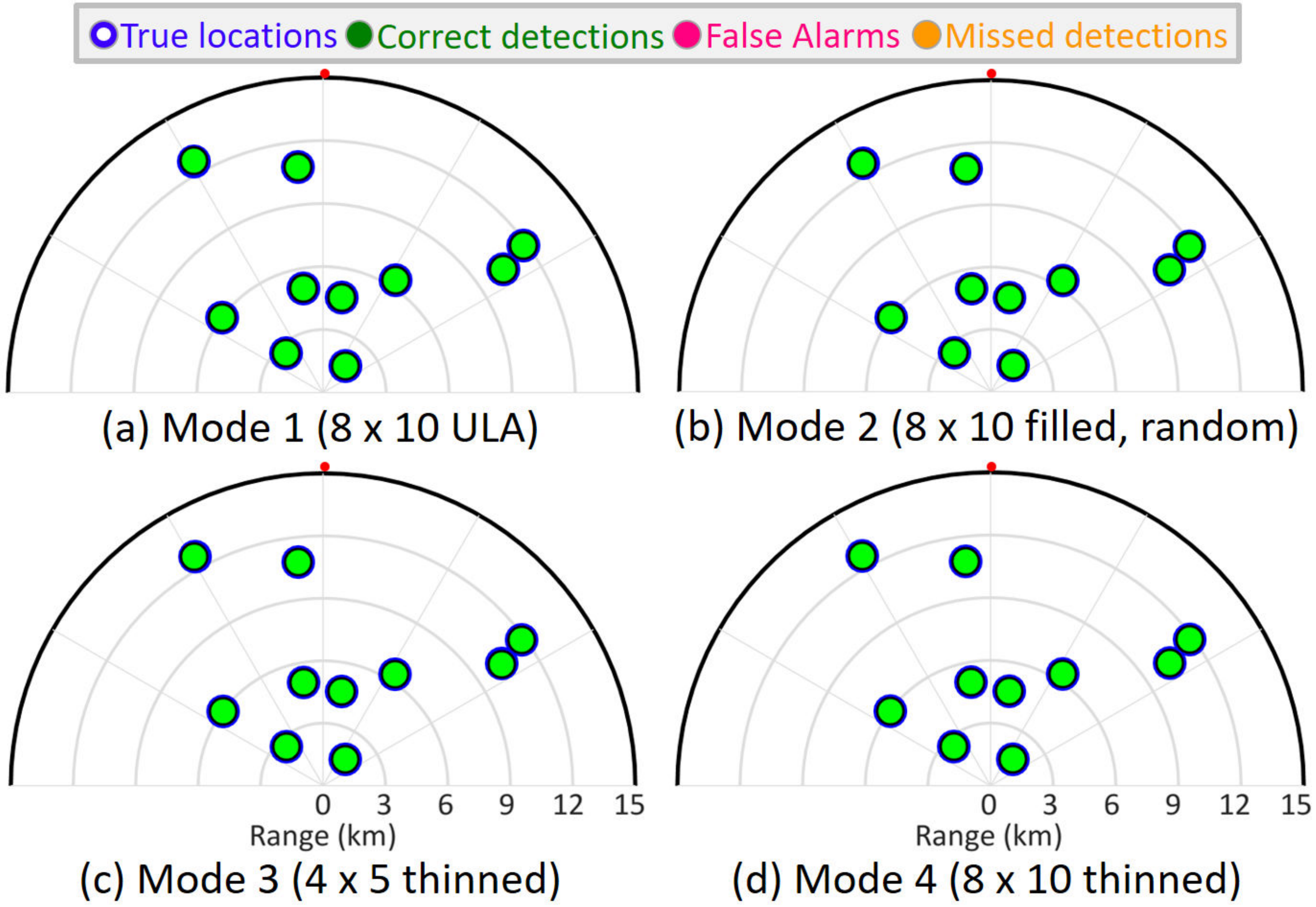}
\caption{Plan Position Indicator (PPI) display of results for Mode 1 and 3. The origin is the location of the radar. The dark dot indicates the north direction relative to the radar. Positive (negative) distances along the horizontal axis correspond to the east (west) of the radar. Similarly, positive (negative) distances along the vertical axis correspond to the north (south) of the radar. The estimated targets are plotted over the ground truth \cite{mishra2016cognitive,cohen2017sub}. \textcopyright 2017 IEEE. Reprinted, with permission, from D. Cohen, K. V. Mishra, D. Cohen, E. Ronen, Y. Grimovich, M. Namer, M. Meltsin, and Y. C. Eldar, ``Sub-Nyquist MIMO radar prototype with Doppler processing,'' in \textit{IEEE Radar Conference}, 2017, pp. 1179-1184.}
\label{fig:sameperf}
\end{SCfigure}
%-----------------------------------------------------------------------------------
\begin{SCfigure}[50]
%\centering
\includegraphics[width=0.5\columnwidth]{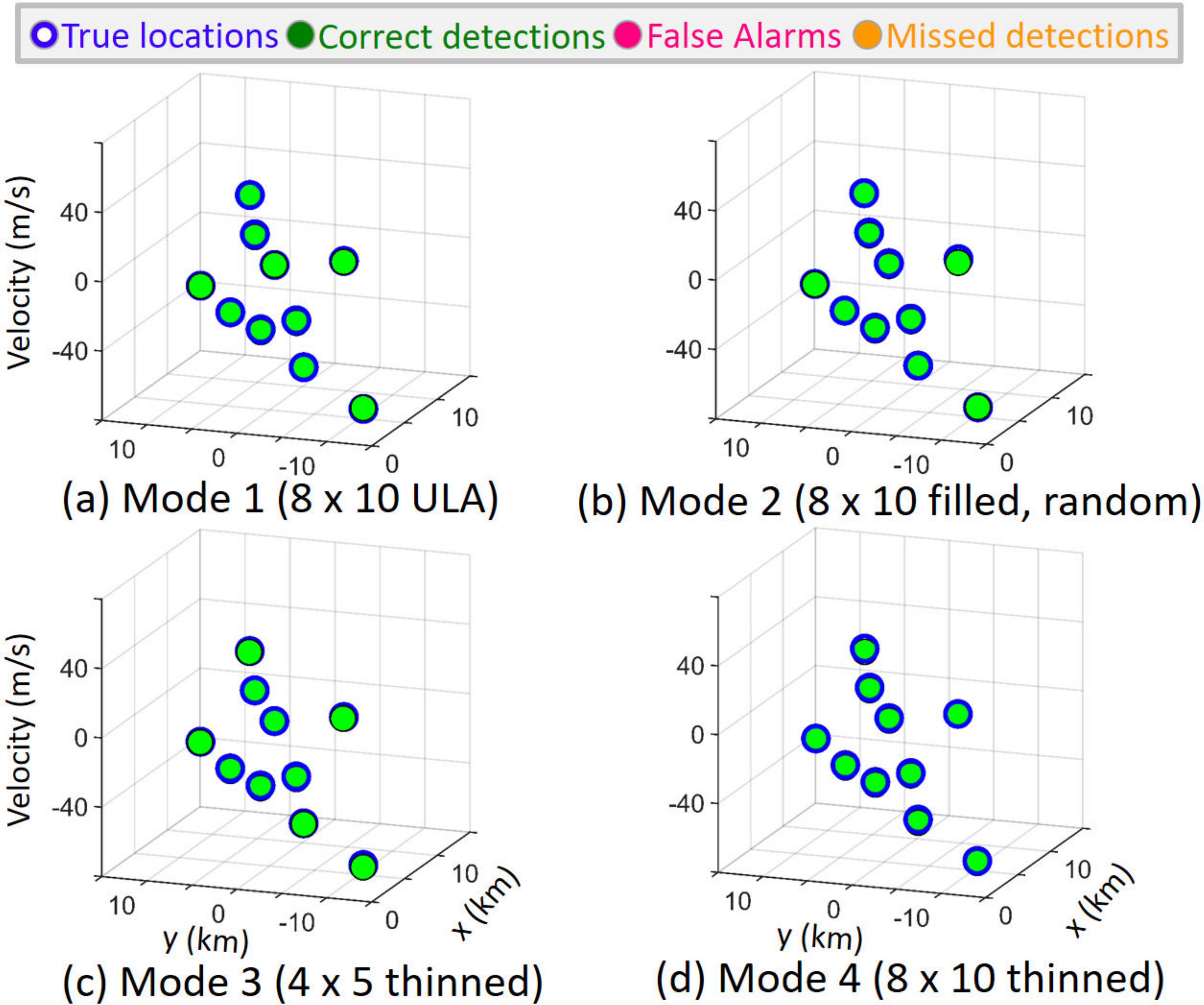}
\caption{Range-Azimuth-Doppler map for the target configurations shown in Fig.~\ref{fig:sameperf}. The lower axes represent Cartesian coordinates of the polar representation of the PPI plots from Fig.~\ref{fig:sameperf}. The vertical axis represents the Doppler spectrum \cite{mishra2016cognitive,cohen2017sub}. \textcopyright 2017 IEEE. Reprinted, with permission, from D. Cohen, K. V. Mishra, D. Cohen, E. Ronen, Y. Grimovich, M. Namer, M. Meltsin, and Y. C. Eldar, ``Sub-Nyquist MIMO radar prototype with Doppler processing,'' in \textit{IEEE Radar Conference}, 2017, pp. 1179-1184.}
\label{fig:sameperf_ddm}
\end{SCfigure}
%-----------------------------------------------------------------------------------

We evaluated the performance of all modes through hardware experiments. We transmitted $P=10$ pulses at a PRF of $100$ $\mu$s and all modes were evaluated against identical target scenarios. In the first experiment, when the angular spacing (in terms of the sine of azimuth) between any two targets was greater than $0.025$ and the signal SNR = $-8$ dB, the recovery performance of the thinned $4\times5$ array in Mode 3 was not worse than Modes 1 and 2. For this experiment, Figs.~\ref{fig:sameperf} and \ref{fig:sameperf_ddm} show the plan position indicator (PPI) plot and range-azimuth-Doppler maps of all the modes. Here, a successful detection (circle with light fill and no boundary) occurs when the estimated target is within one range cell, one azimuth bin and one Doppler bin of the ground truth (circle with dark boundary and no fill); otherwise, the estimated target is labeled as a false alarm (circle with dark fill). When a target remains undetected, we label the ground truth location as a missed detection (circle with hatched fill).

%We next considered a sparse target scene with $L = 10$ targets including two couples of targets with close azimuth dimension, with angular spacing of $0.02$. The SNR of the injected signal was $-5$ dB. Since the angular resolution of Mode 4 is better than the other three modes, all the targets are successfully detected in Mode 4. Mode 1 and 3 produced a false alarm or missed detection as seen in the inset plots of Figs.~\ref{fig:diffperf} and ~\ref{fig:diffperf_ddm}. Mode 2 also shows successful recovery in the sense of our detection criterion. However, relatively better performance of Mode 2 over Modes 1 and 3 is not entirely fortuitous here. Figure ~\ref{fig:arrays} shows that both Tx and Rx array elements in Mode 2 are distributed such that its virtual array is wider than Modes 1 and 3. Thus, the effective angular resolution for Mode 2 could be better than 1 and 3, but still worse than 4.

\begin{SCfigure}[50]
%\centering
\includegraphics[width=0.6\columnwidth]{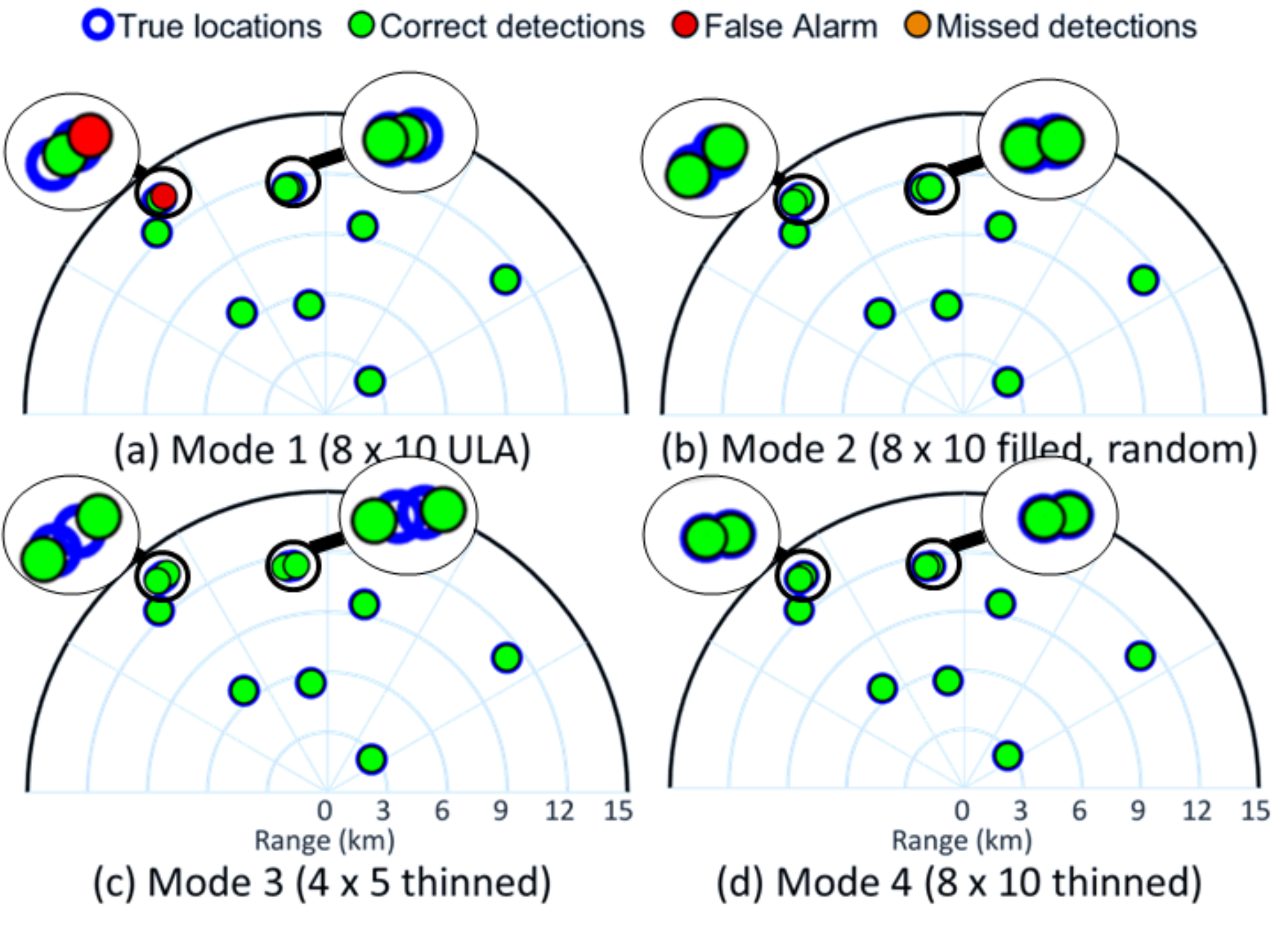}
\caption{Same scenario as in Fig. 5, but only Mode 3 is operating cognitively. All modes have the same overall transmit power per transmitter. The inset plots show the selected region in each PPI display on a magnified scale.}
\label{fig:cogperf}
\end{SCfigure}

\begin{SCfigure}[50]
%\centering
\includegraphics[width=0.6\columnwidth]{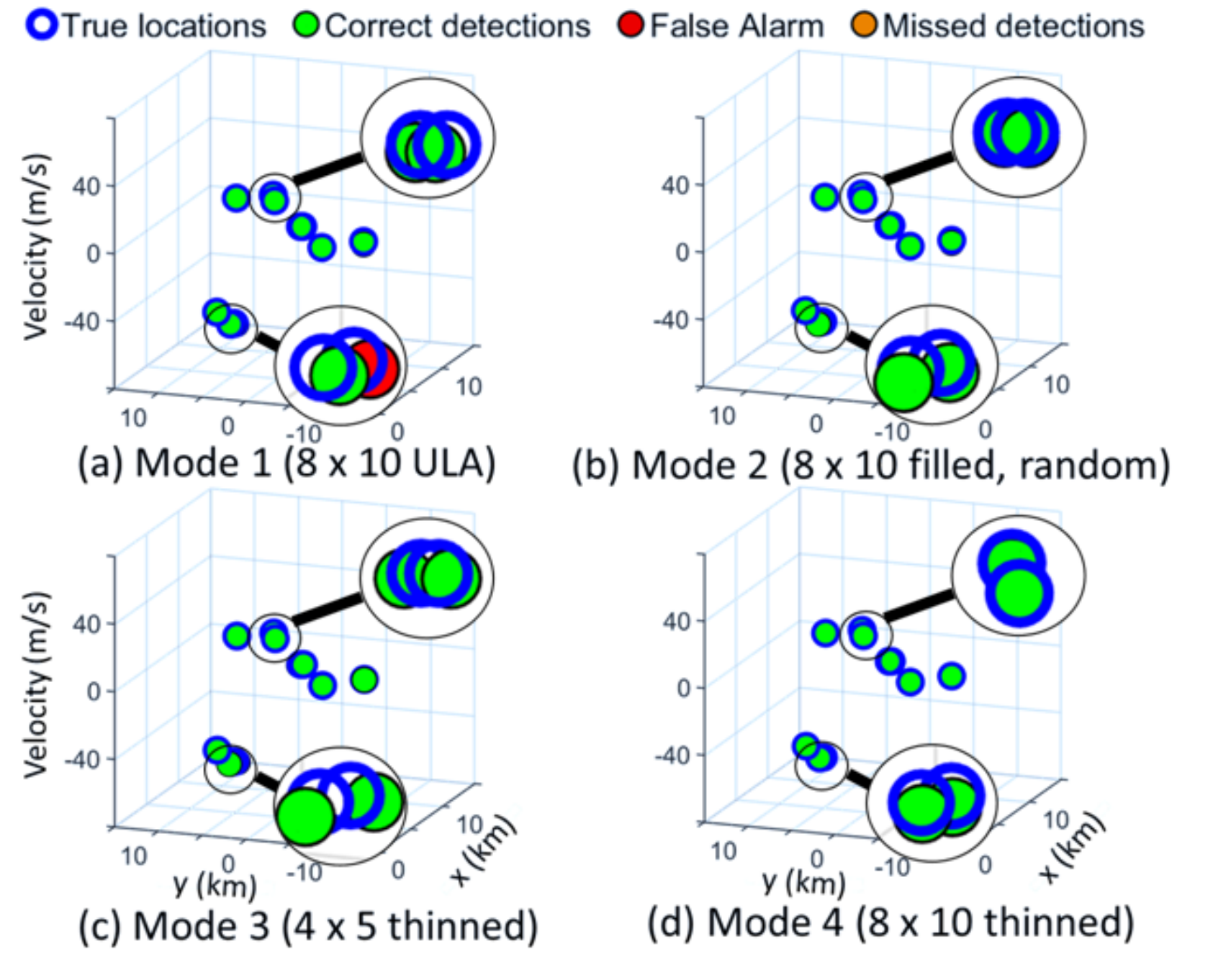}
\caption{Same scenario as in Fig. 6, but only Mode 3 is operating cognitively. All modes have the same overall transmit power per transmitter. The inset plots show the selected region in each map on a magnified scale.}
\label{fig:cogperf_ddm}
\end{SCfigure}

Finally, we considered a high noise scenario with SNR = $-15$ dB. We operated only Mode 3 cognitively and kept all other modes in non-cognitive mode. We noticed that the non-cognitive Nyquist $8\times 10$ Mode 1 array exhibits false alarms while cognitive sub-Nyquist $4\times 5$ Mode 3 array is still able to detect all the targets (Figs.~\ref{fig:cogperf} and ~\ref{fig:cogperf_ddm}), thereby demonstrating robustness to low SNR.
%-----------------------------------------------------------------------------------
%\begin{SCfigure}[50][t]
%%\centering
%\includegraphics[width=0.6\columnwidth]{allmodes_closelyspacedtargets.pdf}
%\caption{Same scenario as in Fig. 5, but for a closely-spaced target scenario. The inset plots show the selected region in each PPI display on a magnified scale.}
%\label{fig:diffperf}
%\end{SCfigure}
%-----------------------------------------------------------------------------------
%\begin{SCfigure}[50][t]
%%\centering
%\includegraphics[width=0.6\columnwidth]{allmodes_closelyspacedtargets_dpplermap.pdf}
%\caption{Same scenario as in Fig. 6, but for a closely-spaced target scenario. The inset plots show the selected region in each map on a magnified scale.}
%\label{fig:diffperf_ddm}
%\end{SCfigure}
%-----------------------------------------------------------------------------------
%-----------------------------------------------------------------------------------
%-----------------------------------------------------------------------------------

\section{Sub-Nyquist SAR}
\label{sec:sar}
\index{sub-Nyquist radar!synthetic aperture radar (SAR)}\index{synthetic aperture radar (SAR)!sub-Nyquist}
Synthetic aperture radar (SAR) and other similar radar techniques were one of the first applications of CS methods (see reviews in \cite{ender2010compressive,cetin2014sparsity}). SAR imaging data are not naturally sparse in the range-time domain. However, they are often sparse in other domains such as wavelet. Our motivation to apply sub-Nyquist methods here is to address the following SAR processing challenge. Among the several algorithms that are available to process SAR data, the range Doppler algorithm (RDA)\index{range Doppler algorithm (RDA)} is most widely used to obtain high-resolution images \cite{barnes2014synthetic}. Its performance is, however, limited by the range cell migration correction (RCMC)\index{range cell migration correction (RCMC)} step which requires oversampled data in order to decouple range and azimuth axes.

Recently, \cite{aberman2017sub} proposed a sub-Nyquist SAR that replaces RDA by a Fourier domain method that achieves non-integer non-constant shifts in the RCMC interpolation via the Fourier series coefficients. This avoids the interpolation step in RCMC and further allows sub-Nyquist sampling following the Fourier-domain analysis presented in previous sections. A similar technique was earlier employed in ultrasound imaging \cite{chernyakova2014fourier} to dramatically reduce sampling and processing rates.

In this section, we present this \textit{Fourier domain RDA} processing as a framework for sub-Nyquist sampling of SAR signals. The first part of the sub-Nyquist algorithm exploits the relationship between the signals before and after RCMC in the Fourier domain. We
show analytically, that a single Fourier coefficient after RCMC can be computed using a small number of Fourier coefficients of the raw data, which translates into low rate sampling as shown in Section~\ref{subsec:delaydoprec}. Having the partial Fourier samples after RCMC, the second part of the algorithm is aimed at solving a 2D CS problem in order to reconstruct the image from the low rate samples. Finally, we then show that cognitive transmission can also be extended to SAR. We end by demonstrating a prototype that we designed and developed to realize concepts of cognitive SAR (CoSAR)\index{sub-Nyquist radar!CoSAR}\index{cognitive radar!synthetic aperture radar (SAR)} \cite{aberman2017sub}.

\subsection{Traditional SAR Processing via RDA}
\label{subsec:rda}
Consider a radar which travels along a path with velocity $\nu$ and transmits a time-limited pulse $h(t)$ at PRI $T$. The pulse has negligible energy at frequencies beyond the bandwidth $B_h/2$. The transmitted pulses are sent from $M$ different locations, $\{\mathbf{x}_m\}_{m=0}^{M-1}$, where $\mathbf{x}_0$ is the origin and $||\mathbf{x}_m-\mathbf{x}_0|| = m|\nu|T$. The
pulses are transmitted into a scene with reflectivity $\sigma(\mathbf{r})$. The received signal,
after coherent demodulation, is given by
\begin{align}
\label{eq:sigmod_sar}
d_m(t) = \int\sigma(\mathbf{r})h(t-2||\mathbf{r}-\mathbf{x}_m||/c)\times w_a(\mathbf{x}_m,\mathbf{r})e^{-j4\pi f_c||\mathbf{r}-\mathbf{x}_m||/c}d\mathbf{r},
\end{align}
where $||\mathbf{r}-\mathbf{x}_m||$ is the distance from the radar to a scatter point and $w_a(\mathbf{x}_m,\mathbf{r})$ is the antenna beam pattern which varies depending on
the SAR operation mode \cite{barnes2014synthetic}. The main goal of SAR data processing is to construct the scene's reflectivity map, $\sigma(\mathbf{r})$, from the raw data. The reflected signal $d_m(t)$ at a point $m$ requires to be sampled at least at the bandwidth $B_h$ as per the Nyquist sampling theorem. The resulting discrete-time signal is $d[n, m] = d_m(nTs)$, with $0 \le n < N = \lfloor T f_s\rfloor$, where $f_s = 1/T_s$ is the sampling rate.

RDA processing consists of the following steps. First the sampled raw data is compressed in the range dimension:
\begin{align}
\label{eq:rc}
s[n,m] = d[n,m]\ast h^{\ast}[-n],
\end{align}
where $h[n]$ is the sampled transmit signal. This data is then transformed to the range-Doppler domain using DFT along the azimuth:
\begin{align}
\label{eq:rct}
S[n,k] = \sum\limits_{m=0}^{M-1}s[n,m]e^{-j2\pi km/M}.
\end{align}
RCMC is applied assuming a far-field approximation. The purpose of RCMC is to compensate for the effect of range cell migration due to the varied satellite-scatterer distance and to correct the hyperbolic behavior of the target trajectories. The RCMC operator can be written as
\begin{align}
\label{eq:rcmc}
\tilde{C}[n,k] = S[n+nak^2,k].
\end{align}
For every Doppler frequency $k$, the range axis is scaled by $1 + ak^2$. The value of $a$ is predetermined depending on the observation mode. For example, in stripmap SAR, $a = \frac{\lambda^2}{8|\nu|T^2M^2}$. This range-variant shift requires values which fall outside the discrete grid. A MF then achieves compression in azimuth via
\begin{align}
\label{eq:ac}
Y[n,k] = \tilde{C}[n,k]e^{-j\pi\frac{k^2}{K_a[n]}},
\end{align}
where $K_a[n]$ is the range dependent azimuth chirp rate. Finally, an inverse DFT in the azimuth direction yields the focused image:
\begin{align}
\label{eq:idft_az}
I[n,m] = \frac{1}{M}\sum\limits_{k=0}^{M-1}Y[n,k]e^{j2\pi mk/M}.
\end{align}

There are two ways to implement RCMC: In the first option, RCMC is performed by range interpolation in the Range-Doppler domain. However, this interpolation is time-consuming and computationally demanding. The second approach involves the assumption that the range cell migration is range invariant, at least over a finite range block. In this case, RCMC is implemented using a DFT, linear phase multiply, and inverse DFT per block. However, this implementation has the disadvantage that blocks should overlap in range, and the efficiency gain may not be worth the added complexity.
\subsection{Fourier domain RDA and sub-Nyquist SAR}
\label{subsec:frda}
\index{sub-Nyquist radar!synthetic aperture radar (SAR)!algorithm}\index{synthetic aperture radar (SAR)!sub-Nyquist!algorithm}\index{range Doppler algorithm!Fourier-domain}
In this section, we introduce a new RDA processing technique implemented in frequency using the Fourier series coefficients of the raw data. This paves the way for substantial reduction in the number of samples in the time-domain interpolation needed to obtain the same image quality and without any assumptions on the signal structure or the invariance of range blocks.

We begin with the continuous version of (\ref{eq:rcmc}):
\begin{align}
\label{eq:rcmc_cont}
C_k(t) = S_k(t(1+ak^2)),
\end{align}
where $S_k(nT_s)=S[n,k]$. The Fourier series coefficients of $C_k(t)$ over the interval $[0, T)$ can be expressed as \cite{aberman2017sub}
\begin{align}
\label{eq:rcmc_f1}
C_k[l] = \frac{1}{T}\int\limits_{0}^{T}S_k(t)q_{k,l}(t),
\end{align}
where $q_{k,l}(t)$ approximates the scaling operation in (\ref{eq:rcmc}). The Fourier coefficients of the continuous-time signals $S_k(t)$ and $q_{k,l}(t)$ are, respectively, $S_k[n]$ and
\begin{align}
Q_{k,l}[n] = \frac{1}{1+ak^2}e^{-j\pi(n+\frac{1}{1+ak^2})}\text{sinc}\left(n + \frac{l}{1+ak^2}\right).
\end{align}
It can be shown that most of the energy of the set $Q_{k,l}[n]$ is concentrated around a specific component $n_{k,l}$.  Thus, for every Doppler frequency $k$, the Fourier series coefficients of the scaled signal, $C_k(t)$, can be calculated as a linear combination of a local
choice of Fourier series coefficients of $S_k(t)$
\begin{align}
\label{eq:rcmc_f}
C_k[l] = \sum\limits_{n\in\nu(k,l)}S_k[n]Q_{k,l}[-n],
\end{align}
where $\nu(k,l)$ is the set of indices that dictate the decay property of $Q_{k,l}[n]$.

Assuming the Fourier series coefficients $D_m[l]$ of the raw data $d_m(t)$ can be acquired directly, the range compression is achieved in the Fourier domain as
\begin{align}
\label{eq:rc_f}
\tilde{D}_m[l] = TD_m[l]H^{\ast}[l],
\end{align}
where $H[l]$ are the Fourier series coefficients of the transmitted pulse $h(t)$. Applying azimuth DFT gives $S_k[l]$ that can be used in (\ref{eq:rcmc_f}) to perform Fourier domain RCMC. The inverse DFT on the coefficients $C_k[l]$ provides the corrected sampled signal after RCMC. One could then proceed with the remaining steps of RDA, i.e., (\ref{eq:ac}) and (\ref{eq:idft_az}), to complete the processing.

The number of Fourier coefficients required can be further reduced if a basis (e.g. wavelet) is found in which the desired image is sparse. Then, the relationship between $C_k[l]$ and the raw data samples $D_m[l]$ can be exploited to solve for the coefficients in the sparse basis using fewer Fourier coefficients. 
%In the Fourier domain RDA, further rate reduction is possible. Define $\mathbf{C} = \{c_k[l]\}_{0\le k<M}^{l \in \beta_k}$ as the partial Fourier coefficients matrix of the corrected signals. The rest of the RDA processing steps can be written in matrix form as
%\begin{align}
%\label{eq:snsar}
%\mathbf{C} = \mathbf{F}^s[\mathbf{B}^{\ast}\odot(\mathbf{IF_M})],
%\end{align}
%where $\mathbf{F}^s=\{\frac{1}{T}e^{-j2\pi lk/T}T_s\}$ is a sampled artial Fourier series coefficients operator, $\mathbf{B}=\{e^{-j\pi\frac{k^2}{K_a[n]}}\}$ is the azimuth compression matrix, $\mathbf{F_M}$ is $M\times M$ DFT matrix and $\mathbf{I}$ is the desired image. In case $\mathbf{I}$ is sparse in some basis, extracting $\mathbf{I}$ from few rows of $\mathbf{C}$ in (\ref{eq:snsar}) in a sub-Nyquist framework by solving the following optimization problem:
%\begin{align}
%\label{eq:snsar_opt1}
% & \underset{\mathbf{I}}{\text{minimize}}\;\; ||\Psi(\mathbf{I})||_1 \nonumber\\
% & \text{subject to} \;\; ||\mathbf{C}_p - \mathbf{F}^s[\mathbf{B}^{\ast}\odot(\mathbf{IF_M})]||^2_{\mathcal{F}} < \epsilon,
%\end{align}
%where $\mathbf{C}_p$ and $\mathbf{F}^s_p$
%are row-undersampled versions of $\mathbf{C}$ and $\mathbf{F}^s$, respectively; $\Psi$ is a sparsifying transform operator; and $\epsilon$ controls the fidelity of the reconstruction to the measured data. 
In \cite{aberman2017sub}, it was suggested to modify fast iterative shrinkage-thresholding algorithm (FISTA) to solve this problem %using Algorithm~\ref{algo:sarfista} which 
and achieve full data reconstruction from the partial measurements with reasonable computational load. %This leads to a sub-Nyquist sampling and processing of SAR signals that also benefits from a reasonable computational load.
%\begin{algorithm}
%\caption{SAR FISTA for sub-Nyquist sampling \cite{aberman2017sub}}\label{algo:sarfista} 
%\begin{algorithmic}[1]
%\qinput Xamples $\mathbf{D}_p = \{d_m[l]\}_{0\le m\le M}^{l\in {\kappa}}$, Measurement matrices $\mathbf{F}^{s}_p$,, $\mathbf{B}$, and $\mathbf{F}_M$
%\qoutput Estimate of sparse coefficients of SAR image, $\mathbf{X}$, such that $\mathbf{I} = \Psi^{-1}(\hat{\mathbf{X}})$
%\State Initialization: $\mathbf{C}_p = \{c_k[l]\}_{0\le k\le M}^{l\in\kappa} \leftarrow \mathbf{D}_p$
%Initialize: $\mathbf{X}^0 = \mathbf{0}$, $\mathbf{X}^1 = \mathbf{0}$, $t_0 = 1$, $t_1 = 1$, $k = 1$, $\lambda_1,\beta\in(0,1)$, $\bar{\lambda}>0$
%\State \textbf{while} not converged \textbf{do}
%\State $\mathbf{Z}^k = \mathbf{X}^k + \frac{t_{k-1}-1}{t_k} (\mathbf{X}^k- \mathbf{X}^{k-1})$
%\State $\mathbf{U}^k = \mathbf{Z}^k - \frac{1}{L_f} \nabla\mathbf{F}(\Psi^{-1}(\hat{\mathbf{X}}))$
%\State Soft thresholding: $\mathbf{X}^{k+1} = \text{soft}\left(\mathbf{U}^k, \frac{\lambda_k}{L_f} \right)$
%\State $t_{k+1} = \frac{1+\sqrt{4t_k^2+1}}{2}$ ($L_f$ is the Lipshitz constant of $\nabla \mathbf{F_M}(I)$)
%\State $\lambda_{k+1} = \text{max}(\beta\lambda_k,\bar{\lambda})$
%\State k=k+1
%\State \textbf{end while}
%$\hat{\mathbf{X}}=\mathbf{X}$
%\end{algorithmic}
%\end{algorithm}
Similar to cognitive pulse Doppler radar (Section~\ref{subsec:cograd}) and cognitive SUMMeR (Section~\ref{subsec:hw_mimo}), sub-Nyquist SAR systems can also be modified to fit cognitive radar requirements and allow for dynamic transmission and reception of several narrow frequency bands. We present the hardware prototype of such a system in the next subsection.

\subsection{Hardware Prototype}
\label{subsec:cosar_hw}
\begin{SCfigure}[50]
%\centering
\includegraphics[width=0.5\columnwidth]{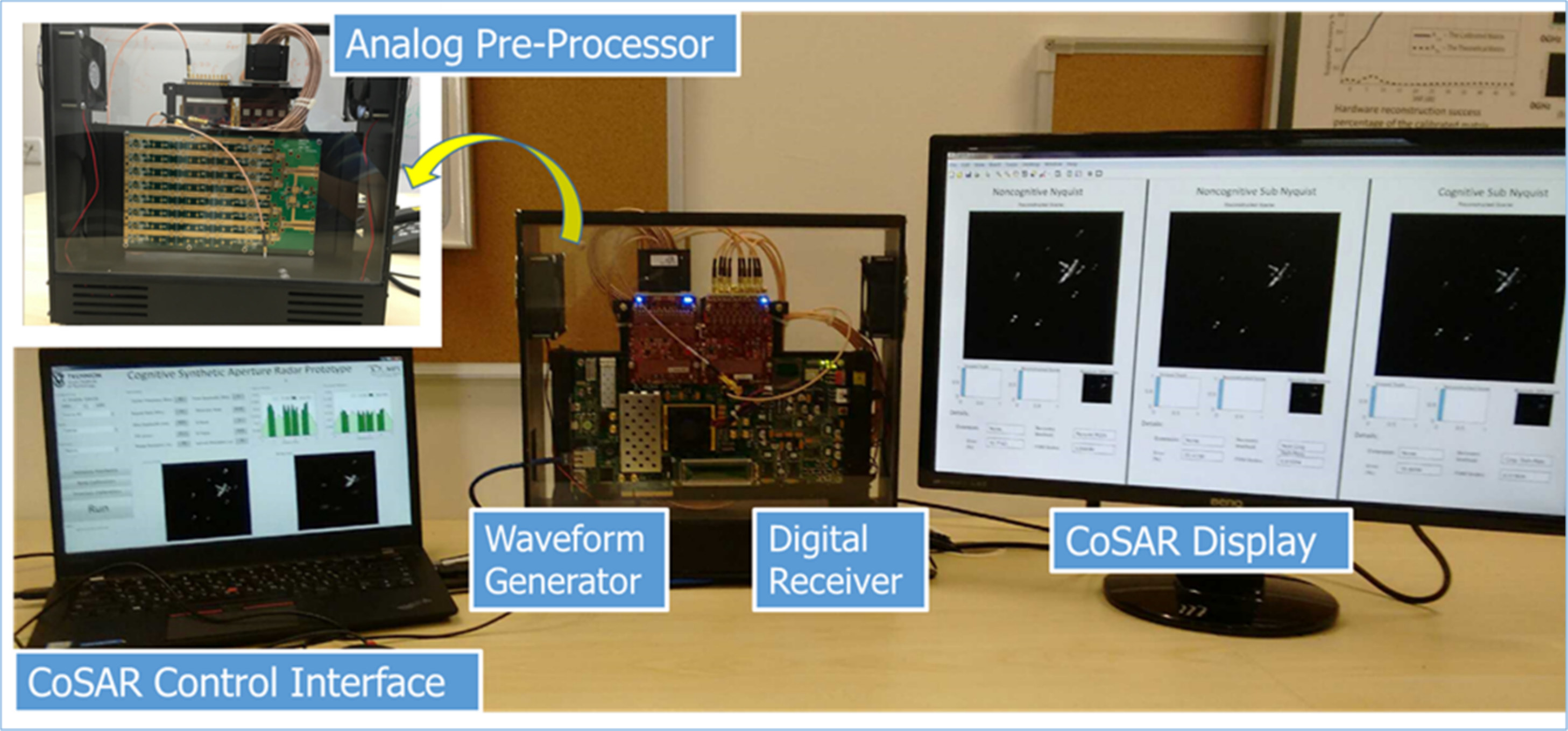}
\caption{Cognitive SAR (CoSAR) prototype and (inset) analog pre-processor.}
\label{fig:cosar_setup}
\end{SCfigure}

\begin{SCfigure}[50]
%\centering
\includegraphics[width=0.5\columnwidth]{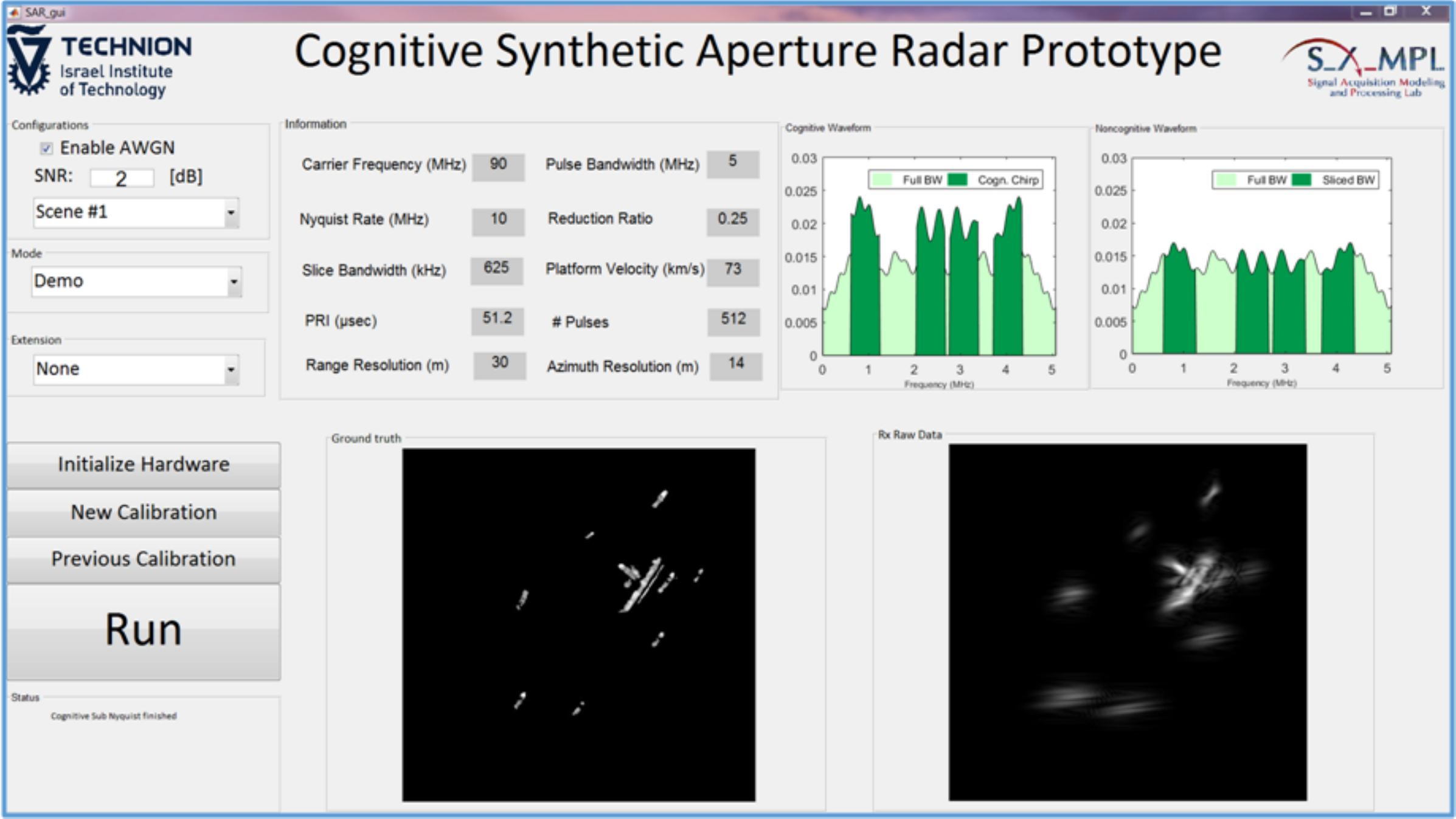}
\caption{CoSAR GUI showing the cognitive and non-cognitive chirp waveforms along with the sampled subbands at top right.}
\label{fig:cosar_gui}
\end{SCfigure}

\index{sub-Nyquist radar!synthetic aperture radar (SAR)!hardware}\index{synthetic aperture radar (SAR)!sub-Nyquist}\index{sub-Nyquist radar!CoSAR!hardware}\index{cognitive radar!synthetic aperture radar (SAR)!hardware}
We designed and developed a hardware prototype of a CoSAR system and evaluated Fourier domain RDA processing in real-time. Figure~\ref{fig:cosar_setup} shows the entire set up. The PRI is $51.2$ $\mu$s and carrier frequency of the signal is $90$ MHz. A control interface (Fig.~\ref{fig:cosar_gui}) activates the prototype which generates the desired I/Q signal and feeds it to the analog pre-processor (inset). The analog pre-processor filters have $30$ dB stopband attenuation in order to filter out interference from neighboring channels. The digital receiver obtains and processes samples at low-rates. The processed image is then shown on the radar display. We used a $5$ MHz cognitive chirp signal whose only $4$ narrow subbands of $625$ kHz bandwidth were sampled and processed by the digital receiver. The Xampling and RCMC are performed at $1/4$th and $1/8$th of the Nyquist rate, respectively.

\begin{SCfigure}[50]
%\centering
\includegraphics[width=0.6\columnwidth]{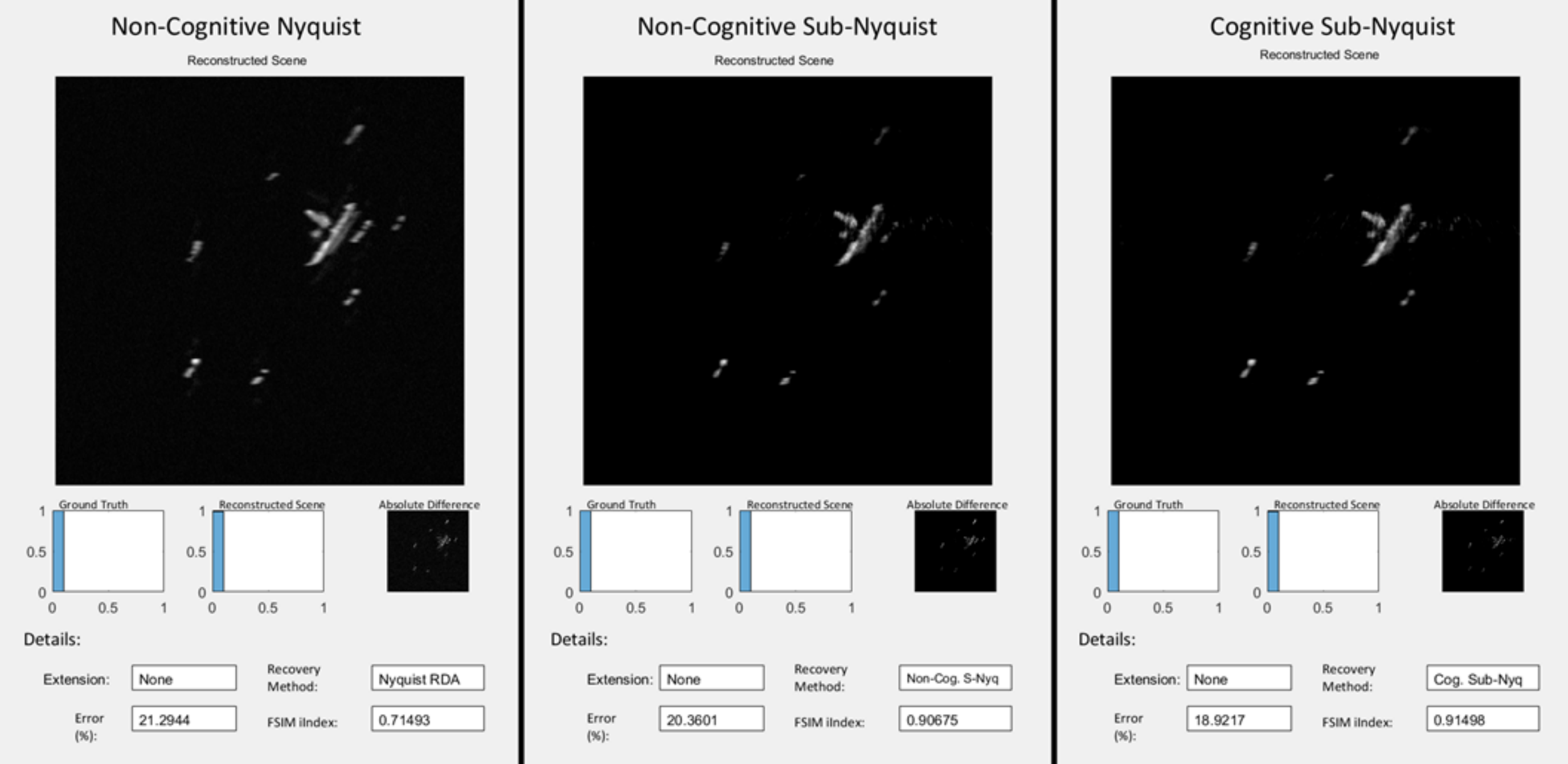}
\caption{Comparison of prototype outputs for an image of a ship.}
\label{fig:cosar_res}
\end{SCfigure}

Similar to the cognitive SUMMeR system, our CoSAR prototype can operate in both cognitive and non-cognitive modes. Figure~\ref{fig:cosar_res} shows results of these modes at Nyquist and sub-Nyquist sampling rates at SNR = $2$dB. The range and cross-range (azimuth) resolutions are $30$ and $10$ m, respectively. When compared with the Nyquist rate of $10$ MHz, the combined sampling rate of the four slices is $2.5$ MHz leading to reduction of rate by $75\%$. We note that CoSAR reconstruction exhibits smaller error than the non-cognitive Nyquist processing in low SNR scenarios despite sampling at a significantly reduced sampling rate. Further, the prototype demonstrates operation of SAR using narrow subbands that can be adaptively changed. This opens up the possibility of spectral coexistence of SAR with other satellite-borne services. 
%-----------------------------------------------------------------------------------
%-----------------------------------------------------------------------------------
%-----------------------------------------------------------------------------------
%-----------------------------------------------------------------------------------

\section{Summary}
\label{sec:summ}
In this chapter, we reviewed sub-Nyquist radar principles, algorithms, prototypes, and applications. Our focus was on pulse-Doppler systems for which sub-Nyquist processing can be individually applied to temporal, Doppler, and spatial domains. Our approach has distinct advantages over several past CS-based designs. The proposed sub-Nyquist radar receivers perform low rate sampling and processing which can be implemented with simple hardware, impose no restrictions on the transmitter, use a CS dictionary that does not scale up with the problem size, and exhibit robustness to clutter and noise. 

We presented colocated MIMO radar as an application where joint spatio-temporal sub-Nyquist processing leads to reduction in antenna elements and savings in signal bandwidth. In SAR imaging, sub-Nyquist processing in the Fourier domain leads to sampling rate reduction without compromising high-quality and high-resolution imaging. We demonstrated that sub-Nyquist receivers lead to the feasibility of cognitive radar which transmits thinned spectrum signals. This development was significant in making the spectral coexistence of radar with a communication service possible. We also extended cognition ideas based on sub-Nyquist processing to MIMO and SAR systems.

Most importantly, we emphasized that sub-Nyquist radars are realizable in hardware for each of the systems described in this chapter. The hardware prototypes were in-house and custom-made using many off-the-shelf components. The systems operate in real-time and their performance is robust to high noise and clutter. We believe that such practical implementations pave the way to delivering the promise of reduced-rate processing in radar remote sensing.

% bibliography
\renewcommand{\refname}{Bibliography}
  \bibliographystyle{IEEEtran}
  %\label{refs}
  \bibliography{main}

\printindex

\end{document}